\documentclass{ws-ijmpa}
\usepackage[super,compress]{cite}
\usepackage{graphicx}
\begin{document}
\markboth{Authors' Names}{Instructions for typing manuscripts (paper's title)}

\renewcommand \baselinestretch{1.30}
\newcommand{\BE}{\begin{equation}}
\newcommand{\EE}{\end{equation}}
\newcommand{\LP}{\lambda \Phi^4}
\newcommand{\dl}{\delta^{(3)}({\bf r})}
\newcommand{\del}{{\mbox{\boldmath $\nabla$}}}
\newcommand{\half}{{\scriptstyle{\frac{1}{2}}}}
\newcommand{\vol}{{\sf V}}
\newcommand{\num}{{\sf N}}

%
\catchline{}{}{}{}{}
%

\title{Michelson-Morley Experiments: at the crossroads of Relativity, Cosmology and Quantum Physics
}

\author{Maurizio Consoli}

\address{Istituto Nazionale di Fisica Nucleare, Sezione di Catania\\ Via S.Sofia 64, 95123, Catania, Italy\\
maurizio.consoli@ct.infn.it}

\author{Alessandro Pluchino}

\address{Dipartimento di Fisica e Astronomia 'E.Majorana', University of Catania\\and Istituto Nazionale di Fisica Nucleare, Sezione di Catania\\Via S.Sofia 64, 95123, Catania, Italy\\alessandro.pluchino@ct.infn.it}

\maketitle


\begin{abstract}
Today, the original Michelson-Morley experiment and its early repetitions at the beginning
of the 20th century are considered as a venerable historical chapter
for which, at least from a physical point of view, there is nothing
more to refine or clarify. The emphasis is now on the modern
versions of these experiments, with lasers stabilized by optical
cavities, that, apparently, have improved by many orders of
magnitude on the limits placed by those original measurements.
Though, in those old experiments light was propagating in gaseous
systems (air or helium at atmospheric pressure) while now, in modern
experiments, light propagates in a high vacuum or inside solid
dielectrics. Therefore, in principle, the difference might not
depend on the technological progress only but also on the different
media that are tested by preventing a straightforward comparison.
Starting from this observation, one can formulate a new theoretical
scheme where the tiny, irregular residuals observed so far, from
Michelson-Morley to the present experiments with optical resonators,
point consistently toward the long sought preferred reference frame
tight to the CMB. The existence of this scheme, while challenging
the traditional `null interpretation', presented in all textbooks
and specialized reviews as a self-evident scientific truth, further
emphasizes the central role of these experiments for Relativity,
Cosmology and Quantum Physics.

\end{abstract}



\section{Introduction}

From the very beginning there are two interpretations of Relativity:
Einstein's Special Relativity  \cite{einstein} and the `Lorentzian'
formulation \cite{lorentz}. Apart
from all historical aspects, the difference could simply be phrased
as follows. In a Lorentzian approach, the relativistic effects
originate from the individual motion of each observer S', S''...with
respect to some preferred reference frame $\Sigma$, a convenient
redefinition of Lorentz' ether. Instead, according to Einstein,
eliminating the concept of the ether leads to interpret the same
effects as consequences of the {\it relative} motion of each pair of
observers S' and S''. This is possible because the basic
quantitative ingredients, namely Lorentz Transformations, have a
crucial group structure and are the same in both formulations. In
the case of one-dimensional motion \footnote{We ignore here the
subtleties related to the Thomas-Wigner spatial rotation which is
introduced when considering two Lorentz transformations along
different directions, see e.g. \cite{ungar, costella, kanevisser}.},
an intuitive representation is given in Fig.1.

\begin{figure}[h]
\begin{center}
\includegraphics[width=10.0 cm]{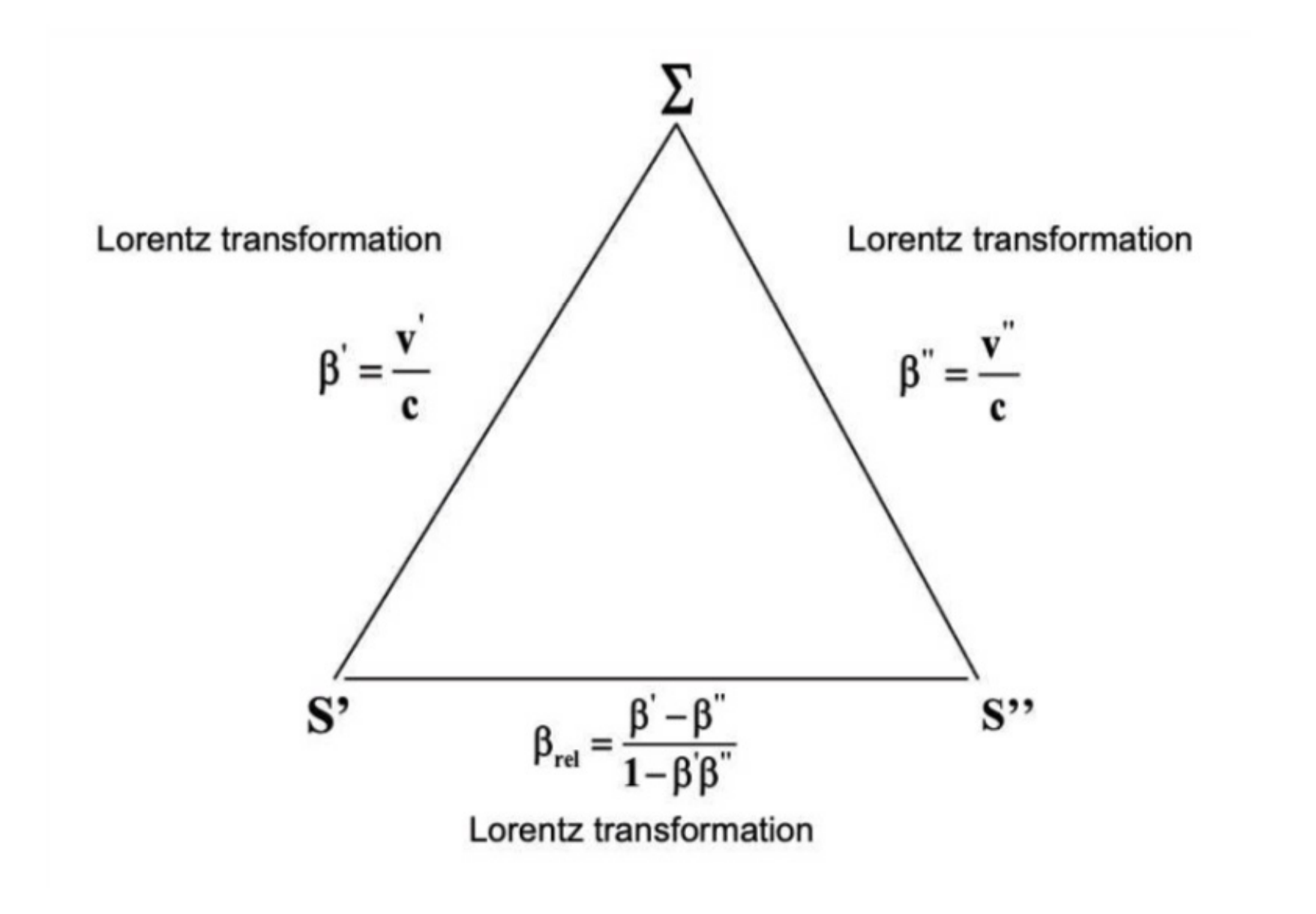}
\end{center}
\caption{\it An intuitive representation of the two interpretations
of Relativity.} 
\label{triangle}
\end{figure}

For this reason, it has been generally assumed that there is a
substantial phenomenological equivalence of the two formulations.
This point of view was, for instance, already clearly expressed by
Ehrenfest in his lecture `On the crisis of the light ether
hypothesis' (Leyden, December 1912) as follows: ``So, we see that
the ether-less theory of Einstein demands exactly the same here as
the ether theory of Lorentz. It is, in fact, because of this
circumstance, that according to Einstein's theory an observer must
observe exactly the same contractions, changes of rate, etc. in the
measuring rods, clocks, etc. moving with respect to him as in the
Lorentzian theory. And let it be said here right away and in all
generality. As a matter of principle, there is no experimentum
crucis between the two theories''. Therefore, by assuming that, in a
Lorentzian perspective, the motion with respect to $\Sigma$ could not be
detected, the usual attitude was to consider the difference between
the two interpretations as a philosophical problem.

However, it was emphasized by Bell \cite{Bell} that adopting the Lorentzian point of view
could be crucial to reconcile hypothetical faster-than-light signals
with causality, as with the apparent non-local aspects of the
Quantum Theory. Indeed, if all reference frames are placed on the
same footing, as in Special Relativity, how to decide of the time
ordering of two events A and B along the world line of a
hypothetical effect propagating with speed $> c$? This ordering can
be different in different frames, because in some frame $S'$ one
could find $t'_A
> t'_B$ and in some other frame $S''$ the opposite $t''_B
> t''_A$. This causal paradox, which is the main reason why superluminal signals are
not believed to exist, disappears in a Lorentzian formulation where
the different views of the two observers become a sort of optical
illusion, like an aberration\footnote{If $S'$ is connected to $\Sigma$ by a Lorentz transformation with parameter $\beta'=v'/c$, by the inverse transformation we can find the time coordinates in $\Sigma$ starting from $x'_A$, $ct'_A$, $x'_B$ and $ct'_B$, namely $cT_A= \gamma'(\beta' x'_A+ ct'_A)$ and $cT_B= \gamma'(\beta' x'_B+ ct'_B)$, with $1/\gamma'= \sqrt{1- (\beta')^2}$. Analogously, for $S''$ and parameter $\beta''=v''/c$, we will find the same values, i.e.  $cT_A= \gamma''(\beta''x''_A+ ct''_A)$ and $cT_B= \gamma''(\beta''x''_B+ ct''_B)$,  with now $1/\gamma''= \sqrt{1- (\beta'')^2}$. Thus no ambiguity is possible, either $T_A > T_B$ or viceversa so that the view in the preferred $\Sigma-$frame becomes the relevant one to decide on causal effects.}.

But the mere logical possibility of $\Sigma$ is not
enough. For a full resolution of the paradox, the $\Sigma-$ frame
should show up through a determination of the kinematic parameters
$\beta'$, $\beta''$... Thus, we arrive to the main point of this
article: the prejudice that, even in a Lorentzian formulation of relativity, the
individual $\beta'$, $\beta''$... cannot be experimentally
determined. This belief derives from the assumption that the
Michelson-Morley type of experiments, from the original 1887 trial
to the modern versions with lasers stabilized by optical cavities,
give `null results', namely that the small residuals found in these
measurements are just typical instrumental artifacts. We recall that
in these precise interferometric experiments, one attempts to detect
in laboratory an `ether-wind', i.e. a small angular dependence of
the velocity of light that might indicate the Earth motion with
respect to the hypothetical $\Sigma$, e.g. the system where the
Cosmic Microwave Background (CMB) is isotropic. While in Special
Relativity, no ether wind can be observed by definition, in a
Lorentzian perspective it is only a `conspiracy' of relativistic
effects  which makes undetectable the individual velocity parameters
$\beta'$, $\beta''$... But the conspiracy works exactly only when
the velocity of light $c_\gamma$ propagating in the various
interferometers coincides with the basic parameter $c$ entering
Lorentz transformations. Therefore, one may ask, what happens if
$c_\gamma \neq c$, for instance when light propagates in air or in gaseous helium as in the old experiments? Starting from this observation, we have formulated a new theoretical scheme \cite{plus,plus2,book,universe} where the small residuals
observed so far, from Michelson-Morley to the present experiments
with optical resonators, point consistently toward the long sought
preferred reference frame tight to the CMB. In this sense, our
scheme is seriously questioning the standard null interpretation of
these experiments which is presented in all textbooks and
specialized reviews as a self-evident scientific truth. In this
article we will review the main results of our extensive work and
also propose further experimental tests.

We emphasize that, besides Relativity, our reinterpretation of the
data intertwines with and influences other areas of contemporary
physics, such as the non-locality of the Quantum Theory, the current
vision of the Vacuum State and Cosmology. These implications are so
important to deserve a preliminary discussion in this Introduction.

\subsection{Relativity and Quantum Non-Locality}

The existence of intrinsically non-local aspects in the Quantum
Theory and the relationship with relativity has been the subject of
a countless number of books and articles, growing more and more
rapidly in recent times, see e.g. \cite{Maudlin,gisin,Bricmont} for
a list of references. The problem dates back to the very early days
of Quantum Mechanics, even before the seminal work of
Einstein-Podolski-Rosen (EPR) \cite{EPR}. Indeed, the basic issue is
already found in Heisenberg's 1929 Chicago Lectures: `` We imagine a
photon represented by a wave packet... By reflection at a
semi-transparent mirror, it is possible to decompose into a
reflected and a transmitted packet...After a sufficient time the two
parts will be separated by any distance desired; now if by
experiment the photon is found, say, in the reflected part of the
packet, then the probability of finding the photon in the other part
of the packet immediately becomes zero. The experiment at the
position of the reflected packet thus exerts a kind of action
(reduction of the wave packet) at the distant point and one sees
that this action is propagated with a velocity greater than that of
light''. After that, Heisenberg, almost frightened by his same
words, feels the need to add the following remark: ``However, it is
also obvious that this kind of action can never be utilized for the
transmission of signals so that it is not in conflict with the
postulates of relativity''.

Heisenberg's final observation is one of the first formulations of
the so called `peaceful coexistence'. Actually, presenting as an
`obvious' fact that this type of effects can never be used to
communicate between observers at a space-like separation sounds more
as a way to avoid the causal paradox, which is present in Special
Relativity, when dealing with faster than light signals. But,
independently of that, this observation expresses a position that
can hardly be considered satisfactory. In fact, if there were really
some `Quantum Information' which propagates with a speed $v_{QI} \gg
c$, could such extraordinary thing be so easily dismissed? Namely,
could we ignore this `something' just because, apparently, it cannot
be efficiently controlled to send `messages'\footnote{Experimental
correlations between spacelike separated measurements can in
principle be explained through hidden influences propagating at a
finite speed $v_{\rm QI} \gg c$  provided $v_{\rm QI}$ is large
enough \cite{salart}. But in ref. \cite{bancal} it is also shown
that for any finite $v_{\rm QI}$, with $c < v_{\rm QI}< \infty$, one
can construct combined correlations to be used for faster-than-light
communication.}? After all, this explains why Dirac, more than forty
years later, was still concluding that ``The only theory which we
can formulate at the present is a non-local one, and of course one
is not satisfied with such a theory. I think one ought to say that
the problem of reconciling quantum theory and relativity is not
solved'' \cite{Dirac}.

But only with Bell's contribution \cite{Bell} the real terms of the
problem were fully understood. He clearly spelled out the local,
realistic point of view. If physical influences must propagate
continuously through space, it becomes unavoidable to complete the
quantum formalism by introducing additional `hidden' variables
associated with the space-time regions in question \footnote{``In
particular, Jordan had been wrong in supposing that nothing was real
or fixed in the microscopic world before observation. For after
observing only one of the two particles the result of subsequently
observing the other (possibly at very remote place) is immediately
predictable. Could it be that the first measurement somehow fixes
what was unfixed or makes real what was unreal, not only for the
near particle but also for the remote one? For EPR that would be an
unthinkable `spooky action at distance'. To avoid such action at
distance one has to attribute, to the space-time regions in
question, real correlated properties in advance of the observation
which predetermine the outcome of these particular observations.
Since these real properties, fixed in advance of the observation,
are not contained in the quantum formalism, that formalism for EPR
is incomplete'' \cite{Bell}.}. But, then, it is possible to derive a
bound on the degree of correlation of physical systems that are no
longer interacting but have interacted in their past. This bound has
been used to rule out experimentally \cite{aspect,clauser,zeilinger}
the class of local, hidden-variable theories which are based on
causal influences propagating at subluminal speed. Experimentally
excluding this class of theories means rejecting a familiar vision
of reality. Thanks to Bell, ``A seemingly philosophical debate about
the nature of physical reality could be settled by an experiment!
...The conclusion is now clear: Einstein's view of physical reality
cannot be upheld'' \cite{aspectbanquet}.

Thus, the importance of Bell's work cannot be underestimated:
``Bell's result combined with the EPR argument was not a `no hidden
variables theorem' but a non-locality theorem, the impossibility of
hidden variables being only one step in a two-step argument...It
means that some action at a distance exists in Nature, even though
it does not say what this action consists of'' \cite{Bricmont}.  It
was this awareness to give him the perception that ``... we have an
apparent incompatibility, at the deepest level, between the two
fundamental pillars of contemporary theory'' \cite{Bell}, namely
Quantum Theory and Special Relativity. This inspired his view where
the existence of the preferred $\Sigma-$frame would free ourselves
from the no-signalling argument to dispose of the causality paradox.

\subsection{Relativity and the Vacuum State}

A frequent objection to the idea of relativity with a preferred
frame is that, after all, Quantum Mechanics is not a fundamental
description of the world. What about, if we started from a more
fundamental Quantum Field Theory (QFT)? In this perspective, the
issue of the preferred frame can be reduced to find a particular,
logical step that prevents to deduce that Einstein Special
Relativity, with no preferred frame, is the {\it physically
realized} version of relativity. This is the version which is always
assumed when computing S-matrix elements for microscopic processes.
However, what one is actually using is the machinery of Lorentz
transformations whose first, complete derivation dates back,
ironically, to Larmor and Lorentz who were assuming the existence of
a fundamental state of rest (the ether).

Our point, discussed in \cite{epjc,dedicated,foop,zizzi}, is that
there is indeed a particular element which has been missed so far
and which concerns the assumed Lorentz invariance of the vacuum state.
Even though one is still using the Latin word `vacuum', which means
empty, here we are dealing with the lowest energy state. According
to the present view, this is not trivially empty but is determined
by the condensation process of some elementary quanta
\footnote{Before our work, the idea that the phenomenon of vacuum
condensation could produce `conceptual tensions' with both Special
and General Relativity, was discussed by Chiao \cite{Chiao}: ``The
physical vacuum, an intrinsically nonlocal ground state of a
relativistic quantum field theory, which possesses certain
similarities to the ground state of a superconductor... This would
produce an unusual `quantum rigidity' of the system, associated with
what London called the `rigidity of the macroscopic wave
function'... The Meissner effect is closely analog to the Higgs
mechanism in which the physical vacuum also spontaneously breaks
local gauge invariance ''\cite{Chiao}.}. Namely the energy is
minimized when these quanta, such as Higgs particles,
quark-antiquark pairs, gluons... macroscopically occupy the same quantum state, i.e.
the zero-3-momentum state \footnote{In the physically relevant case
of the Standard Model, the phenomenon of vacuum condensation can be
summarized by saying that ``What we experience as empty space is
nothing but the configuration of the Higgs field that has the lowest
possible energy. If we move from field jargon to particle jargon,
this means that empty space is actually filled with Higgs particles.
They have Bose condensed'' \cite{thooft}.}. Thus,
if the condensation process singles out a certain reference frame
$\Sigma$, the fundamental question is how to reconcile this picture
with the basic postulate of axiomatic QFT: the exact Lorentz
invariance of the vacuum \cite{cpt}. This postulate, meaning that
the vacuum state must remain unchanged under Lorentz boost, should not be confused with the condition that only local scalars (as the Higgs
field, or the gluon condensate, or the chiral condensate...) acquire a non-zero vacuum expectation value.

To make this evident, let us introduce the reference vacuum state
 $|\Psi^{(0)}\rangle$, appropriate to the observer at rest in the
$\Sigma-$frame singled out by the condensation process, and the
corresponding vacuum states $| \Psi'\rangle$, $| \Psi''\rangle$,..
appropriate to moving observers $S'$, $S''$,... By assuming that these different vacua are generated by
Lorentz boost unitary operators $U'$, $U''$...acting on
$|\Psi^{(0)}\rangle$, say $| \Psi'\rangle= U' |\Psi^{(0)}\rangle$,
$| \Psi''\rangle= U'' |\Psi^{(0)}\rangle$... For any Lorentz-scalar
operator ${G}$, such that $G=(U')^\dagger G U'=(U'')^\dagger G
U''...$, it follows trivially \BE \langle
{G}\rangle_{\Psi^{(0)}}=\langle {G}\rangle_{\Psi'}=\langle
{G}\rangle_{\Psi''}=..\EE However, this by no means implies
$|\Psi^{(0)}\rangle=| \Psi'\rangle =| \Psi''\rangle$... To this end,
one should construct explicitly the three boost generators $L_{0i}$
(with i=1,2,3) and show that $ L_{0i} |\Psi^{(0)}\rangle =0$. But,
in four space-time dimensions, the explicit construction of these
operators, and of the corresponding Poincar\'e algebra \footnote
{This means an operatorial representation of the 10 generators
$P_\mu$ and $L_{\mu\nu}$ ($\mu$, $\nu$= 0, 1, 2, 3), where $P_\mu$
describe the space-time translations and $L_{\mu\nu}=-L_{\nu\mu}$
the space rotations and Lorentz boosts, with commutation relations $
[P_\mu,P_\nu]=0$, $[L_{\mu\nu}, P_\rho]= i\eta_{\nu\rho}P_\mu -
i\eta_{\mu\rho}P_\nu$ and $[L_{\mu\nu}, L_{\rho\sigma}]=
-i\eta_{\mu\rho}L_{\nu\sigma}+ i\eta_{\mu\sigma}L_{\nu\rho}
-i\eta_{\nu\sigma}L_{\mu\rho}+i\eta_{\nu\rho}L_{\mu\sigma}$ where
$\eta_{\mu\nu}={\rm diag}(1,-1,-1,-1)$ is the Minkowski tensor. A
 Lorentz-invariant vacuum has to be annihilated by all 10
generators.} is only known for the free-field case through the
simple Wick-ordering prescription relatively to the free-field
vacuum $|0\rangle$. In an interacting theory, the construction is
implemented order by order in perturbation theory. Therefore, in the
presence of non-perturbative phenomena (such as Spontaneous Symmetry
Breaking, chiral symmetry breaking, gluon condensation...) where the
physical vacuum $ |\Psi^{(0)}\rangle$ cannot be constructed from the
free-field vacuum $ |0\rangle$ order by order in perturbation
theory, proving the Lorentz invariance of the vacuum represents an
insurmountable problem. In this situation, with a Lorentz-invariant
interaction, the resulting theory will still be Lorentz covariant
but, with a non-Lorentz-invariant vacuum, there would be a preferred
reference frame \footnote{To our knowledge,in four space-time
dimensions, a non-perturbative analysis of a Lorentz-invariant
vacuum has been attempted by very few authors. In the case of
non-linear field theories with $P(\Phi(x))$ interactions, such as
$\Phi^4(x)$, this was discussed by Segal \cite{segal}. He considered
a suitable generalization of the standard Wick ordering $:P(\Phi) :$
relative to $|0\rangle$, say $::P(\Phi)::$, such that $\langle
\Psi^{(0)}|::P(\Phi)::|\Psi^{(0)}\rangle=0$ in the true vacuum
state. His conclusion was that $::P(\Phi)::$ is not well-defined
until the physical vacuum is known, but, at the same time, the
physical vacuum also depends on the definition given for
$::P(\Phi)::$. From this type of circularity Segal concluded that,
in general, in such a nonlinear QFT, the physical vacuum will {\it
not} be invariant under the full Lorentz symmetry of the underlying
Lagrangian density.}.

\subsection{Relativity and the CMB}

Finally, some remarks about the physical nature of the hypothetical
$\Sigma-$frame. A natural candidate is the reference system where
the temperature of the CMB looks exactly isotropic or, more
precisely, where the CMB kinematic dipole \cite{yoon} vanishes. This
dipole is in fact a direct consequence of the motion of the Earth
($\beta=V/c)$ \BE T(\theta)={{T_o\sqrt{1-\beta^2}}\over{1- \beta
\cos \theta} } \EE Accurate observations with satellites in space
\cite{smoot} have shown that the measured temperature variations
correspond to a motion of the solar system described by an average
velocity $V\sim 370$ km/s, a right ascension $\alpha \sim 168^o$ and
a declination $\gamma\sim -7^o$, pointing approximately in the
direction of the Leo constellation. This means that, if one sets
$T_o \sim $ 2.7 K and $\beta\sim 0.00123$, there are angular
variations of a few millikelvin \BE \label{CBR}\Delta T^{\rm
CMB}(\theta) \sim T_o \beta \cos\theta = \pm 3.3 ~{\rm mK} \EE which
represent by far the dominant contribution to the CMB anisotropy.

Could the reference system with vanishing CMB dipole represent the
fundamental preferred frame for relativity as in the original
Lorentzian formulation? The standard answer is that one should not
confuse these two concepts. The CMB is a definite medium and, as
such, sets a rest frame where the dipole anisotropy is zero. Our
motion with respect to this system has been detected but, by itself,
this is not in contradiction with Special Relativity. Though, to
good approximation, this kinematic dipole arises by combining the
various forms of peculiar motion which are involved (rotation of the
solar system around the center of the Milky Way, motion of the Milky
Way toward the center of the Local Group, motion of the Local Group
of galaxies in the direction of the Great Attractor...)
\cite{smoot} . Thus, if one could switch-off the local
inhomogeneities that produce these peculiar forms of motion, it is
natural to imagine a global frame of rest associated with the
Universe as a whole. A vanishing CMB dipole could then just {\it
indicate} the existence of this fundamental system $\Sigma$ that we
may conventionally decide to call `ether' but the cosmic radiation
itself would not {\it coincide} with this form of ether.

This is why, to discriminate between the two concepts,
Michelson-Morley type of experiments become crucial. Detecting a
small angular dependence of the velocity of light in the Earth
laboratory, and correlating this angular dependence with the Earth
cosmic motion, would provide the missing link with the logical
arguments from Quantum Non-Locality \footnote{``Non-Locality is most
naturally incorporated into a theory in which there is a special
frame of reference. One possible candidate for this special frame of
reference is the one in which the CMB is isotropic. However, other
than the fact that a realistic interpretation of quantum mechanics
requires a preferred frame and the CMB provides us with one, there
is no readily apparent reason why the two should be linked''
\cite{Hardy}.} and with the idea of a condensed vacuum which selects
a particular reference frame through the macroscopic occupation of
the same zero 3-momentum state. More generally a non-null
interpretation of the Michelson-Morley experiments would resolve the
puzzle of a world endowed with a fundamental space and a fundamental
time whose existence, otherwise, would remain forever hidden to us.

After this general Introduction, we will start by reviewing in
Sect.2 the basic ingredients for a modern analysis of the
Michelson-Morley experiments. Then we will summarize in Sects.3 and
4 our re-analysis \cite{plus,plus2,book,universe} of the classical
experiments and in Sect.5 the corresponding treatment of the present
experiments with optical resonators.  As a matter of fact, once the
small residuals are analyzed in our scheme, the long sought
$\Sigma-$frame tight to the CMB is naturally emerging. Sect.6 will
finally contain a summary and our conclusions.

\section{A modern view of the `ether-drift'  experiments}

The Michelson-Morley experiments are also called `ether-drift'
experiments because they were designed to detect the drift of the
Earth in the ether by observing a dragging of light associated with
the Earth cosmic motion. Today, experimental evidence for both the
undulatory and corpuscular aspects of radiation has substantially
modified the consideration of an underlying ethereal medium, as
support of the electromagnetic waves, and its logical need for the
physical theory. Besides, Lorentz Transformations forbid dragging
and the irregular behavior of the small observed residuals is
inconsistent with the smooth time modulations that one would expect
from the Earth rotation. Therefore, at first sight, the idea of
detecting a non-null effect seems hopeless.

However, as anticipated, dragging is only forbidden if the velocity
of light in the interferometers is the same parameter $c$ of Lorentz
transformations. For instance, when light propagates in a gas, the
sought effect of a preferred system $\Sigma$ could be due to the
small fraction of refracted light. Obviously, this would be much
smaller than classically expected but, in view of the extraordinary
precision of the interferometers, it could still be measurable. In
addition, the idea of smooth time modulations of the signal reflects
the traditional identification of the local velocity field, which
describes the drift, with the projection of the global Earth motion
at the experimental site. This identification is equivalent to a
form of regular, laminar flow where global and local velocity fields
coincide. Instead, depending on the nature of the physical vacuum,
the two velocity fields could only be related indirectly, as it
happens in turbulent flows, so that numerical simulations would be
needed for a consistent statistical description of the data.

In the following, we will summarize the scheme of
refs.\cite{plus,plus2,book,universe} starting with the case of light
propagating in gaseous media, as for the classical experiments.

\subsection{Basics of the ether-drift experiments}

In the classical measurements in gases (Michelson-Morley, Miller,
Tomaschek, Kennedy, Illingworth, Piccard-Stahel,
Michelson-Pease-Pearson, Joos) \cite{mm}-\cite{joos}, one was
measuring the fringe shifts produced by a rotation of the
interferometer. Instead, in modern experiments, with lasers
stabilized by optical cavities, see e.g. \cite{applied} for a
review, one measures frequency shifts.
\begin{figure}[h]
\begin{center}
\includegraphics[width=7.0 cm]{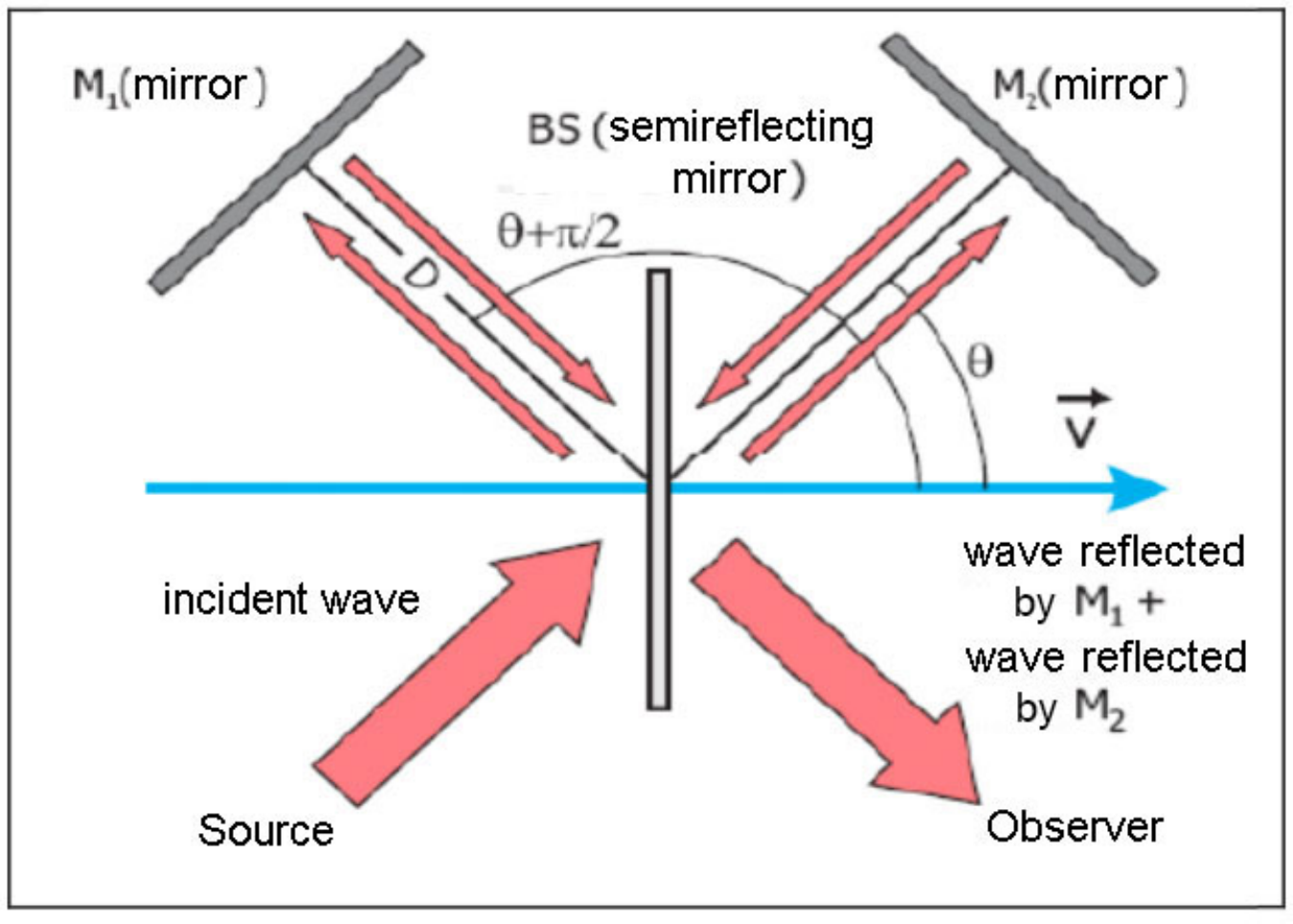}
\end{center}
\caption{\it A schematic illustration of the Michelson
interferometer. Note that, by computing the transit times as in
Eq.(\ref{deltaT}), we are assuming the validity of Lorentz
transformations so that the length of a rod does not depend on its
orientation in the frame $S$ where it is at rest.}
\label{Michinterferometer}
\end{figure}
The modern experiments adopt a different technology but, in the end,
have exactly the same scope: searching for an anisotropy of the
two-way velocity of light $\bar{c}_\gamma(\theta)$ which is the only
one that can be measured unambiguously \BE \bar{c}_\gamma(\theta)=
{{ 2  c_\gamma(\theta) c_\gamma(\pi + \theta) }\over{
c_\gamma(\theta) + c_\gamma(\pi + \theta) }} \EE By introducing its
anisotropy \BE \Delta\bar{c}_\theta =
\bar{c}_\gamma(\pi/2+\theta)-\bar{c}_\gamma(\theta) \EE there is a
simple relation with the time difference $\Delta t(\theta)$ for
light propagation back and forth along perpendicular rods of length
$D$. In fact, by assuming the validity of Lorentz transformations,
the length of a rod does not depend on its orientation, in the $S$
frame where it is at rest, see Fig.\ref{Michinterferometer}, and one
finds,
\begin{figure}[h]
\begin{center}
\includegraphics[width=7.0 cm]{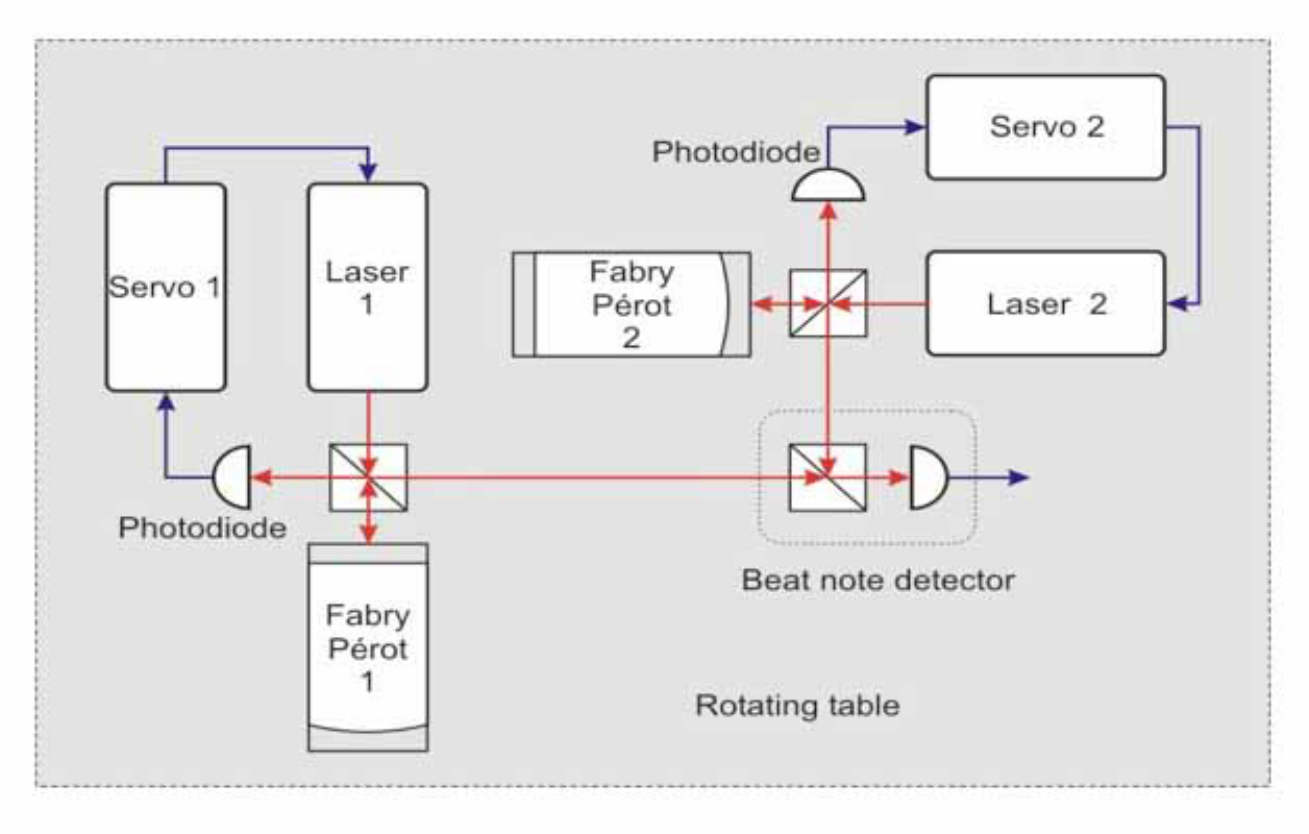}
\caption{\it The scheme of a modern ether-drift experiment. The
light frequencies are first stabilized by coupling the lasers to
Fabry-Perot optical resonators. The frequencies $\nu_1$ and $\nu_2$
of the resonators are then compared in the beat note detector which
provides the frequency shift $\Delta \nu(\theta)=\nu_1(\pi/2 +
\theta) -\nu_2(\theta)$. For a review, see e.g. \cite{applied}. }
\label{Fig.apparatus}
\end{center}
\end{figure}
\begin{equation}
\label{deltaT} \Delta t(\theta)=
{{2D}\over{\bar{c}_\gamma(\theta)}}-
{{2D}\over{\bar{c}_\gamma(\pi/2+\theta)}} \sim
{{2D}\over{c}}~{{\Delta \bar{c}_\theta } \over{c}}
\end{equation}
(where, in the last relation, we have assumed that light propagates
in a medium of refractive index ${\cal N}=1 + \epsilon$, with
$\epsilon\ll 1$). This gives directly the fringe patterns ($\lambda$
is the light wavelength)
\begin{equation}
\label{newintro} {{\Delta \lambda(\theta)}\over{\lambda}} \sim
{{2D}\over{\lambda}} ~{{\Delta \bar{c}_\theta } \over{c}}
\end{equation}
which were measured in the classical experiments.

In modern experiments, on the other hand, a possible anisotropy of
$\bar{c}_\gamma(\theta)$ would show up through the relative
frequency shift, i.e. the beat signal, $\Delta\nu(\theta)$ of two
orthogonal optical resonators, see Fig.\ref{Fig.apparatus}. Their
frequency
\begin{equation}
\label{nutheta0}
  \nu (\theta)= {{ \bar{c}_\gamma(\theta) m}\over{2L}}
\end{equation}
is proportional to the two-way velocity of light within the
resonator through an integer number $m$, which fixes the cavity
mode, and the length of the cavity $L$ as measured in the laboratory
$S'$ frame. Therefore, once the length of a cavity in its rest frame
does not depend on its orientation, one finds
\begin{equation}
\label{bbasic2}
 {{\Delta \nu(\theta) }\over{\nu_0}}  \sim
      {{\Delta \bar{c}_\theta } \over{c}}
\end{equation}
where $\nu_0$ is the reference frequency of the two resonators.


\subsection{The limit of refractive index ${\cal N}= 1
+\epsilon$ }

Let us consider light propagating in a medium which is close to the
ideal vacuum limit, i.e. whose refractive index is ${\cal N}= 1
+\epsilon$ with $\epsilon \ll 1$. The medium (e.g. a gas) fills an
optical cavity at rest in a frame $S$ which moves with velocity $v$
with respect to the hypothetical $\Sigma$. Now, by assuming i) that
the velocity of light is exactly isotropic when $S\equiv \Sigma$ and
ii) the validity of Lorentz transformations, then any anisotropy in
$S$ should vanish identically either for $v = 0$ or for the ideal
vacuum case ${\cal N} = 1$ when the velocity of light $c_\gamma$
coincides with the basic parameter $c$ of Lorentz transformations.
Thus, one can expand in powers of the two small parameters
$\epsilon$ and $\beta=v/c$. By taking into account that, by its very
definition, the two-way velocity $\bar{c}_\gamma(\theta)$ is
invariant under the replacement $\beta \to -\beta$ and that, for any
fixed $\beta$, is also invariant under the replacement $\theta \to
\pi +\theta$, to lowest non-trivial level ${\cal
O}(\epsilon\beta^2)$, one finds the general expression
\cite{plus,foop}
\begin{eqnarray}
\label{legendre}
       \bar{c}_\gamma(\theta) \sim {{c}\over{ {\cal N} }} \left[1- \epsilon~\beta^2
\sum^\infty_{n=0}\zeta_{2n}P_{2n}(\cos\theta)
  \right]
\end{eqnarray}
Here, to take into account invariance under $\theta \to \pi
+\theta$, the angular dependence has been given as an infinite
expansion of even-order Legendre polynomials with arbitrary
coefficients $\zeta_{2n}={\cal O}(1)$. In Einstein's Special
Relativity, where there is no preferred reference frame, these
$\zeta_{2n}$ coefficients should vanish identically. In a Lorentzian
approach, on the other hand, there is no reason why they should
vanish {\it a priori} \footnote{As anticipated, for Lorentz, only a
conspiracy of effects prevents to detect the motion with respect to
the ether, which, however different might be from ordinary matter,
is nevertheless endowed with a certain degree of substantiality. For
this reason, in his view, ``it seems natural not to assume at
starting that it can never make any difference whether a body moves
through the ether or not''\cite{electron}.}.

By leaving out the first few $\zeta$'s as free parameters in the
fits, Eq.(\ref{legendre}) could already represent a viable form to
compare with experiments. Still, one can further sharpen the
predictions by exploiting one more derivation of the $\epsilon \to
0$ limit with a preferred frame. This other argument is based on the
effective space-time metric $g^{\mu\nu}=g^{\mu\nu}({\cal N})$ which,
through the relation $g^{\mu\nu}p_\mu p_\nu=0$, describes light
propagation in a medium of refractive index ${\cal N}$. For the
quantum theory, a derivation of this metric from first principles
was given by Jauch and Watson \cite{jauch} who worked out the
quantization of the electromagnetic field in a dielectric. They
noticed that the procedure introduces unavoidably a preferred
reference frame, the one where the photon energy spectrum does not
depend on the direction of propagation, and which is ``usually taken
as the system for which the medium is at rest''. However, such an
identification reflects the point of view of Special Relativity with
no preferred frame. Instead, one can adapt their results to the case
where the angle-independence of the photon energy is only valid when
both medium and observer are at rest in some particular frame
$\Sigma$.

In this perspective, let us consider two identical optical cavities,
namely cavity 1, at rest in $\Sigma$, and cavity 2, at rest in $S$,
and denote by $\pi_\mu\equiv ( {{E_\pi}\over{c}},{\bf \pi}) $ the
light 4-momentum for $\Sigma$ in his cavity 1 and by $p_\mu\equiv (
{{E_p}\over{c}},{\bf p})$ the corresponding light 4-momentum for
$S$ in his cavity 2. Let us also denote by $g^{\mu\nu}$ the
space-time metric that $S$ uses in the relation $g^{\mu\nu}p_\mu
p_\nu=0$ and by
\begin{equation}
\label{metricsigma}\gamma^{\mu\nu}={\rm diag}({\cal N}^2,-1,-1,-1)
\end{equation}
the metric used by $\Sigma$ in the relation
$\gamma^{\mu\nu}\pi_\mu\pi_\nu=0$ and which gives an isotropic
velocity $c_\gamma=E_\pi/|{\bf \pi}|={{c}\over{{\cal N}}}$.

Now, Special Relativity was formulated, more than a century ago, by
assuming the perfect equivalence of two reference systems in uniform
translational motion. Instead, with a preferred frame $\Sigma$, as
far as light propagation is concerned, this physical equivalence is
only assumed in the ideal ${\cal N}=1$ limit. As anticipated, for
${\cal N}\neq 1$, where light gets absorbed and then re-emitted, the
fraction of refracted light could keep track of the particular
motion of matter with respect to $\Sigma$ and produce, in the frame
$S$ where matter is at rest, an angular dependence of the velocity
of light. Equivalently, assuming that the solid parts of cavity 2
are at rest in a frame $S$, which is in uniform translational motion
with respect to $\Sigma$, no longer implies that the medium which
stays inside, e.g. the gas, is in a state of thermodynamic equilibrium\footnote{ Think for instance of the collective interaction of a
gaseous medium with the CMB radiation or with hypothetical dark
matter in the Galaxy. However weak this interaction may be, it would
mimic a non-local thermal gradient that could bring the gas out of
equilibrium. The advantage of the following analysis is that it only
uses symmetry properties without requiring a knowledge of the
underlying dynamical processes. }. Thus, one should keep an open
mind and exploit the implications of the basic condition
\begin{equation} \label{limitingintro} g^{\mu\nu}({\cal N}=1)=
\gamma^{\mu\nu}({\cal N}=1)=\eta^{\mu\nu}\end{equation} where
$\eta^{\mu\nu}$ is the Minkowski tensor. This standard equality
amounts to introduce a transformation matrix, say $A^{\mu}_{\nu}$,
such that
\begin{equation}
\label{vacuum}
g^{\mu\nu}=A^{\mu}_{\rho}A^{\nu}_{\sigma}\eta^{\rho\sigma}=\eta^{\mu\nu}
\end{equation}
This relation is strictly valid for ${\cal N}=1$. However, by
continuity, one is driven to conclude that an analogous relation
between $g^{\mu\nu}$ and $\gamma^{\mu\nu}$ should also hold in the
$\epsilon \to 0$ limit. The only subtlety is that relation
(\ref{vacuum}) does not fix uniquely $A^{\mu}_{\nu}$. In fact, one
can either choose the identity matrix, i.e.
$A^{\mu}_{\nu}=\delta^{\mu}_{\nu}$, or a Lorentz transformation,
i.e. $A^{\mu}_{\nu}=\Lambda^{\mu}_{\nu}$. Since for any finite $v$
these two matrices cannot be related by an infinitesimal
transformation, it follows that $A^{\mu}_{\nu}$ is a two-valued
function in the $\epsilon \to 0$ limit. Therefore, in principle, there are two solutions. If
$A^{\mu}_{\nu}$ is the identity matrix, we find a first solution
\begin{equation}\left[g^{\mu\nu}({\cal N})\right]_1=
\gamma^{\mu\nu}\sim \eta^{\mu\nu} + 2\epsilon \delta^\mu_0
\delta^\nu_0\end{equation} while, if $A^{\mu}_{\nu}$ is a Lorentz
transformation, we find the other solution
\begin{equation} \label{2intro} \left[g^{\mu\nu}({\cal
N})\right]_2= \Lambda^{\mu}_{\rho}
\Lambda^{\nu}_{\sigma}\gamma^{\rho\sigma} \sim \eta^{\mu\nu} +
2\epsilon v^\mu v^\nu
\end{equation} $v^\mu$ being the dimensionless
$S$ 4-velocity, $v^\mu\equiv(v^0,{\bf v}/c)$ with $v_\mu v^\mu=1$.

Notice that with the former choice, implicitly adopted in Special
Relativity to preserve isotropy in all reference systems also for
${\cal N} \neq 1$, one is introducing a discontinuity in the
transformation matrix for any $\epsilon \neq 0$. Indeed, the whole
emphasis on Lorentz transformations depends on enforcing
Eq.(\ref{vacuum}) for $A^{\mu}_{\nu}=\Lambda^{\mu}_{\nu}$ so that
$\Lambda^{\mu \sigma}\Lambda^{\nu}_{\sigma}=\eta^{\mu\nu}$ and the
Minkowski metric applies to all equivalent frames.

On the other hand, with the latter solution, by replacing in the
relation $p_\mu p_\nu g^{\mu\nu}=0$, the photon energy now depends
on the direction of propagation. Then, by defining the light
velocity $c_\gamma(\theta)$ from the ratio $E_p/|{\bf p}|$, one
finds \cite{plus,foop} \BE \label{oneway0}
       c_\gamma(\theta) \sim {{c} \over{{\cal N}}}~\left[
       1- 2\epsilon \beta \cos\theta -
       \epsilon \beta^2(1+\cos^2\theta)\right]
\EE and a two-way velocity
\begin{eqnarray}
\label{twoway00}
       \bar{c}_\gamma(\theta)&=&
       {{ 2  c_\gamma(\theta) c_\gamma(\pi + \theta) }\over{
       c_\gamma(\theta) + c_\gamma(\pi + \theta) }}
       \sim {{c} \over{{\cal N}}}~\left[1-\epsilon\beta^2\left(1 +
       \cos^2\theta\right) \right]\equiv {{c}\over{\bar{\cal N}(\theta)}}
\end{eqnarray}
here $\theta$ is the angle between ${\bf v}$ and $\bf p$  (as
defined in the $S$ frame) and
\BE
\label{nbartheta}\bar{\cal N}(\theta)\sim {\cal N}
~[1+\epsilon\beta^2 (1 + \cos^2\theta)] \EE
Eq.(\ref{twoway00}) corresponds to setting in Eq.(\ref{legendre})
$\zeta_0=4/3$, $\zeta_{2}= 2/3$ and all $\zeta_{2n}=0$ for $n
> 1$ and can be considered a modern version of Maxwell's original
calculation \cite{maxwell}. It represents a definite, alternative
model for the interpretation of experiments performed close to the
ideal vacuum limit $\epsilon = 0$, such as in gaseous systems, and
will be adopted in the following \footnote{A conceptual detail
concerns the relation of the gas refractive index ${\cal N}$, as
defined in the $\Sigma-$frame through Eq.(\ref{metricsigma}), to the
experimental quantity ${\cal N}_{\rm exp}$ which is extracted from
measurements of the two-way velocity in the Earth laboratory. By
assuming a $\theta-$dependent refractive index as in
Eq.(\ref{nbartheta}) one should thus define ${\cal N}_{\rm exp}$ by
an angular average, i.e. $ {\cal N}_{\rm exp} \equiv \langle
\bar{\cal N}(\theta) \rangle_\theta= {\cal N}  ~\left[1
+{{3}\over{2}} ({\cal N} -1)\beta^2\right]$. One can then determine
the unknown value ${\cal N}  \equiv {\cal N}(\Sigma)$ (as if the
container of the gas were at rest in $\Sigma$), in terms of the
experimentally known quantity ${\cal N}_{\rm exp}\equiv{\cal N}({\rm
Earth})$ and of $v$. As discussed in refs.
\cite{plus}-\cite{universe}, for $v\sim 370$ km/s, the resulting
difference $|{\cal N}_{\rm exp} - {\cal N}| \lesssim 10^{-9}$ is
well below the experimental accuracy on ${\cal N}_{\rm exp}$ and,
for all practical purposes, can be ignored.}.

\section{A first look at the classical experiments}

From Eq.(\ref{twoway00}) we find  a fractional
anisotropy
\begin{equation} \label{bbasic2new} {{\Delta \bar{c}_\theta }
\over{c}}
={{\bar{c}_\gamma(\pi/2+\theta)-\bar{c}_\gamma(\theta)}\over{c}}\sim
     \epsilon~(v^2/c^2)
       \cos2\theta \end{equation}
which produces a fringe pattern
\begin{equation} \label{newintro1} {{\Delta
\lambda(\theta)}\over{\lambda}}= {{2D}\over{\lambda}} ~{{\Delta
\bar{c}_\theta } \over{c}}\sim {{D}\over{\lambda}}~
2\epsilon~{{v^2}\over{c^2}}\cos 2\theta
\end{equation}
\begin{figure}[h]
\begin{center}
\includegraphics[width=7.5 cm]{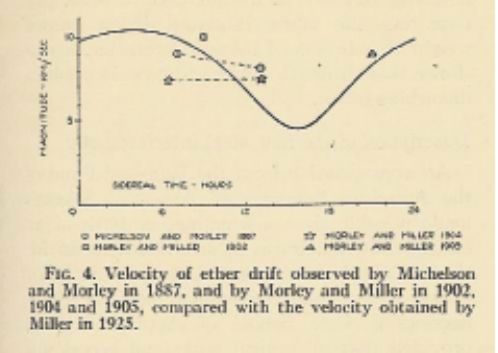}
\end{center}
\caption{\it The observable velocity reported by
Miller \cite{miller} for various experiments.} \label{miller}
\end{figure}
The dragging of light in the Earth frame is then described as a
pure 2nd-harmonic effect, which is periodic in the range $[0,\pi]$,
as in the classical theory (see e.g. \cite{kennedy}). However, its amplitude
\begin{equation} \label{a2new}
 A_2={{D}\over{\lambda}} ~2\epsilon ~{{{v}^2}\over{c^2}}
\end{equation}
is now much smaller being suppressed by the factor $2\epsilon$
relatively to the classical value. This was traditionally reported
for the orbital velocity of 30 km/s  as \BE \label{a2class} A^{\rm
class}_2={{ D}\over{\lambda}} (\frac{30~{\rm km/s}}{c})^2\EE This
difference could then be re-absorbed into an {\it observable}
velocity which is related to the {\it kinematical} velocity $v$
through the gas refractive index
\begin{equation}  \label{vobs} v^2_{\rm obs} \sim 2\epsilon v^2 \end{equation}
so that \BE A_2 = {{D}\over{\lambda}}  ~{{{v}^2_{\rm
obs}}\over{c^2}} \EE Thus, this  $v_{\rm obs}$ is the small velocity
traditionally extracted from the measured amplitude $A^{\rm EXP}_2$
in the classical analysis of the experiments
\begin{equation}  \label{vobs2} v_{\rm obs} \sim 30 ~{\rm km/s} ~\sqrt{\frac {A^{\rm EXP}_2}{A^{\rm class}_2}}\end{equation}
see e.g. Fig.\ref{miller}.

We can now understand the pattern observed in the classical
experiments. For instance, in the original Michelson-Morley
experiment, where ${{D}\over{\lambda}}\sim 2\cdot 10^7$, the
classically expected amplitude was $A^{\rm class}_2\sim$ 0.2. But
the experimental amplitude measured in the six sessions was $A^{\rm
EXP}_2 = 0.01\div 0.02 $. This corresponds to an average anisotropy
${{|\Delta\bar{c}_\theta|_{\rm exp}}\over{c}}\sim 4 \cdot 10^{-10}$
and was originally interpreted in terms of a velocity $v_{\rm obs}
\sim 8 $ km/s. However, for an experiment in air at room temperature
and atmospheric pressure where $\epsilon\sim 2.8\cdot 10^{-4}$, this
observable velocity would now correspond to a true kinematic value
$v \sim 340$ km/s which would fit well with the cosmic motion
indicated by the CMB dipole anistropy. Therefore, the importance of
the issue requires to sharpen the analysis of the old experiments,
starting from the early 1887 trials.

\subsection{The 1887 Michelson-Morley experiment in Cleveland}

\begin{figure}
\centerline{\includegraphics[width=6.0cm]{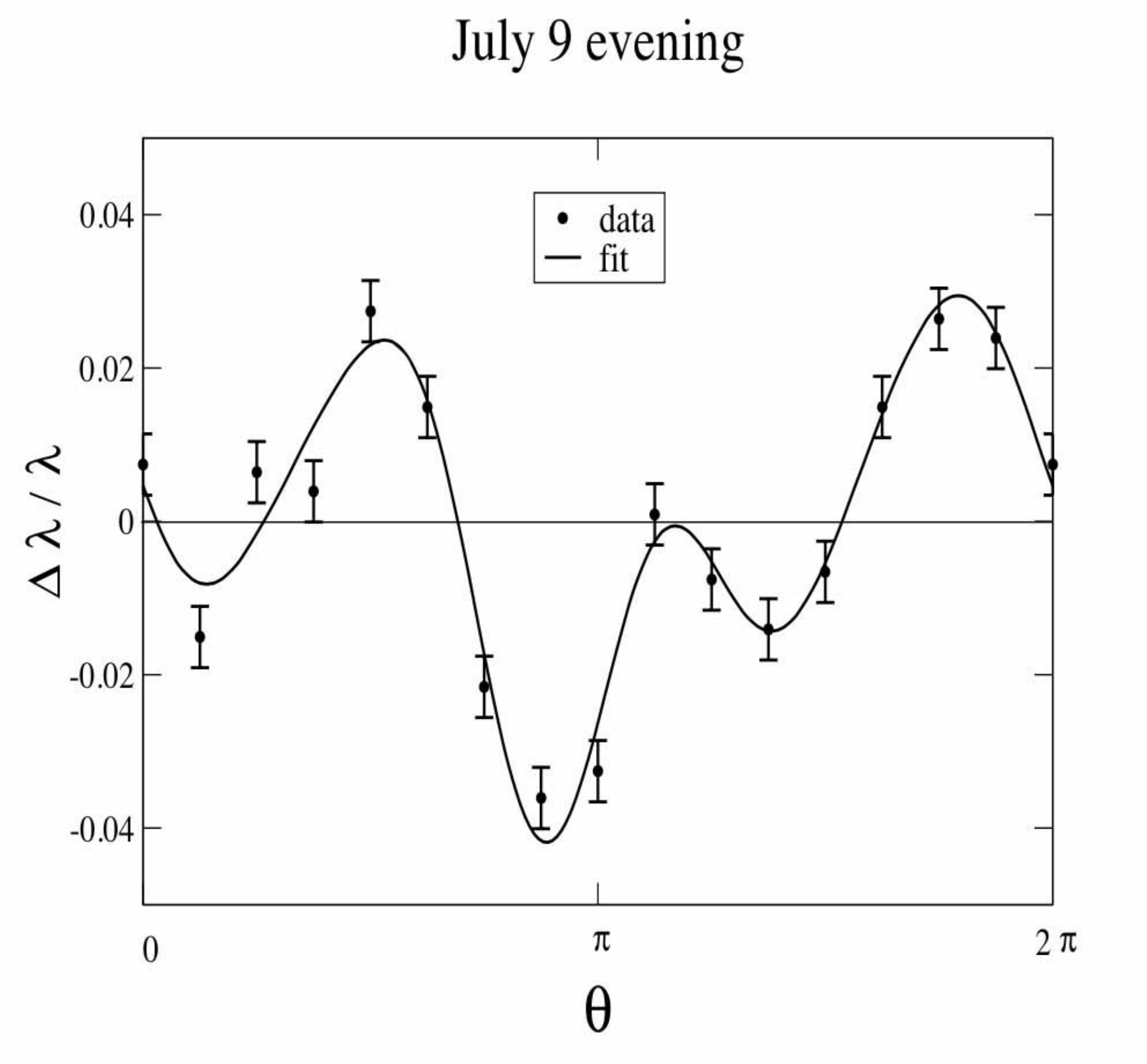}} \caption{\it
The fringe shifts for the session of July 9 evening. The fit is
performed by including terms up to fourth harmonics. The figure is
taken from ref.\cite{cimento}.}\label{july9}
\end{figure}

\begin{figure}
\centerline{\includegraphics[width=6.0cm]{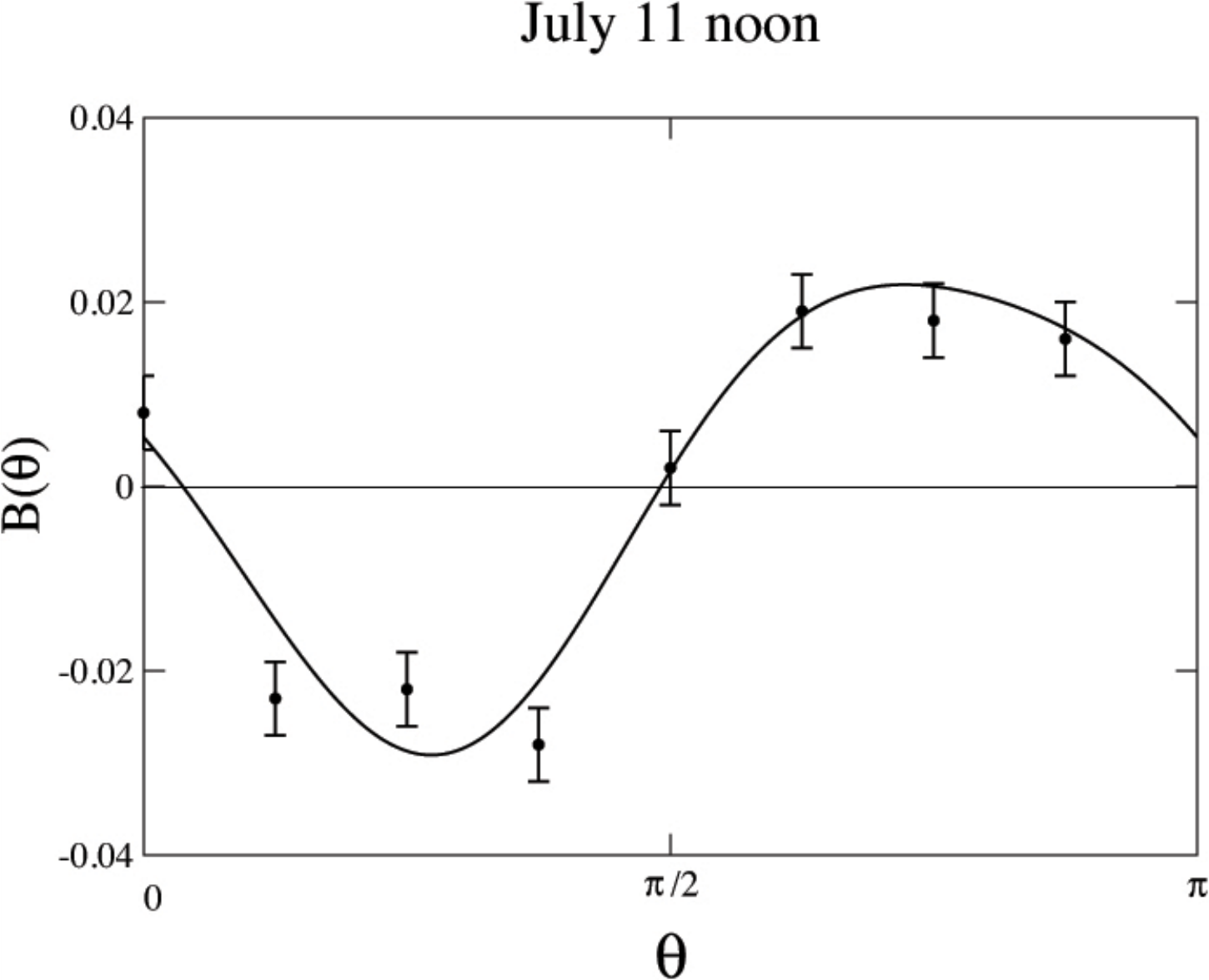}} \caption{\it A
fit to the even combination $B(\theta)$ Eq.(\ref{even}). The second
harmonic amplitude is ${A}^{\rm EXP}_2=0.025 \pm 0.005$ and the
fourth harmonic is ${A}^{\rm EXP}_4=0.004 \pm 0.005$. The figure is
taken from ref.\cite{cimento}.}\label{fit11july}
\end{figure}

The precision of the Michelson-Morley apparatus \cite{mm} was
extraordinary, about $\pm 0.004$ of a fringe. For all details, we
address the reader to our book \cite{book}. Here, we just limit
ourselves to quote from Born \cite{bornbook}. When discussing the
classically expected fringe shift upon rotation of the apparatus by
90 degrees, namely $2 A^{\rm class}_2\sim$ 0.4, he says explicitly: ``Michelson
was certain that the one-hundredth part of this displacement  would
still be observable'' (i.e. 0.004). As a check, the Michelson-Morley fringe shifts were recomputed in
refs.\cite{cimento,plus,book} from the original article \cite{mm}, see Table 1. These data were then analyzed in a Fourier expansion, see e.g.
Fig.\ref{july9} (note that a 1st-harmonic has to be present in the data due to the arrangement of the mirrors needed to have fringes of finite width, see \cite{miller,hicks}). One can thus extract amplitude and phase of the 2nd-harmonic component by fitting the even combination of fringe shifts
 \BE \label{even}  B(\theta)=
 {{\Delta \lambda (\theta) +\Delta \lambda (\pi
 +\theta)}\over{2\lambda}}\EE see Fig.\ref{fit11july}.

\begin{table}[htb]
\tbl{
The fringe shifts ${{\Delta \lambda(i)}\over{\lambda}}$
for all noon (n.) and evening (e.) sessions of the Michelson-Morley
experiment. The angle of rotation is defined as
$\theta={{i-1}\over{16}} ~2\pi$. The Table is taken from
ref.\cite{plus}.
}
{\begin{tabular}{@{}clllllll@{}} \toprule
i ~~     &  July 8 (n.) & July 9 (n.) &
July 11 (n.)
       &  July 8 (e.) & July 9 (e.) & July 12 (e.)     \\
\hline
1 ~~     & $-$0.001 & +0.018 &+0.016& $-$0.016& +0.007& +0.036 \\
2 ~~     & +0.024 & $-$0.004 &$-$0.034& +0.008& $-$0.015& +0.044 \\
3~~      & +0.053 & $-$0.004 &$-$0.038& $-$0.010& +0.006& +0.047 \\
4 ~~    & +0.015 & $-$0.003 &$-$0.066& +0.070& +0.004& +0.027 \\
5 ~~     & $-$0.036 & $-$0.031 &$-$0.042& +0.041& +0.027& $-$0.002 \\
6 ~~     & $-$0.007 & $-$0.020 &$-$0.014& +0.055& +0.015& $-$0.012 \\
7 ~~     & +0.024 & $-$0.025 &+0.000& +0.057& $-$0.022& +0.007 \\
8 ~~     & +0.026 & $-$0.021 &+0.028& +0.029& $-$0.036& $-$0.011 \\
9 ~~     & $-$0.021 & $-$0.049 &+0.002& $-$0.005& $-$0.033& $-$0.028 \\
10~~     & $-$0.022 & $-$0.032 &$-$0.010& +0.023& +0.001& $-$0.064 \\
11~~     & $-$0.031 & +0.001 &$-$0.004& +0.005& $-$0.008& $-$0.091 \\
12~~     & $-$0.005 & +0.012 &+0.012& $-$0.030& $-$0.014& $-$0.057 \\
13~~     & $-$0.024 & +0.041 &+0.048& $-$0.034& $-$0.007& $-$0.038 \\
14~~     & $-$0.017 & +0.042 &+0.054& $-$0.052& +0.015& +0.040 \\
15~~     & $-$0.002 & +0.070 &+0.038& $-$0.084& +0.026& +0.059 \\
16~~     & +0.022 & $-$0.005 &+0.006& $-$0.062& +0.024& +0.043 \\
\botrule
\end{tabular}}
\label{fringes}
\end{table}

The 2nd-harmonic amplitudes for the six experimental sessions are
reported in Table 2.  Due to their statistical consistency, one can
compute the mean and variance and obtain $A^{\rm EXP}_2 \sim 0.016
\pm 0.006$. This value is consistent with an observable velocity \BE
\label{vobsmm} v_{\rm obs} \sim 8.5 ~^{+1.7}_{-2.2}~~~{\rm km/s} \EE
in complete agreement with Miller, ses Fig.\ref{miller}. In this
sense, our re-analysis supports the claims of Hicks and Miller. The
fringe shifts were much smaller than expected but in two
experimental sessions (11 July noon and 12 July evening), the
second-harmonic amplitude is non-zero at the 5$\sigma$ level and in
other two sessions (July 9 noon and July 8 evening) is non-zero at
the 3$\sigma$ level.
\begin{table}[htb]
\tbl{The 2nd-harmonic amplitudes for the six experimental
sessions of the Michelson-Morley experiment. The table is taken from
ref.\cite{plus}.} 
{\begin{tabular}{@{}cl@{}} \toprule
SESSION       & ~~~~~~      $A^{\rm EXP}_2$   \\
\hline
July 8  (noon) & $0.010 \pm 0.005$  \\
July 9  (noon) & $0.015 \pm 0.005$   \\
July 11 (noon) & $0.025 \pm 0.005$    \\
July 8  (evening) & $0.014 \pm 0.005$  \\
July 9  (evening) &$0.011 \pm 0.005$   \\
July 12 (evening) & $0.024 \pm 0.005$  \\
\botrule
\end{tabular}}
\end{table}
\begin{figure}
\centerline{\includegraphics[width=5.0 cm]{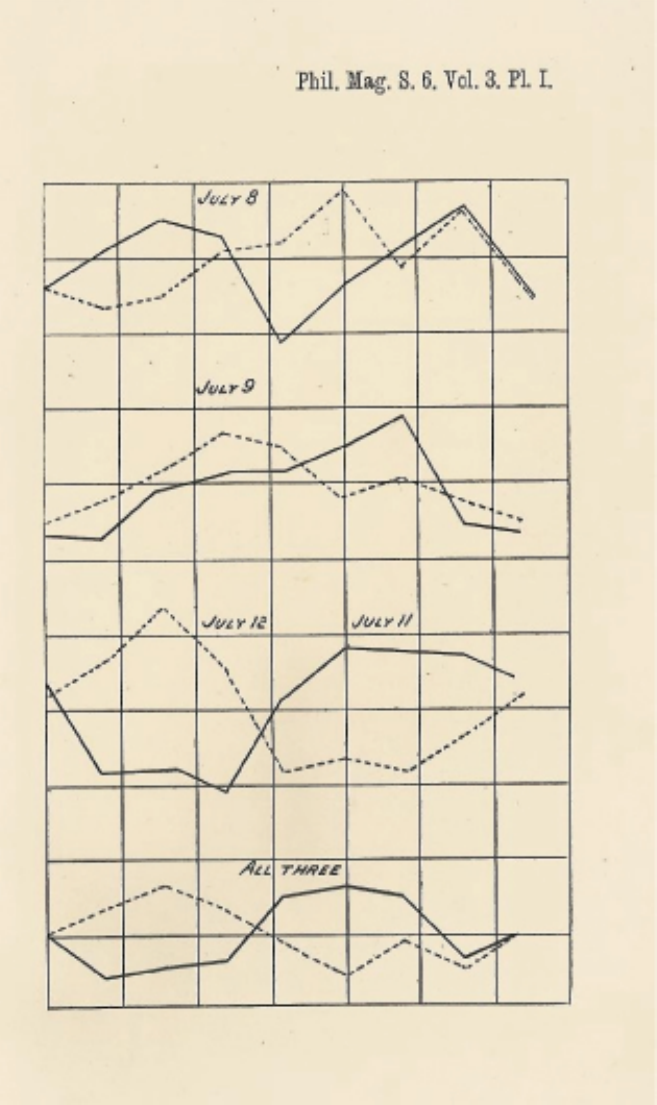}} \caption{\it
The  even combination of fringe shifts $B(\theta)$ for the
Michelson-Morley data as reported by Hicks \cite{hicks}. Solid and
dashed lines refer respectively to noon and evening observations.
Compare the solid curve of July 11th with the analogous curve in
Fig.\ref{fit11july}. }\label{figurehicks}
\end{figure}
As such, the average measured amplitude $A^{\rm EXP}_2 \sim 0.016$,
although much smaller than the classical expectation $A^{\rm
class}_2\sim$ 0.2, was not completely negligible. Thus it is natural
to ask: should this value be interpreted as a typical instrumental
artifact (a ``null result'') or could also indicate a genuine
ether-drift effect? Of course, this question is not new and, in the
past, greatest experts have raised objections to the standard null
interpretation of the data. This point of view was well summarized
by Miller in 1933 \cite{miller} as follows: ``The brief series of
observations (by Michelson and Morley) was sufficient to show
clearly that the effect did not have the anticipated magnitude.
However, and this fact must be emphasized, {\it the indicated effect
was not zero}''. The same conclusion had already been obtained by
Hicks in 1902 \cite{hicks}: ``The data published by Michelson and
Morley, instead of giving a null result, show distinct evidence for
an effect of the kind to be expected''. There was a 2nd-harmonic
effect whose amplitude, however, was substantially smaller than the
classical expectation (see Fig.\ref{figurehicks}).

\begin{figure}[h]
\centerline{\includegraphics[width=6.5cm]{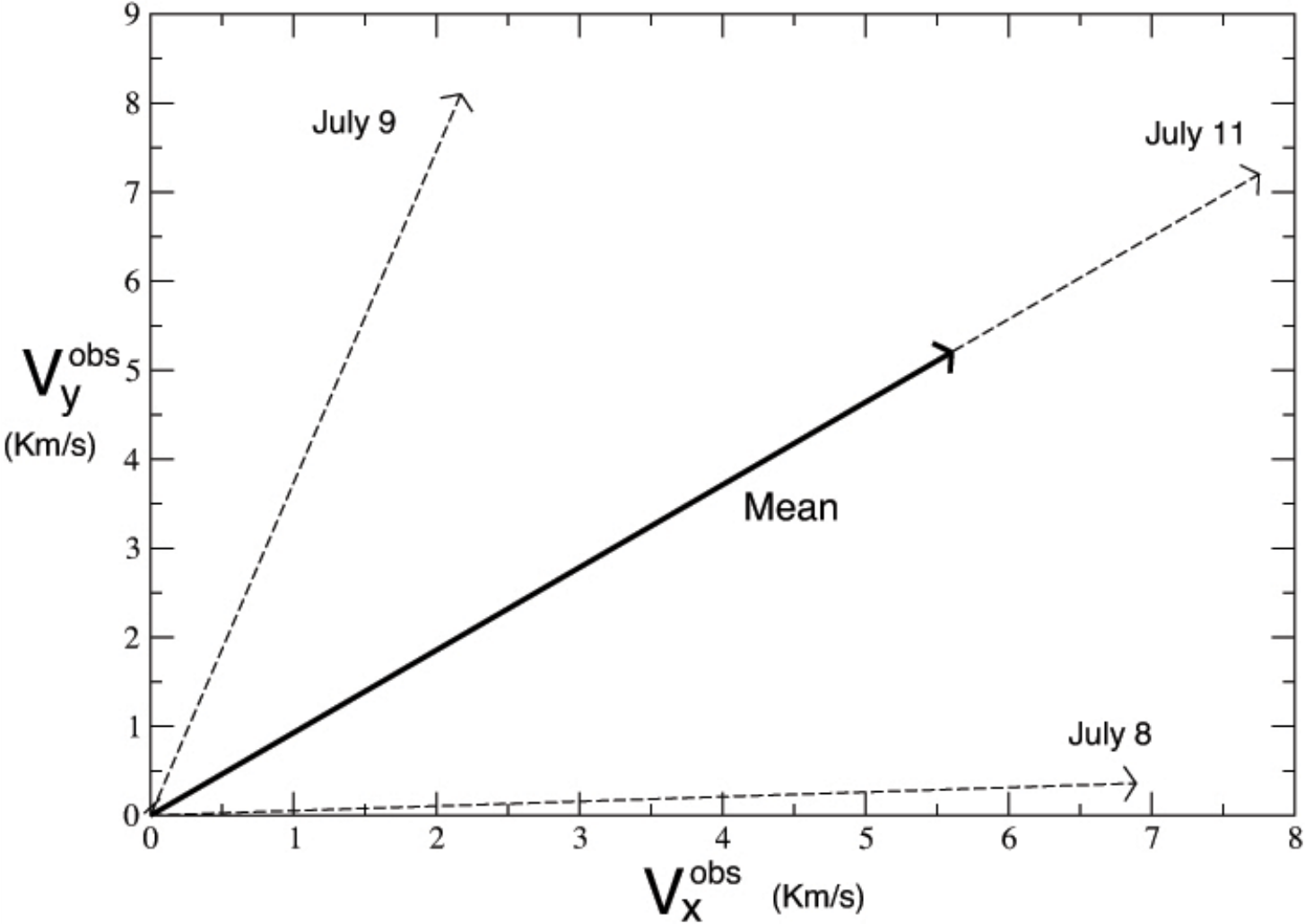}} 
\caption{\it
The observable velocities for the three noon sessions of the
Michelson-Morley experiment and their mean. The x-axis corresponds
to $\theta_2=0^o\equiv 360^o$ and the y-axis to $\theta_2=270^o$.
Statistical uncertainties of the various determinations are not
reported. All individual directions could also be reversed by
180$^o$. The figure is taken from \cite{plus}. }\label{directions}
\end{figure}

Thus the real point about the Michelson-Morley data does {\it not}
concern the small magnitude of the amplitude but the sizeable
changes in the `azimuth', i.e in the phase $\theta_2$ of the
2nd-harmonic which gives the direction of the drift in the plane of
the interferometer. By performing observations at the same hour on
consecutive days (so that variations in the orbital motion are
negligible) one expects that this angle should remain the same
within the statistical errors. Now, by taking into account that, in
a 2nd-harmonic effect, the angle is always defined up to $\pm
180^o$, one choice for the experimental $\theta_2-$values is $357^o
\pm 14^o$, $285^o \pm 10^o$ and $317^o \pm 8^o$ respectively for the
noon sessions of July 8th, 9th and 11th. For this assignment, the
individual velocity vectors $v_{\rm
obs}(\cos\theta_2,-\sin\theta_2)$ and their mean are shown in Fig.
\ref{directions}. As a consequence, directly averaging the
amplitudes of the individual sessions is considerably different from
first performing the vector average of the data and then computing
the resulting amplitude. In the latter case, the average amplitude
is reduced from 0.016 to about 0.011, with a central value of the
observable velocity which decreases from 8.5 km/s to 7 km/s.

This irregular character of the observations has always represented a
strong argument to interpret the small observed residuals as typical
instrumental effects.
\\
\\

\subsection{Further insights: Miller vs. Piccard-Stahel}

To get further insights we have compared two other sets of
measurements, namely Miller's observations \cite{miller} and those
performed by Piccard and Stahel \cite{piccard3}. Miller's large
interferometer had an optical path of about 32 metres and was
installed on top of Mount Wilson. His most extensive observations
were made in blocks of ten days around April 1, August 1, September
15, 1925, and later on around February 8, 1926, with a total number
of 6402 turns of the interferometer \cite{miller}. The result of his
1925 measurements, presented at the APS meeting in Kansas City on
December 1925, was confirming his original claim of 1921, namely
``there is a positive, systematic ether-drift effect, corresponding
to a relative motion of the Earth and the ether, which at Mt. Wilson
has an apparent velocity of 10 km/s''.

Being aware that Miller's previous 1921 announcement of a non-zero
ether-drift of about 9 km/s, if taken seriously, could undermine the
basis of Einstein's relativity (Miller's results ``carried a mortal
blow to the theory of relativity''), Piccard and Stahel realized a
precise apparatus with a small optical path of 280 cm that could be
carried on board of a free atmospheric balloon (at heights of 2500
and 4500 m) to check the dependence on altitude. In this first
series of measurements thermal disturbances were so strong that they
could only set an upper limit of about 9 km/s to the magnitude of
any ether-drift. However, after this first series of trials, precise
observations were performed on dry land in Brussels and on top of
Mt.Rigi in Switzerland (at an height of 1800 m).

Despite the optical path was much shorter than the size of the
instruments used at that times in the United States, Piccard and
Stahel were convinced that the precision of their measurements was
higher because spurious disturbances were less important. In
particular, with respect to the traditional direct observation, the
fringe shifts were registered by photographic recording. Also, for
thermal insulation, the interferometer was surrounded either by a
thermostat filled with ice or by an iron enclosure where it could be
possible to create a vacuum. This last solution was considered after
having understood that the main instability in the fringe system was
due to thermal disturbances in the air of the optical arms (rather
than to temperature differences in the solid parts of the
apparatus). However, very often the interference fringes were put
out of order after few minutes by the presence of residual bubbles
of air in the vacuum chamber. For this reason, they finally decided
to run the experiment at atmospheric pressure with the ice
thermostat which, by its great heat capacity, was found to stabilize
the temperature in a satisfactory way.

We have thus considered the compatibility of these two experiments.
Miller was always reporting his observations by quoting separately
the amplitude and the phase of the individual sessions. In this way,
as shown in Fig.\ref{miller}, the average observable velocity,
obtained from a classical interpretation of his data, was $v_{\rm
obs}\sim 8.4 \pm 2.2$ km/s. Piccard and Stahel were instead first
performing a vector average of the data and, since the phase was
found to vary in a completely arbitrary way, were quoting the much
smaller value $v_{\rm obs}\sim (1.5\div 1.7)$ km/s. For this reason,
their measurements are traditionally considered a definite
refutation of Miller.

\begin{figure}
\begin{center}
\includegraphics[bb= 0 0 5000 590,
angle=0,scale=0.25]{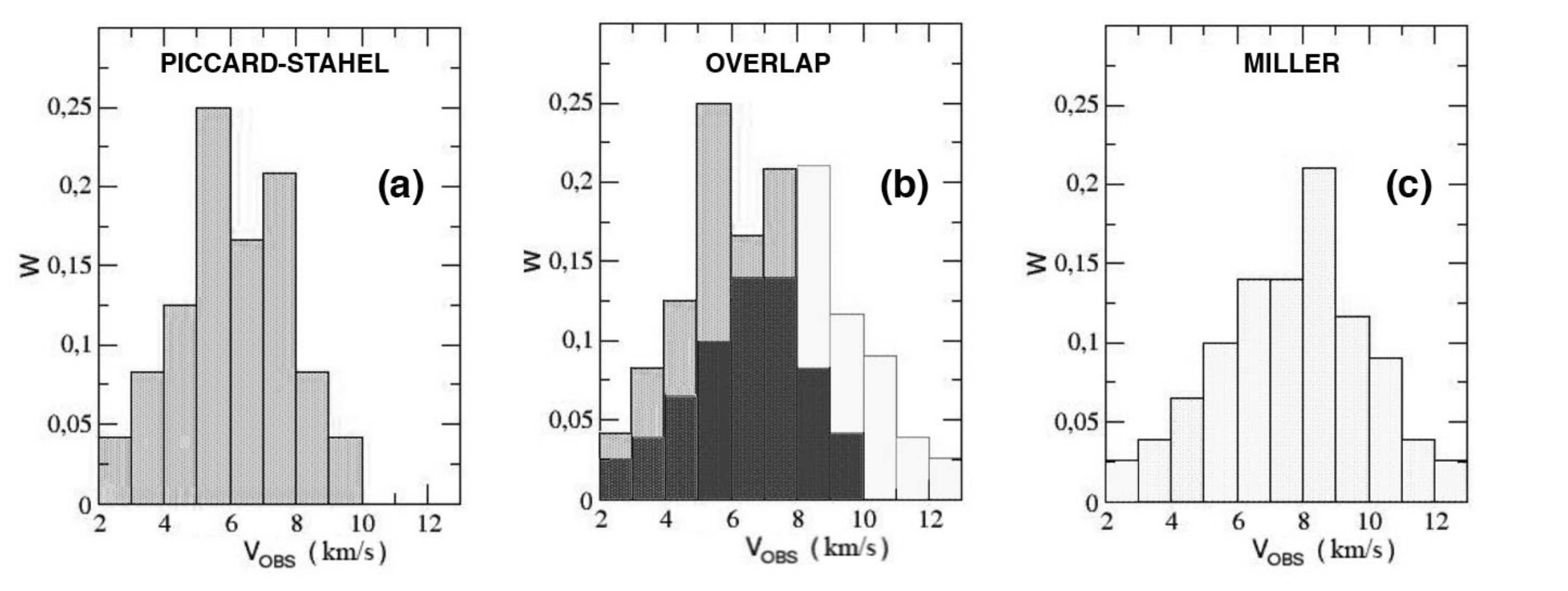}
\end{center}
\caption{\it {We report in panel (a) the probability histogram $W$
for the observable velocity obtained from the 24 individual
amplitudes reported by Piccard and Stahel in \cite{piccard3}. In
panel (c) we report the analogous histogram obtained from Miller's
Figure 22d in \cite{miller}. In both cases, the vertical
normalization is to a unit area. Finally, in panel (b) we report the
overlap of the two histograms. The area of the overlap is 0.645.
This gives a consistency between the two experiments of about
$64\%$. The figure is taken from \cite{book}.} } \label{overlap}
\end{figure}

But suppose that the ether-drift phenomenon has an intrinsic
non-deterministic nature, which induces random fluctuations in the
direction of the local drift. In this case, a vector average of the
data from various sessions would completely obscure the physical
information contained in the individual observations. For this
reason, a meaningful comparison with Miller requires to apply his
same procedure to the Piccard-Stahel data. Namely, first summarizing
each measurement into a definite pair $(A^{\rm EXP}_2, \theta^{\rm
EXP}_2)$ for amplitude and azimuth, and then computing the magnitude
of the observable velocity from the measured amplitudes. With this
different procedure, the Piccard and Stahel observable velocity, at
the 75$\%$ CL, becomes now much larger, namely
\begin{equation}~~~~~~~~~~~~ v_{\rm obs}= 6.3^{+1.5}_{-2.0} ~{\rm
km/s}  ~~~~~~~~~~~~~~~~~~({\rm Piccard-Stahel})\end{equation} and is
now compatible with Miller's results. For a more refined test, we
constructed probability histograms by considering the large set of
measurements reported by Miller in Figure 22d of \cite{miller} and
the 24 individual amplitudes reported by Piccard and Stahel in
\cite{piccard3}, see Fig.\ref{overlap}.
From the area of the overlap, the consistency of the two experiments
can be estimated to be about 64$\%$ which is a quite high level. At
the same time, since the agreement is restricted to the region
$v_{\rm obs} < 9 $ km/s, Miller's higher values are likely affected
by systematic disturbances. This would confirm Piccard and Stahel's
claim that their apparatus, although of a smaller size, was more
precise.

Therefore, summarizing, there is a range of observable velocity, say
$v_{\rm obs} \sim 6.0 \pm 2.0 $ km/s, where the results of the three
experiments we have considered, namely Michelson-Morley, Miller and
Piccard and Stahel, overlap consistently. This common range is
obtained from the 2nd-harmonic amplitudes measured in a plenty of
experimental sessions performed at different sidereal times and in
different laboratories. As such, to a large extent, it should also
be independent of spurious systematic effects. On the basis of
Eq.(\ref{vobs}), this range corresponds to a true kinematic velocity
$v \sim 250 \pm 80 $ km/s which could reasonably fit with the
projection of the Earth velocity within the CMB at the various
laboratories. Truly enough, this is only a first, partial view which
must be supplemented by a deeper understanding of the observed
random variability of the phase.

\section{Going deeper into the ether-drift phenomenon}

The traditional expectation that an ether drift should precisely
follow the smooth modulations induced by the Earth rotation, derives
from the identification of the {\it local} velocity field which
describes the drift in the plane of the interferometer, say
$v_\mu(t)$, with the corresponding projection of the {\it global}
Earth motion, say $\tilde v_\mu(t)$. By comparing with the motion of
a macroscopic body in a fluid, this identification is equivalent to
assume a form of regular, laminar flow, where global and local
velocity fields coincide. Depending on the nature of the fluid, this
assumption may be incorrect.

\begin{figure}
\centerline{\includegraphics[width=6.5cm]{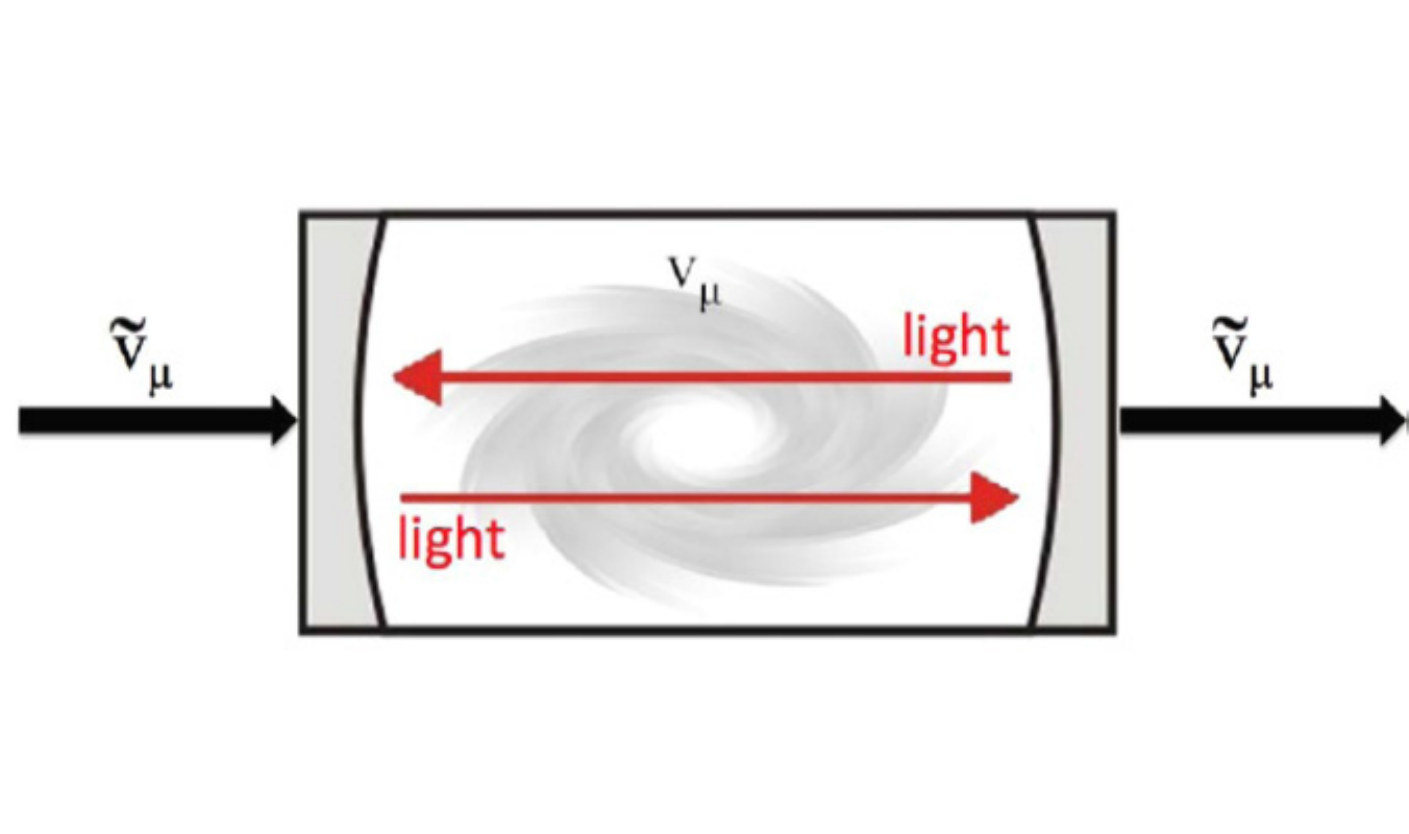}} \caption{\it
The propagation of light in an optical cavity. It is emphasized
that, independently of its particular name (physical vacuum,
ether...) and differently from the solid parts of the apparatus, the
underlying substrate is {\it not} completely entrained with the
Earth motion. Thus, in general, its state of motion $v_\mu(t)$ is
different from $\tilde v_\mu(t)$.} \label{CAVITY2}
\end{figure}

To formulate an alternative model of ether drift, in refs.
\cite{plus,plus2,book,universe}, we started from Maxwell's original
view \cite{maxwell} of light as a {\it wave process} which takes
place in some substrate: ``We are therefore obliged to suppose that
the medium through which light is propagated is something distinct
from the transparent media known to us''. He was calling the
underlying substrate `ether' while, today, we prefer to call it
`physical vacuum'. However, this is irrelevant. The essential point
for the propagation of light, e.g. inside an optical cavity, is
that, differently from the solid parts of the apparatus, this
physical vacuum is {\it not} completely entrained with the Earth
motion see Fig.\ref{CAVITY2}.

Thus, to explain the irregular character of the data, one could try
to model the state of motion of the vacuum substrate as in a
turbulent fluid \cite{chaos,physica} or, more precisely, as in a
fluid in the limit of zero viscosity. Then, the simple picture of a
laminar flow is no more obvious due to the subtlety of the
infinite-Reynolds-number limit, see e.g. Sect. 41.5 in Vol.II of
Feynman's lectures \cite{feybook}. In fact, beside the laminar
regime where $v_\mu(t)=\tilde v_\mu(t)$, there is also another
solution where $v_\mu(t)$ becomes a continuous but nowhere
differentiable velocity field \cite{onsager,eyink} \footnote{The
idea of the physical vacuum as an underlying stochastic medium,
similar to a turbulent fluid, is deeply rooted in basic foundational
aspects of both quantum physics and relativity. For instance, at the
end of XIX century, the last model of the ether was a fluid full of
very small whirlpools (a `vortex-sponge') \cite{whittaker}. The
hydrodynamics of this medium was accounting for Maxwell equations
and thus providing a model of Lorentz symmetry as emerging from a
system whose elementary constituents are governed by Newtonian
dynamics. In a different perspective, the idea of a quantum ether,
as a medium subject to the fluctuations of the uncertainty
relations, was considered by Dirac \cite{dirac}. More recently, the
model of turbulent ether has been re-formulated by Troshkin
\cite{troshkin} (see also \cite{puthoff} and \cite{tsankov}) within
the Navier-Stokes equation, by Saul \cite{saul} within Boltzmann's
transport equation and in \cite{pla12} within Landau's
hydrodynamics. The same picture of the vacuum (or ether) as a
turbulent fluid was Nelson's \cite{nelson} starting point. In
particular, the zero-viscosity limit gave him the motivation to
expect that ``the Brownian motion in the ether will not be smooth''
and, therefore, to conceive the particular form of kinematics at the
base of his stochastic derivation of the Schr\"odinger equation. A
qualitatively similar picture is also obtained by representing
relativistic particle propagation from the superposition, at short
time scales, of non-relativistic particle paths with different
Newtonian mass \cite{kleinert}. In this formulation, particles
randomly propagate (as in a Brownian motion) in an underlying
granular medium which replaces the trivial empty vacuum
\cite{jizba}.}.

\begin{figure}
\centerline{\includegraphics[width=14cm]{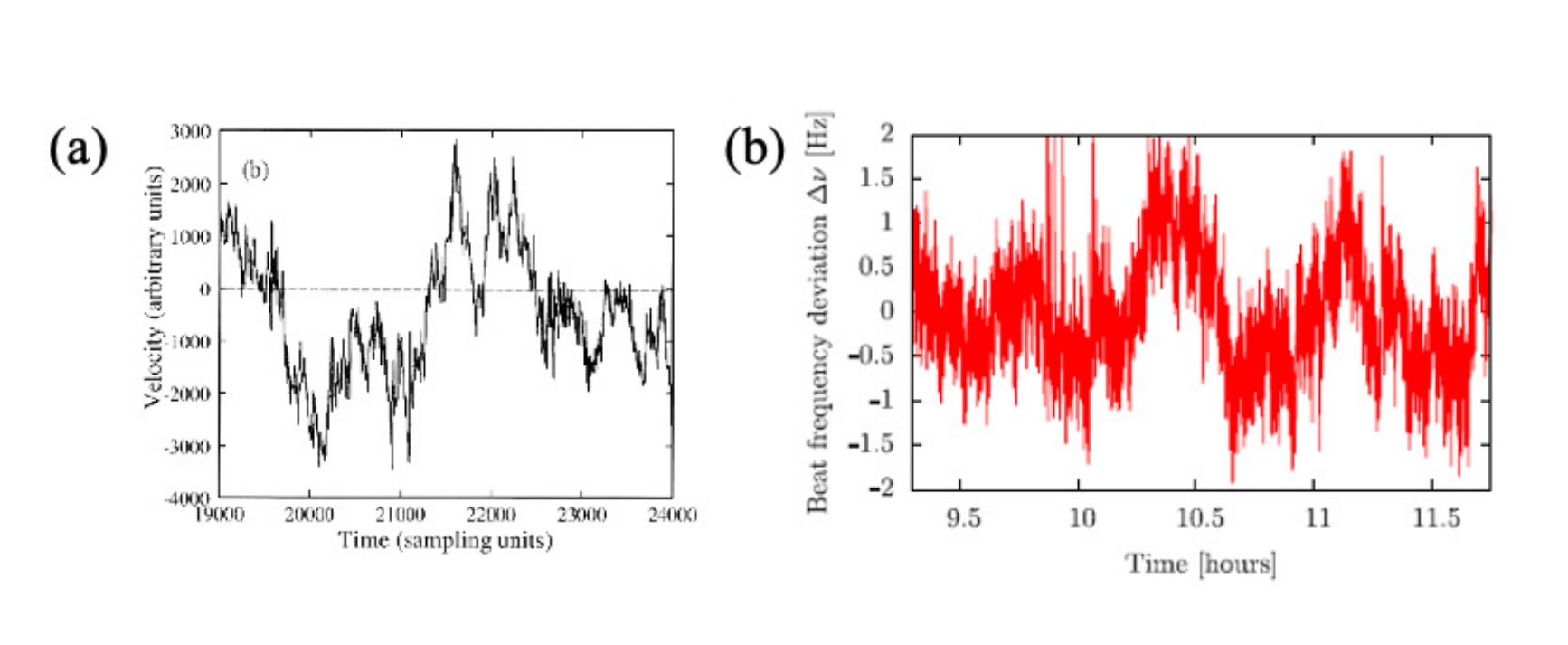}}
\caption{\it We compare two different signals. Panel a) reports the
turbulent velocity field measured in a wind tunnel \cite{frisch}.
Panel b) reports the instantaneous frequency shift measured with
vacuum optical resonators in ref.\cite{schiller2015}. For the
adopted laser frequency $\nu_0=2.8\cdot 10^{14}$ Hz a $\Delta
\nu=\pm 1$ Hz corresponds to a fractional value $\Delta \nu/\nu_0$
of about $\pm 3.5 \cdot 10^{-15}$.} \label{turbulence_signal}
\end{figure}

Together with these theoretical arguments, the analogy with a
turbulent flow finds support in modern ether drift experiments where
one measures the frequency shifts of two optical resonators. To this
end, consider Fig.\ref{turbulence_signal}. Panel a) reports the
turbulent velocity field measured in a wind tunnel \cite{frisch}. No
doubt, this is a genuine signal, not noise. Panel b) reports instead
the instantaneous frequency shift measured with vacuum optical
cavities in ref.\cite{schiller2015}. So far, this other signal is
interpreted as spurious noise.

\begin{figure}
\centerline{\includegraphics[width=13cm]{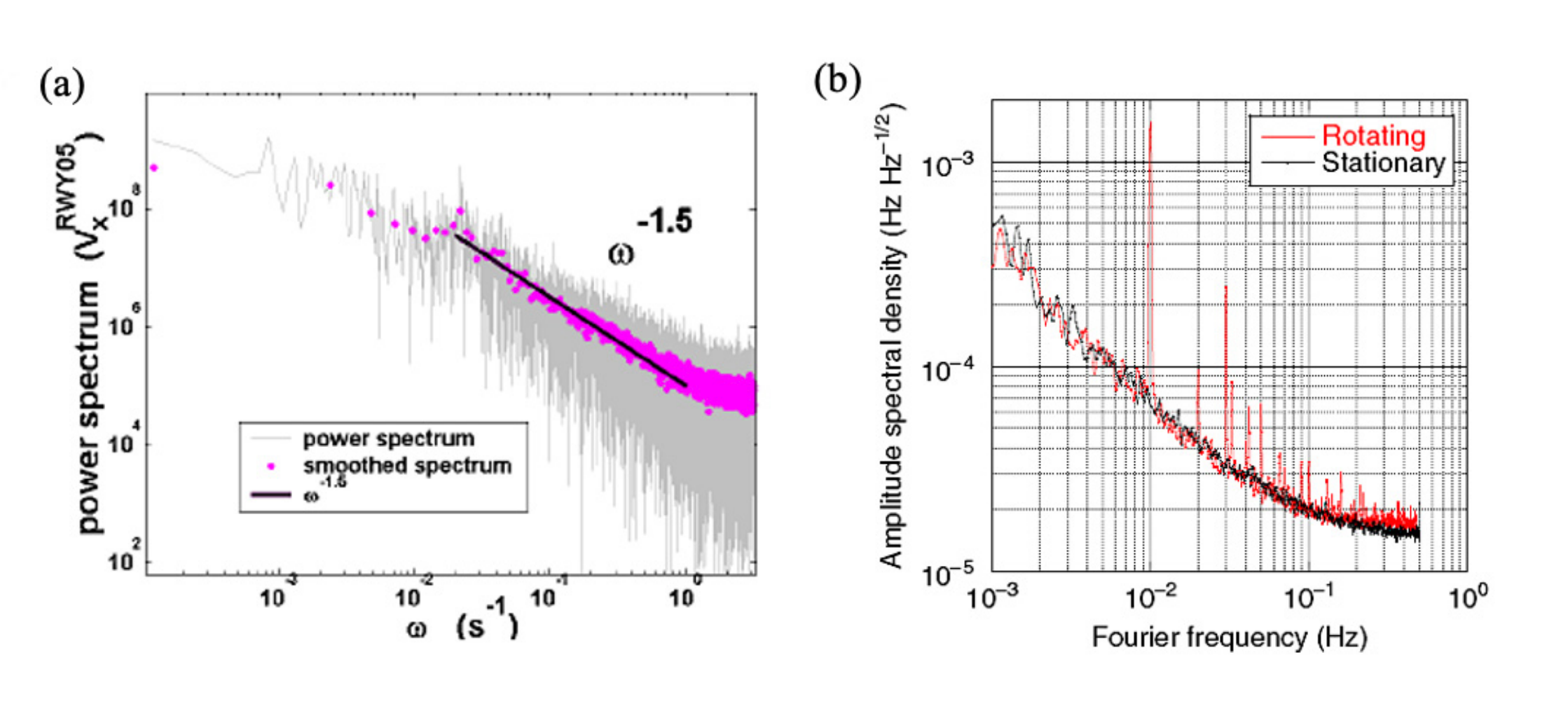}}
\caption{\it We compare two different signals. Panel a) reports the
power spectrum $S(\omega)\sim \omega^{-1.5} $ of the wind turbulence
measured at the Florence Airport \cite{rizzo}. Panel b) shows the
spectral amplitude $\sqrt {S(\omega)}\sim \omega^{-0.7}$ of the
frequency shift measured by Nagel et al. \cite{nagelnature}. Above
some minimal frequency the two curves reach a flat plateau. This
corresponds to the maximum integration time beyond which the signal
ceases to behave as a pure white-noise.} \label{turbulence_spectrum}
\end{figure}

Consider now Fig.\ref{turbulence_spectrum}. Panel a) shows the power
spectrum $S(\omega)\sim \omega^{-1.5} $ of the wind turbulence
measured at the Florence Airport \cite{rizzo}. No doubt, this is a
physical signal. Panel b) shows the spectral amplitude $\sqrt
{S(\omega)}\sim \omega^{-0.7}$ of the frequency shift measured by
Nagel et al. \cite{nagelnature}. Again, this latter signal is
interpreted as spurious noise.

Clearly these are just analogies but, very often, physical
understanding proceeds by analogies. We have
thus exploited the idea that the irregular signal observed in ether-drift
experiments has a fundamental stochastic nature as when turbulence,
at small scales, becomes statistically homogeneous and isotropic. With such an irregular signal numerical simulations are needed for a consistent description of the data. Therefore, for a check, one
should first extract from the data the (2nd-harmonic) phase and
amplitude and concentrate on the latter which is positive definite
and remains non-zero under any averaging procedure. When measured at
different times, this amplitude will anyhow exhibit modulations
that, though indirectly, can provide information on the underlying
cosmic motion.

To put things on a quantitative basis, let us assume the set of
kinematic parameters $(V,\alpha,\gamma)_{\rm CMB}$ for the Earth
motion in the CMB, a latitude $\phi$ of the laboratory and a given
sidereal time $\tau=\omega_{\rm sid}t$ of the observations (with
$\omega_{\rm sid}\sim {{2\pi}\over{23^{h}56'}}$). Then, for short-time
observations of a few days, where the only time dependence is due to the Earth rotation, 
a simple application of spherical trigonometry \cite{nassau} gives the projections
in the $(x,y)$ plane of the interferometer
\begin{equation} \label{tnassau2}
   \tilde{v}_x(t)  = \tilde{v}(t)\cos\tilde{\theta}_2(t)= V\left[ \sin\gamma\cos \phi -\cos\gamma
      \sin\phi \cos(\tau-\alpha)\right]
\end{equation} 
\begin{equation} \label{tnassau3}
     \tilde{v}_y(t)= \tilde{v}(t)\sin\tilde{\theta}_2(t)= V\cos\gamma\sin(\tau-\alpha) \end{equation}
with a magnitude 
\begin{equation} \label{tprojection}
      \tilde {v}(t) =V |\sin z(t)| 
\end{equation}
and
\begin{equation} \label{tnassau1}
      \cos z(t)= \sin\gamma\sin \phi + \cos\gamma
      \cos\phi \cos(\tau-\alpha)
\end{equation}
As for the signal, let us also re-write Eq.(\ref{bbasic2new}) as
\begin{equation} \label{basictext}
     {{\Delta \bar{c}_\theta(t) } \over{c}}
    \sim
 \epsilon {{v^2(t) }\over{c^2}}\cos 2(\theta
-\theta_2(t)) \end{equation}  where $v(t)$ and $\theta_2(t)$ now
indicate respectively the magnitude and direction of the {\it local} drift
in the same $(x,y)$ plane of the interferometer. This can also be
re-written as
\begin{equation}  \label{amplitude10old} {{\Delta \bar{c}_\theta(t) } \over{c}}\sim
2{S}(t)\sin 2\theta +
      2{C}(t)\cos 2\theta \end{equation} with \begin{equation}
       2C(t)= \epsilon~ {{v^2_x(t)- v^2_y(t)  }
       \over{c^2}}~~~~~~~2S(t)=\epsilon ~{{2v_x(t)v_y(t)  }\over{c^2}}
\end{equation}  and $v_x(t)=v(t)\cos\theta_2(t)$,
$v_y(t)=v(t)\sin\theta_2(t)$. 

In an analogy with a turbulent flow, the requirement of statistical isotropy means that the local quantities $v_x(t)$ and $v_y(t)$, which determine the
observable properties of the drift, are very irregular functions that differ non trivially from their smooth, global counterparts $\tilde{v}_x(t)$ and $\tilde{v}_y(t)$, and can only be simulated numerically. To this end, a representation in terms of random Fourier series
\cite{onsager,landau,fung} was adopted in
refs.\cite{plus,plus2,book,universe} in a simplest
uniform-probability model, where the kinematic parameters of the
global $\tilde v_\mu(t)$ are just used to fix the boundaries for
the local random $v_\mu(t)$. The basic ingredients are summarized in
the Appendix.

In this model, the functions $S(t)$ and $C(t)$ have the
characteristic behaviour of a {\it white-noise} signal with vanishing
statistical averages $\langle C(t)\rangle_ {\rm stat}=0$ and
$\langle S(t)\rangle_ {\rm stat}=0$ at {\it any} time $t$ and
whatever the global cosmic motion of the Earth. One can then understand the
observed irregular behaviour of the fringe shifts
\begin{equation} \label{fringe} {{\Delta
\lambda(\theta;t)}\over{\lambda}} = {{2D}\over{\lambda}} \left[2
S(t) \sin 2\theta  + 2 C(t)\cos 2\theta \right]
\end{equation}
In fact, their averages would be non vanishing just because the
statistics is finite. Otherwise with more and more observations one
would find
\begin{equation} \label{averageold}\langle{{\Delta
\lambda(\theta;t)}\over{\lambda}}\rangle_ {\rm stat}=
{{2D}\over{\lambda}} \left[2\sin 2\theta~\langle S(t)\rangle_ {\rm
stat} + 2\cos 2\theta~\langle C(t)\rangle_ {\rm stat} \right] \to 0
\end{equation} In particular, the direction $\theta_2(t)$ of the local drift, defined by
the relation $\tan2\theta_2(t)= S(t)/C(t)$, would vary randomly with
no definite limit.

We have then checked the model by comparing with the amplitudes.
Here we have first to consider the theoretical amplitude $\tilde
A_2(t)$ associated with the global motion \BE \label{smoothfinal}
\tilde A_2(t) \sim {{D }\over{\lambda}} \cdot 2\epsilon\cdot {{V^2
\sin^2z(t) } \over{c^2 }} \EE and then the amplitude $A_2(t)$
associated with the local non differentiable velocity components
$v_x(t)$ and $v_y(t)$, Eqs.(\ref{vx}) and (\ref{vy}) of the Appendix
\BE \label{irregularfinal}  A_2(t) \sim {{D }\over{\lambda}} \cdot
2\epsilon\cdot {{v^2_x(t) + v^2_y(t) } \over{c^2 }} \EE Clearly, the
latter will exhibit sizeable fluctuations and be very different from
the smooth $\tilde A_2(t)$. However, as shown in the Appendix, the
relation between $\tilde A_2(t)$ and the statistical average
$\langle A_2(t) \rangle_{\rm stat}$  is extremely simple \BE
\label{amplitude10001finalintro} \langle A_2(t)\rangle_{\rm
stat}={{D }\over{\lambda}}~\cdot 2\epsilon \cdot{{ \langle v^2_x(t)+
v^2_y(t)\rangle_{\rm stat} } \over{c^2}}\sim {{\pi^2 } \over{18
}}\cdot \tilde A_2(t)\EE so that, by averaging the amplitude at different sidereal
times, one can obtain the crucial information on the angular
parameters $\alpha$ and $\gamma$.

Altogether, the amplitudes of those old measurements can thus be
interpreted in terms of three different velocities: a) as $6 \pm 2$
km/s in a classical picture b) as $250 \pm 80$ km/s, in a
modern scheme and in a smooth picture of the drift c) as $340 \pm
110$ km/s, in a modern scheme but now allowing for irregular
fluctuations of the signal. In fact, by replacing
Eq.(\ref{smoothfinal}) with Eq.(\ref{amplitude10001finalintro}),
from the same data, one would now obtain kinematical velocities
which are larger by a factor $\sqrt{18/{\pi^2}} \sim 1.35$. In this
third interpretation, the range of velocity agrees much better with the CMB value of 370 km/s.

To illustrate the agreement of our scheme with all classical
measurements, we address to our book \cite{book} where a detailed
description is given of the experiments by Morley-Miller
\cite{morley}, Miller \cite{miller}, Kennedy \cite{kenconference},
Illingworth \cite{illingworth}, Tomaschek \cite{tomaschek1} and
Piccard-Stahel \cite{piccard3}. Instead, here, we will only consider
the two most precise experiments that, traditionally, have been
considered as definitely ruling out Miller's claims for a non-zero ether drift. Namely the Michelson-Pease-Pearson (MPP)
observations at Mt. Wilson and the experiment performed in 1930 by
Joos in Jena \cite{joos}. In particular, the latter remains
incomparable among the classical experiments. To have an idea,
Sommerfeld, being aware that the residuals in the Michelson-Morley
data were not entirely negligible, concluded that only ``After its
repetition at Jena by Joos, the negative result of Michelson's
experiment can be considered as definitely established'' (A.
Sommerfeld, Optics). However, there is again a subtlety because, as
we shall see, Joos' experiment was {\it not} performed in the same
conditions as the other experiments we have previously considered.

\subsection{Reanalysis of the MPP experiment}

To re-analyze the Michelson-Pease-Pearson (MPP) experiment, we first
observe that no numerical results are reported in the original
articles \cite{mpp,mpp2}. Instead, for more precise indications, one
should look at Pease's paper \cite{pease}. There, one learns that
they concentrated on a purely differential type of measurements.
Namely, they were first statistically averaging the fringe shifts at those sidereal times that,
according to Miller, were corresponding to maxima and minima of the
ether-drift effect. Then, they were forming the difference \BE
\label{delta} \delta (\theta)= \langle {{\Delta\lambda(\theta;t_{\rm
max}) }\over{\lambda}} \rangle_{\rm stat} - \langle
{{\Delta\lambda(\theta;t_{\rm min}) }\over{\lambda}} \rangle_{\rm
stat} \EE which are the only numbers reported by Pease. These
$\delta-$values have a maximal magnitude of $\pm 0.004$ and this is
also the order of magnitude of the experimental amplitude $A^{\rm
EXP}_2\sim$ 0.005 that is usually reported \cite{shankland} for the
MPP experiment when comparing with the much larger expected
classical amplitudes $A^{\rm class}_2 \sim 0.45$ or $A^{\rm class}_2
\sim 0.29$ for optical paths of eighty-five or fifty-five feet
respectively. Now, our stochastic, isotropic model predicts exactly
zero statistical averages for vector quantities such as the fringe
shifts, see Eq(\ref{averageold}). Therefore, it would be trivial to
reproduce the small $\delta$-values in Eq.(\ref{delta}) in a numerical simulation
with sufficiently high statistics. We have thus decided to compare
instead with the only basic experimental session reported
by Pease \cite{pease} (for optical path of fifty-five feet) which
indicates a 2nd-harmonic amplitude $A^{\rm EXP}_2 \sim
0.006$. By comparing with the classical prediction for 30 km/s,
namely $A^{\rm class}_2 \sim 0.29$, this amplitude corresponds to an
observable velocity $v_{\rm obs} \sim$ 4.3 km/s but to a much larger
value on the basis of Eq.(\ref{irregularfinal}).

\begin{figure}
\begin{center}
\includegraphics[scale=0.45]{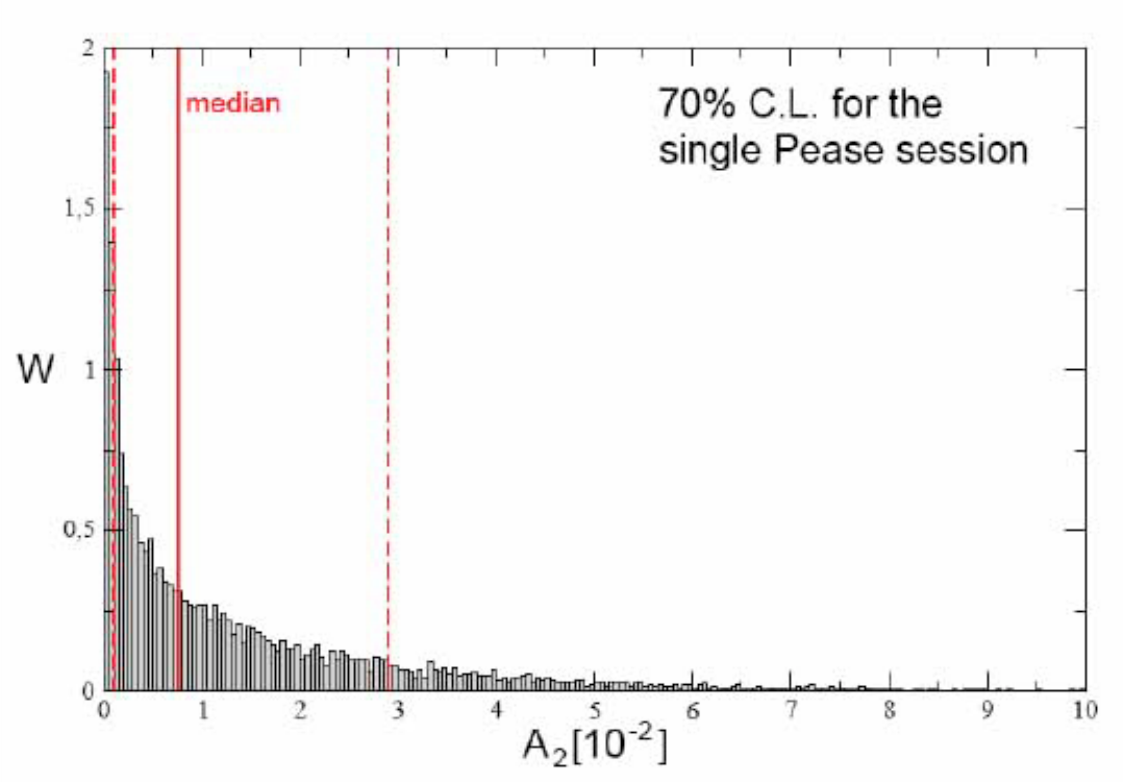}
\end{center}
\caption{ {\it The histogram W of a numerical simulation of 10,000
instantaneous amplitudes for the single session of January 13, 1928,
reported by Pease \cite{pease} . The vertical normalization is to a
unit area. We show the median and the 70$\%$ CL. The limits on the
random Fourier components Eqs.(\ref{vx}) and (\ref{vy}) of the
Appendix were fixed by inserting the CMB kinematical parameters in
Eq.(\ref{isot}).}} \label{pease}
\end{figure}
Since we are dealing with a single measurement, to obtain a better
understanding of its probability content, we have performed a direct
numerical simulation by generating 10,000 values of the amplitude at
the same sidereal time 5:30 of the MPP Mt. Wilson observation. The
CMB kinematical parameters were used to bound the random Fourier
components of the stochastic velocity field Eqs.(\ref{vx}) and
(\ref{vy}) of the Appendix. The resulting histogram is reported in
Fig.\ref{pease}. From this histogram one obtains a mean simulated
amplitude $\langle A^{\rm simul}_2 \rangle \sim$ 0.014. This
corresponds to replace the value of the scalar velocity
$\tilde{v}(t) \sim$ 370 km/s Eq.(\ref{projection}), at the sidereal
time of Pease's observation, in the relation for the statistical
average of the amplitude \BE \label{amplitude10004}
       \langle A_2(t)\rangle_{\rm stat}={{2\epsilon D
}\over{\lambda}}~ {{ \langle  v^2_x(t)+ v^2_y(t)\rangle_{\rm stat} }
\over{c^2}} \sim (1.6\cdot 10^4) \cdot {{\pi^2 } \over{18 }} \cdot{{
\tilde{v}^2(t)}
       \over{c^2 }} \sim ~ 0.009 \cdot {{
\tilde{v}^2(t)}
       \over{\rm (300~km/s)^2 }}\EE
 In the above relation we have replaced
$D/\lambda \sim 2.9\cdot 10^{7}$ (for optical path of fifty-five
feet) and $\epsilon \sim 2.8\cdot 10^{-4}$.

Note that the median of the amplitude distribution is about 0.007.
As a consequence, the value $A^{\rm EXP}_2 \sim 0.006$ lies well
within the 70$\%$ Confidence Limit. Also, the probability content
becomes large at very small amplitudes \footnote{Strictly speaking,
for a more precise description of the data, one should fold the
histogram with a smearing function which takes into account the
finite resolution $\Delta$ of the apparatus. This smearing would
force the curve to bend for $A_2 \to 0$ and tend to some limit which
depends on $\Delta$. Nevertheless, this refinement should not modify
substantially the probability content around the median which is
very close to $A_2= 0.007$.} and there is a long tail extending up
to about $A_2 \sim 0.030$.

The wide interval of amplitudes corresponding to the 70$\%$ C. L.
indicates that, in our stochastic model, one could account for
single observations that differ by an order of magnitude, say from
0.003 to 0.030. Thus, beside the statistical vanishing of vector
quantities, this is another crucial difference with a purely
deterministic model of the ether-drift. In this traditional view, in
fact, within the errors, the amplitude can vary at most by a factor
$r=(\tilde{v}_{\rm max}/\tilde{v}_{\rm min})^2$ where $\tilde{v}_{\rm max}$ and $\tilde{v}_{\rm
min}$ are respectively the maximum and minimum of the projection of
the Earth velocity Eq.(\ref{tprojection}). Since, for the known types
of cosmic motion, one finds $r\sim 2$, the observation of such large
fluctuations in the data would induce to conclude, in a
deterministic model, that there is some systematic effect which
affects the measurements in an uncontrolled way. With an ether drift
of such irregular nature, it then becomes understandable the MPP
reluctance to quote the individual results and instead report those
particularly small combinations in Eq.(\ref{delta}) obtained by
averaging and further subtracting large samples of data. This
general picture of a highly irregular phenomenon is also confirmed
by our reanalysis of Joos' experiment in the following subsection.

\subsection{Joos' experiment}

We will only give a brief description of Joos' 1930 experiment
\cite{joos} and address to our book \cite{book} for more details.
Its sensitivity was about 1/3000 of a fringe, the fringes were
recorded photographically with an automatic procedure and the
optical system was enclosed in a hermetic housing. As reported by
Miller \cite{miller,miller34}, it has been traditionally believed
that the measurements were performed in vacuum. In his article,
however, Joos is not clear on this particular aspect. Only when
describing his device for electromagnetic fine movements of the
mirrors, he refers to the condition of an evacuated apparatus
\cite{joos}. Instead, Swenson \cite{swensonbook,loyd2} declares that
Joos' fringe shifts were finally recorded with optical paths placed
in a helium bath. Therefore, we have decided to follow Swenson's
explicit statements and assumed the presence of gaseous helium at
atmospheric pressure.

\begin{figure}
\begin{center}
\includegraphics[width=7.0 cm]{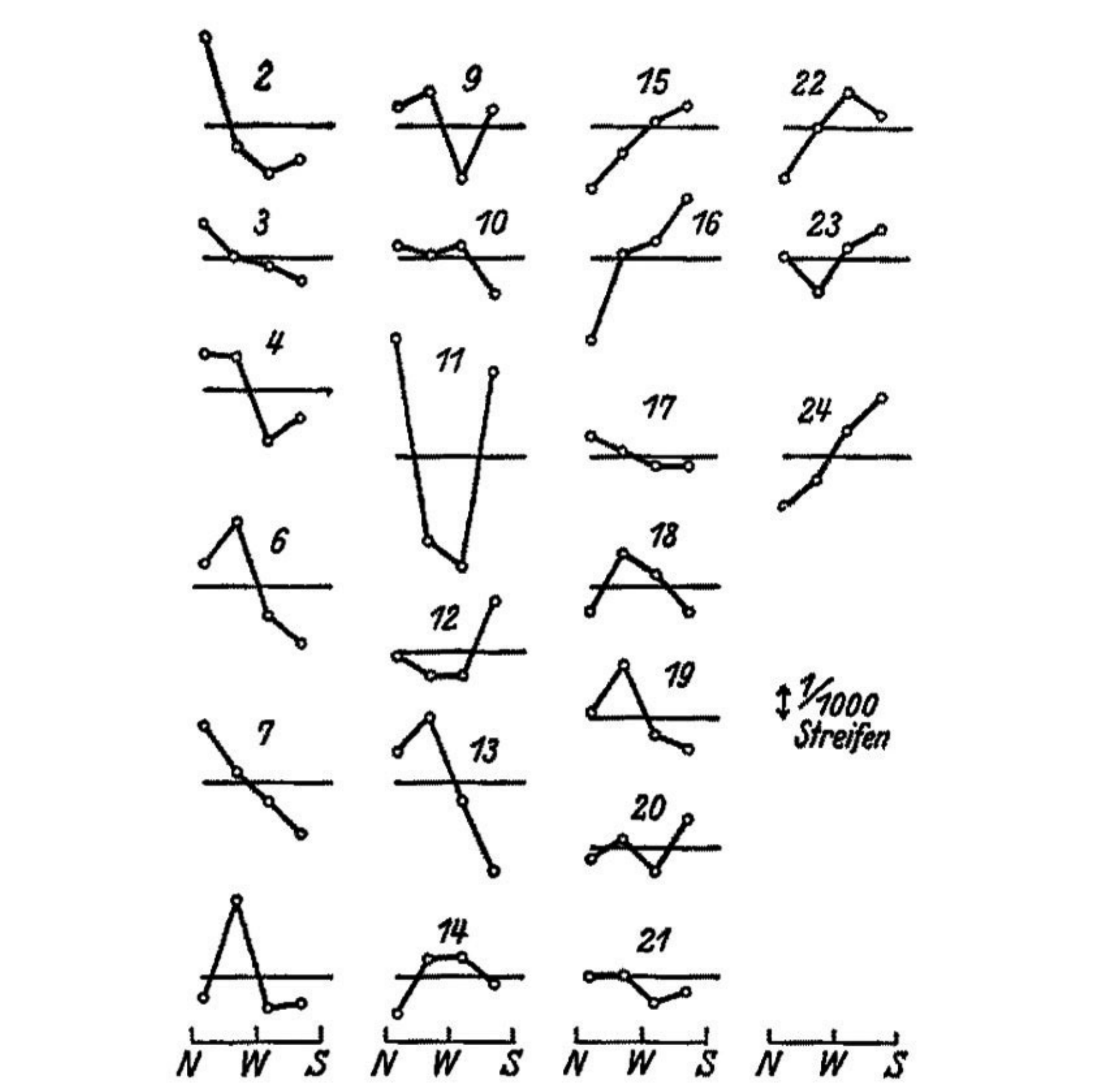}
\end{center}
\caption{\it The fringe shifts reported by Joos \cite{joos}. The
yardstick corresponds to 1/1000 of a wavelength.}
\label{fringe-joos}
\end{figure}

From Eq.(\ref{amplitude10001finalintro}), by replacing $D/\lambda=
3.75 \cdot 10^7$ and the refractive index  ${\cal N}_{\rm
helium}\sim$ 1.000033 for gaseous helium, an average daily
projection of the cosmic Earth velocity $\tilde v(t)=V|\sin
z(t)|\sim$ 330 km/s (appropriate for a Central-Europe laboratory)
would provide the same amplitude as classically expected for the
much smaller observable velocity of 2 km/s. We can thus understand
the substantial reduction of the fringe shifts observed by Joos,
with respect to the other experiments in air.
\begin{figure}
\begin{center}
\includegraphics[width=8.0 cm]{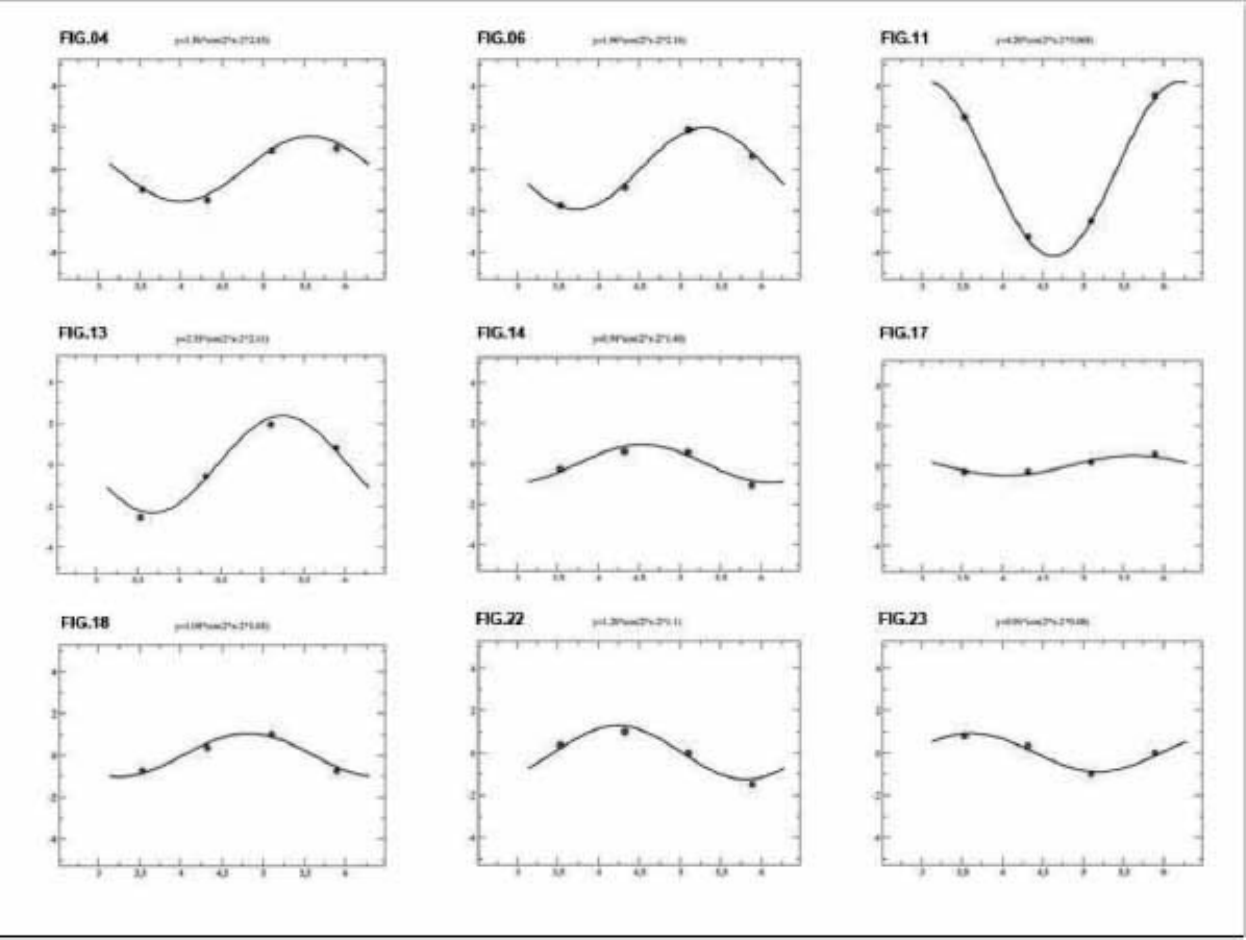}
\end{center}
\caption{\it Some 2nd-harmonic fits to Joos' data. The figure is
taken from ref.\cite{book}.} \label{collage}
\end{figure}

\begin{figure}
\begin{center}
\includegraphics[width=7.0 cm]{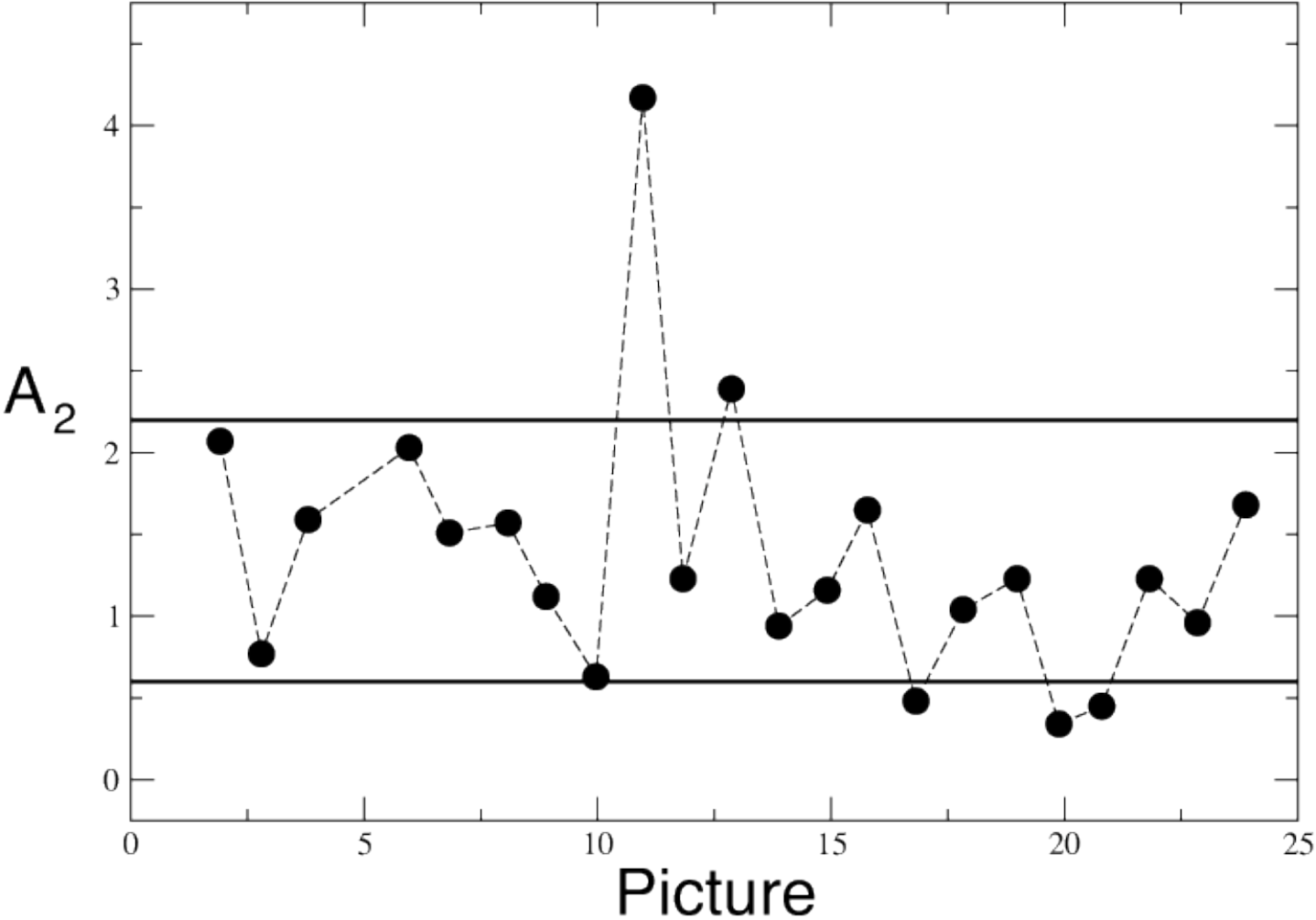}
\end{center}
\caption{\it Joos' 2nd-harmonic amplitudes, in units $10^{-3}$. The
vertical band between the two lines corresponds to the range $(1.4
\pm 0.8)\cdot10^{-3}$. The uncertainty of each value is about $\pm
3\cdot10^{-4}$. The figure is taken from ref.\cite{plus}.}
\label{joosamplitudes}
\end{figure}
The data were taken at steps of one hour during the sidereal day and
two observations (1 and 5) were finally deleted by Joos with the
motivations that there were spurious disturbances, see
Fig.\ref{fringe-joos}. From this picture, Joos adopted 1/1000 of a
wavelength as upper limit and deduced the bound $v_{\rm obs}
\lesssim 1.5$ km/s. To this end, he was comparing with the classical
expectation that, for his apparatus, a velocity of 30 km/s should
have produced a 2nd-harmonic amplitude of 0.375 wavelengths. Though,
since it is apparent that some fringe displacements were certainly
larger than 1/1000 of a wavelength, we have performed 2nd-harmonic
fits to Joos' data, see Fig.\ref{collage}. The resulting amplitudes
are reported in Fig.\ref{joosamplitudes}.

\begin{figure}[h]
\begin{center}
\includegraphics[width=7.5 cm]{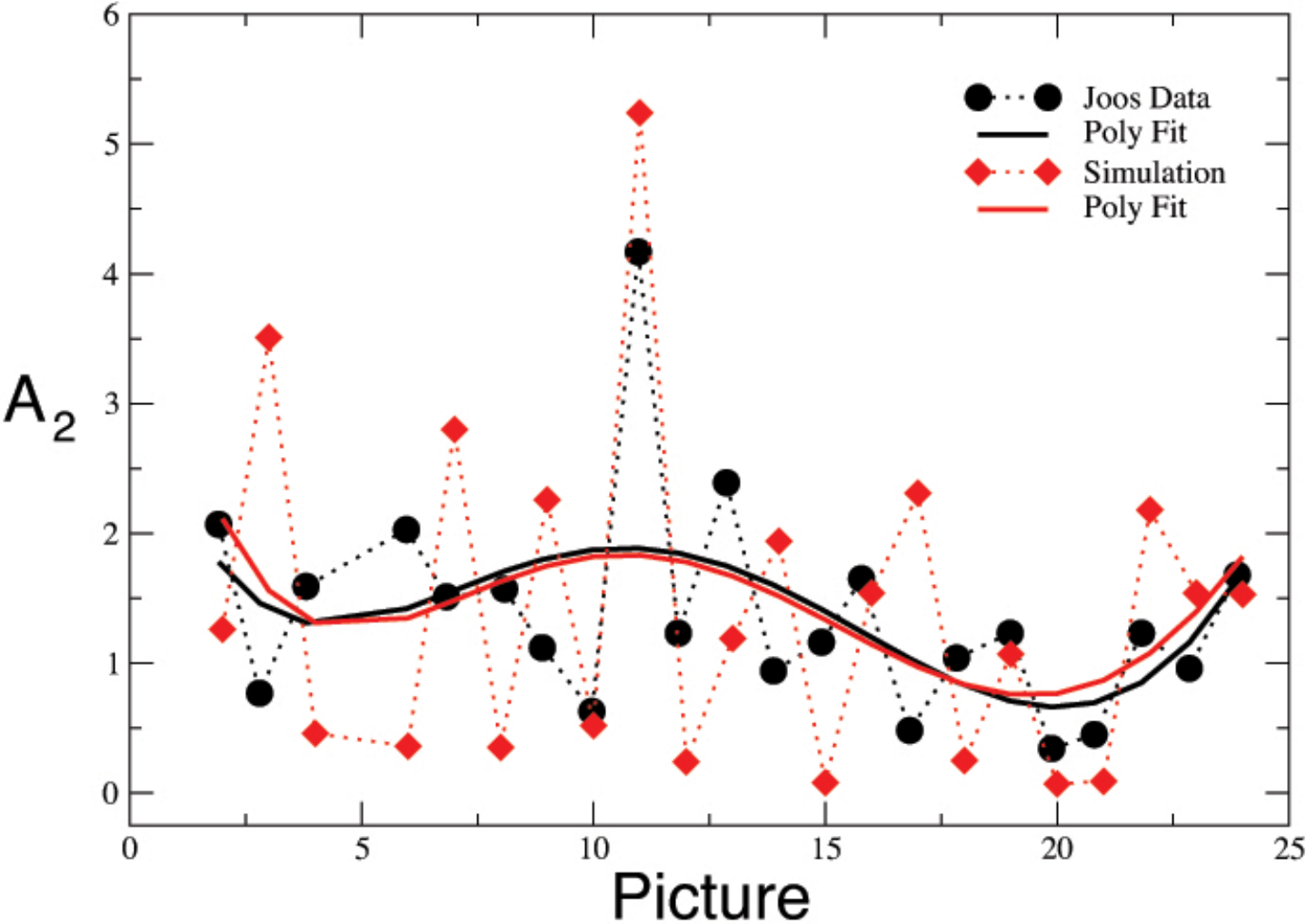}
\end{center}
\caption{\it Joos' 2nd-harmonic amplitudes, in units $10^{-3}$
(black dots), are compared with a single simulation (red diamonds)
at the same sidereal times of Joos' observations. Two 5th-order
polynomial fits to the two sets of values are also shown. The figure
is taken from ref.\cite{plus} .
} \label{joos-comparison}
\end{figure}

\begin{figure}[h]
\begin{center}
\includegraphics[width=7.5 cm]{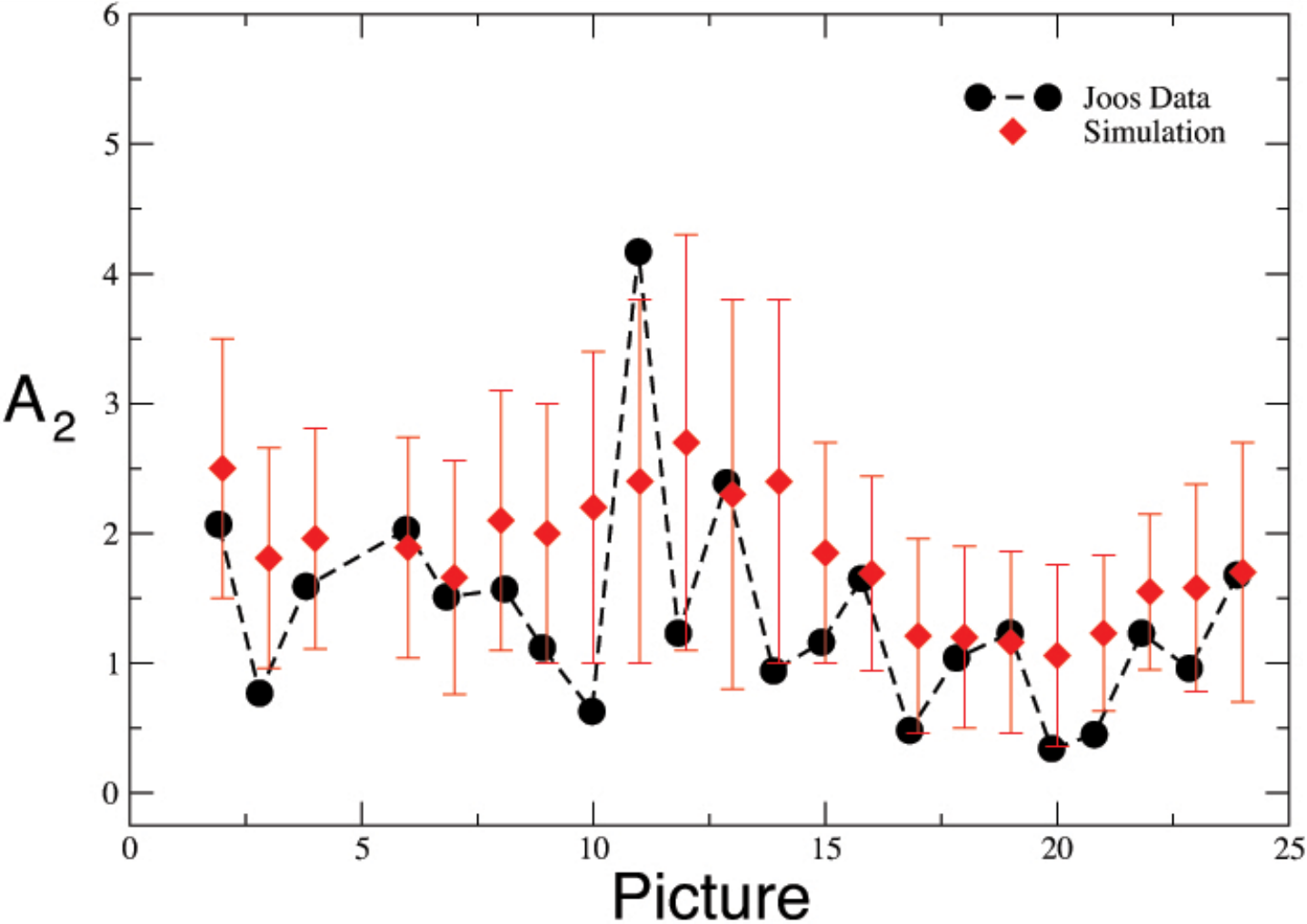}
\end{center}
\caption{\it Joos' 2nd-harmonic amplitudes in units $10^{-3}$ (black
dots) are now compared with a simulation where one averages ten
measurements, performed on 10 consecutive days, at the same sidereal
times of Joos' observations (red diamonds). The change of the
averages observed by varying the parameters of the simulation was
summarized into a central value and a symmetric error. The figure is
taken from ref.\cite{plus}.} \label{joos-comparison-errors}
\end{figure}

We note that a 2nd-harmonic fit to the large fringe shifts in
picture 11 has a very good  chi-square, comparable and often better
than other observations with smaller values, see Fig.\ref{collage}.
Therefore, there is no reason to delete the observation n.11. Its
amplitude, however, $ (4.1 \pm 0.3) \cdot 10^{-3}$ is abot ten times
larger than the average amplitude $ (0.4 \pm 0.3) \cdot 10^{-3}$
from the observations 20 and 21. This difference cannot be
understood in a smooth model of the drift where, as anticipated, the
projected velocity squared at the observation site can at most
differ by a factor of two, as for the CMB motion at typical
Central-Europe latitude where $(\tilde v)_{\rm min} \sim 250$ km/s
and $(\tilde v)_{\rm max} \sim 370$ km/s. To understand these
characteristic fluctuations, we have thus performed various
numerical simulations of these amplitudes \cite{plus,book} in the
stochastic model described in the Appendix and using the kinematical
parameters $(V,\alpha,\gamma)_{\rm CMB}$ to place the limits on the
random velocity component Eqs.(\ref{vx}) and (\ref{vy}).  
\\
\\
Two simulations are shown in Figs.\ref{joos-comparison} and
\ref{joos-comparison-errors} (the corresponding numerical values are
reported in \cite{plus,book}).
\\
\\

We want to emphasize two aspects. First, Joos' average amplitude
$\langle A^{\rm EXP}_2\rangle= (1.4 \pm 0.8)\cdot 10^{-3}$ when
compared with the classical prediction for his interferometer
$A^{\rm class}_2={{D}\over{\lambda}} {{(30 {\rm km/s})^2
}\over{c^2}}\sim 0.375$ gives indeed an observable velocity $v_{\rm
obs}\sim (1.8\pm 0.5)$ km/s very close to the $1.5$ km/s value
quoted by Joos. But, when comparing with our prediction in the
stochastic model Eq.(\ref{amplitude10001finalintro}) one would now
find a true kinematical velocity $\tilde v= 305^{+85}_{-100}$ km/s.
Second, when fitting with Eqs.(\ref{nassau1}) and (\ref{projection})
the smooth black curve of the Joos data  in
Fig.\ref{joos-comparison} one finds \cite{plus} a right ascension
$\alpha({\rm fit-Joos})= (168 \pm 30)$ degrees and an angular
declination $\gamma({\rm fit-Joos})= (-13 \pm 14)$ degrees which are
consistent with the present values $\alpha({\rm CMB}) \sim$ 168
degrees and $\gamma({\rm CMB}) \sim -$7 degrees. This confirms that,
when studied at different sidereal times, the measured amplitude can
provide precious information on the angular parameters.

\subsection{Summary of all classical experiments}

A comparison with all classical experiments is finally shown in
Table \ref{summary}.

\begin{table}[htb]
\tbl{The average 2nd-harmonic amplitudes of classical
ether-drift experiments. These were extracted from the original
papers by averaging the amplitudes of the individual observations
and assuming the direction of the local drift to be completely
random (i.e. no vector averaging of different sessions). These
experimental values are then compared with the full statistical
average Eq.(\ref{amplitude10001finalintro}) for a projection 250
km/s $\lesssim \tilde{v} \lesssim$ 370 km/s of the Earth motion in the CMB
and refractivities $\epsilon=2.8\cdot10^{-4}$ for air and
$\epsilon=3.3\cdot10^{-5}$ for gaseous helium. The experimental
value for the Morley-Miller experiment is taken from the observed
velocities reported in Miller's Figure 4, here our Fig.\ref{miller}.
The experimental value for the Michelson-Pease-Pearson experiment
refers to the only known session for which the fringe shifts are
reported explicitly \cite{pease} and where the optical path was
still fifty-five feet. The symbol $\pm ....$ means that the
experimental uncertainty cannot be determined from the available
informations. The table is taken from ref.\cite{universe}.}
{\begin{tabular}{@{}cllll@{}} \toprule
Experiment &gas
&~~~~$A^{\rm EXP}_2$ &~~~ ${{2D}\over{\lambda}}$~~~~& ~~~ $\langle A_2(t)\rangle_{\rm stat} $   \\
\hline
Michelson(1881)               & air     &$ (7.8 \pm....)\cdot10^{-3}$     &~~~$4\cdot 10^6  $   &$(0.7 \pm 0.2)\cdot 10^{-3}$  \\
Michelson-Morley(1887)   & air & $(1.6 \pm 0.6)\cdot 10^{-2 }$&~~~$4\cdot 10^7$ & $(0.7 \pm 0.2)\cdot 10^{-2}$ \\
Morley-Miller(1902-1905)   & air & $(4.0 \pm 2.0)\cdot 10^{-2 }$&~~~$1.12\cdot 10^8$ &$ (2.0 \pm 0.7) \cdot10^{-2}$\\
Miller(1921-1926)  & air& $(4.4 \pm 2.2)\cdot 10^{-2 }$ & ~~  $1.12\cdot 10^8$ &$(2.0 \pm 0.7) \cdot10^{-2} $ \\
Tomaschek (1924) & air & $(1.0\pm 0.6)\cdot 10^{-2 }  $ &~~~$3\cdot 10^7$& $ (0.5 \pm 0.2) \cdot10^{-2} $\\
Kennedy(1926)  & helium & ~~~$<0.002$&~~~$7 \cdot 10^6$&$ (1.4 \pm 0.5)\cdot10^{-4}  $\\
Illingworth(1927) & helium & $ (2.2 \pm 1.7)\cdot 10^{-4}  $  &~~~$7 \cdot 10^6$ &$ (1.4 \pm 0.5)\cdot10^{-4}$ \\
Piccard-Stahel(1928)          &air & $(2.8 \pm 1.5)\cdot10^{-3}$  &~~~$1.28 \cdot 10^7$& $(2.2 \pm 0.8)\cdot10^{-3}$\\
Mich.-Pease-Pearson(1929) & air& $(0.6 \pm...)\cdot10^{-2}$  &~~~$5.8  \cdot 10^7$& $(1.0 \pm 0.4)\cdot10^{-2}$\\
Joos(1930)  &helium&$(1.4 \pm 0.8)\cdot 10^{-3 }   $  & ~~ $7.5 \cdot 10^7$&$(1.5 \pm 0.6)\cdot10^{-3}$\\
\botrule
\end{tabular}}
\label{summary}
\end{table}

Note the substantial difference with the analogous summary Table I
of ref.\cite{shankland} where those authors were comparing with the
classical amplitudes  Eq.(\ref{a2class}) and emphasizing the much
smaller magnitude of the experimental fringes. Here, is just the
opposite. In fact, our theoretical statistical averages are often
{\it smaller} than the experimental results indicating, most likely,
the presence of systematic effects in the measurements.

At the same time, by adopting Eq.(\ref{amplitude10001finalintro}),
we find $\tilde v_{\rm exp} \sim 418 \pm 62 $ km/s from all
experiments in air and $\tilde v_{\rm exp} \sim 323 \pm 70 $ km/s
from the two experiments in gaseous helium, with a global average
$\langle\tilde v_{\rm exp} \rangle\sim 376 \pm 46 $ km/s which
agrees well with the 370 km/s from the CMB observations.  Even more,
from the two most precise experiments of Piccard-Stahel (Brussels
and Mt.Rigi) and Joos (Jena), we find two determinations, $\tilde
v_{\rm exp}= 360^{+85}_{-110} $ km/s and $\tilde v_{\rm exp}=
305^{+85}_{-100} $ km/s respectively, whose average $\langle \tilde
v\rangle \sim 332^{+60}_{-80} $ km/s reproduces to high accuracy the
projection of the CMB velocity at a typical Central-Europe latitude.

\subsection{The intriguing role of temperature}

As anticipated in Sect.2 (see footnote $^k$), symmetry arguments
can successfully describe a phenomenon regardless of the physical
mechanisms behind it. The same is true here with our relation
${{|\Delta\bar{c}_\theta|}\over{c}}\sim \epsilon (v^2/c^2)$ in
Eq.(\ref{bbasic2new}). It works but does {\it not} explain the ultimate origin of
the small effects observed in the gaseous systems. For instance,
as a first mechanism, we considered the possibility of different
polarizations in different directions in the dielectric, depending
on its state of motion. But, if this works in weakly bound gaseous
matter, the same mechanism should also work in a strongly bound
solid dielectric, where the refractivity is $({\cal N}_{\rm solid}
-1)= O(1)$, and thus produce a much larger
${{|\Delta\bar{c}_\theta|}\over{c}}\sim ({\cal N}_{\rm solid} -1)
(v^2/c^2)\sim 10^{-6} $. This is in contrast with the Shamir-Fox
\cite{fox} experiment in perspex where the observed value was
smaller by orders of magnitude.

We have thus re-considered \cite{epl,plus2,book} the traditional
thermal interpretation \cite{joos2,shankland} of the observed
residuals. The idea was that, in a weakly bound system as a gas, a
small temperature difference $\Delta T^{\rm gas}(\theta)$ in the air of the two optical arms produces
a difference in the refractive index and a $(\Delta\bar{c}_\theta/c) \sim \epsilon_{\rm gas} \Delta T^{\rm
gas}(\theta)/T$, where $T\sim$ 300 K is the absolute temperature of
the laboratory \footnote{The starting point is the Lorentz-Lorenz
equation for the molecular polarizability in the ideal-gas
approximation (as for air or gaseous helium at atmospheric
pressure), see \cite{plus2,book} for the details.}. Miller was aware\cite{miller,miller34} that his results could be due to a $\Delta T^{\rm
gas}(\theta) \lesssim$ 1 mK but objected
that casual changes of the ambiance temperature would largely cancel
when averaging over many measurements. Only temperature effects with
a definite periodicity would survive. For a quantitative estimate, by
averaging over all experiments in Table \ref{summary} we found
$\langle\tilde v_{\rm exp} \rangle\sim 376 \pm 46 $ km/s. Therefore,
by comparing Eq.(\ref{amplitude10001finalintro}) with the
corresponding form for a thermal light anisotropy, we find  \BE
\label{thermalrelation} {{|\Delta\bar{c}_\theta|  }
       \over{c }} \sim \epsilon_{\rm gas}{{
\pi^2 }
       \over{18 }} {{ \langle\tilde v_{\rm exp}
\rangle^2 }
       \over{c^2 }} \sim \epsilon_{\rm gas}{{|\Delta T^{\rm gas}(\theta)| }
       \over{T}} \EE
and a value \cite{plus2,book} $|\Delta T^{\rm gas}(\theta)| \sim (0.26
\pm 0.07)$ mK \footnote{Note that in Eq.(\ref{thermalrelation}) the
gas refractivity drops out. The old estimates \cite{joos2,shankland}
of about 1 mK, based on the relation $\epsilon_{\rm gas} \Delta
T^{\rm gas}(\theta)/T\sim (v^2_{\rm Miller}/2c^2)$, with $v_{\rm
Miller} \sim 10$ km/s, were slightly too large.}.

This motivates the following two observations. First, after a
century from those old measurements, in a typical room-temperature laboratory environment,
a stability at the level of a fraction of millikelvin is still state of the art, see \cite{farkas,zhaoa,trusov}.
This would support the idea that we are dealing with a non-local
effect that places a fundamental limit.

Second, as for possible dynamical explanations, we mentioned in footnote
$^k$ a collective interaction of the gaseous
system with hypothetical dark matter in the Galaxy or with the CMB
radiation. For the consistency with the velocity of 370 km/s, the
latter hypothesis seems now more plausible. In this interpretation,
these interactions would be so weak that, on average, the induced
temperature differences in the optical paths are only 1/10 of the
whole $\Delta T^{\rm CMB}(\theta)$ in Eq.(\ref{CBR}).

Nevertheless, whatever its precise origin, this thermal explanation
can help intuition. In fact, it can explain the {\it quantitative}
reduction of the effect in the vacuum limit where $\epsilon_{\rm
gas} \to 0$ and the {\it qualitative} difference with solid
dielectric media where temperature differences of a fraction
of millikelvin cannot produce any appreciable deviation from
isotropy in the rest frame of the medium.

Admittedly, the idea that small modifications of gaseous matter,
produced by the tiny CMB temperature variations, can be detected by
precise optical measurements in a laboratory, while certainly
unconventional, has not the same implications of a genuine
preferred-frame effect due to the vacuum structure. Still, this
thermal explanation of the small residuals in gases, very recently
reconsidered by Manley \cite{manley}, has a crucial importance. In
fact, it implies that if a tiny, but non-zero, fundamental signal
could definitely be detected in vacuum then, with very precise
measurements, the {\it same} universal signal should also show up in
a {\it solid dielectric} where a disturbing $\Delta T$ of a fraction
of millikelvin becomes irrelevant. Detecting such `non-thermal' light
anisotropy, for the same cosmic motion indicated by the CMB
observations, is thus necessary to confirm the idea of a fundamental
preferred frame.

\section{The modern ether-drift experiments}

Searching for a `non-thermal' light anisotropy, which could be
detected with light propagating in vacuum and/or in solid
dielectrics, we will now compare with the modern experiments
\cite{applied} where ${{\Delta\bar{c}_\theta}\over{c}}\sim
{{\Delta\nu(\theta)}\over{\nu_0}}$ is now extracted from the
frequency shift of two optical resonators, see
Fig.\ref{Fig.apparatus}. The particular type of laser-cavity
coupling used in the experiments is known in the literature as the
Pound-Drever-Hall system \cite{pound,PDH}, see Black's tutorial
article \cite{black} for a beautiful introduction. A laser beam is
sent into a Fabry-Perot cavity which acts as a filter. Then, a part
of the output of the cavity is fed back to the laser to suppress its
frequency fluctuations. This method provides a very narrow bandwidth
and has been crucial for the precision measurements we are going to
describe.

The first application to the ether-drift experiments was realized by
Brillet and Hall in 1979 \cite{brillet}. They were comparing the
frequency of a CH$_4$ reference laser (fixed in the laboratory) with
the frequency of a cavity-stabilized He-Ne laser ($\nu_0\sim
8.8\cdot 10^{13}$ Hz) placed on a rotating table. Since the
stabilizing optical cavity was placed inside a vacuum envelope, the
measured shift $\Delta \nu(\theta)$ was  giving a measure of the
anisotropy of the velocity of light in vacuum. The short-term
stability of the cavity-laser system was found to be about $\pm$ 20
Hz for a 1-second measurement, and comparable to the stability of
the reference CH$_4$ laser. It was also necessary to correct the
data for a substantial linear drift of about 50 Hz/s.

By grouping the data in blocks of 10-20 rotations they found a
signal with a typical amplitude $|\Delta \nu|\sim$ 17 Hz (or a
relative level $10^{-13}$) and with a phase $\theta_2(t)$ which was
randomly varying. Therefore, by increasing the statistics and
projecting along the axis corresponding to the Earth cosmic velocity
obtained from the first CMB observations \cite{gorenstein}, the
surviving average effect was substantially reduced down to about
$\pm 1$ Hz. Finally, by further averaging over a period of about 200
days, the residual ether-drift effect was an average frequency shift
$\langle\Delta \nu\rangle=$0.13 $\pm$ 0.22 Hz, i.e. about 100 times
smaller than the instantaneous $|\Delta \nu|$.

Since the 1979 Brillet-Hall article, substantial improvements have
been introduced in the experiments. Just to have an idea, in
present-day measurements \cite{crossed,schiller2015} with vacuum
cavities the typical magnitude of the instantaneous fractional
signal $|\Delta \nu|/\nu_0$ has been lowered from $10^{-13}$ to
$10^{-15}$, the linear drift from 50 Hz/s to about 0.05 Hz/s and,
after averaging over many observations, the best limit which is
reported is a residual $ \langle{{\Delta\nu}\over{\nu_0}}\rangle
\lesssim 10^{-18}$ \cite{schiller2015}, i.e. about 1000 times
smaller than the instantaneous $10^{-15}$ signal.

The assumptions behind the analysis of the data, however, are
basically unchanged. In fact, a genuine ether drift is always
assumed to be a regular phenomenon depending deterministically on
the Earth cosmic motion and averaging more and more observations is
considered a way of improving the accuracy. But, as emphasized in
Sect.4, the classical experiments indicate genuine
physical fluctuations that are {\it not} spurious noise but,
instead, express how the cosmic motion of the Earth is actually seen
in a detector. For this reason, we will first consider the
instantaneous signal and try to understand if it can admit a
physical interpretation.

\subsection{A $10^{-9}$ refractivity for the vacuum (on the Earth surface)}

\begin{figure}[h]
\begin{center}
\includegraphics[width=7.0 cm]{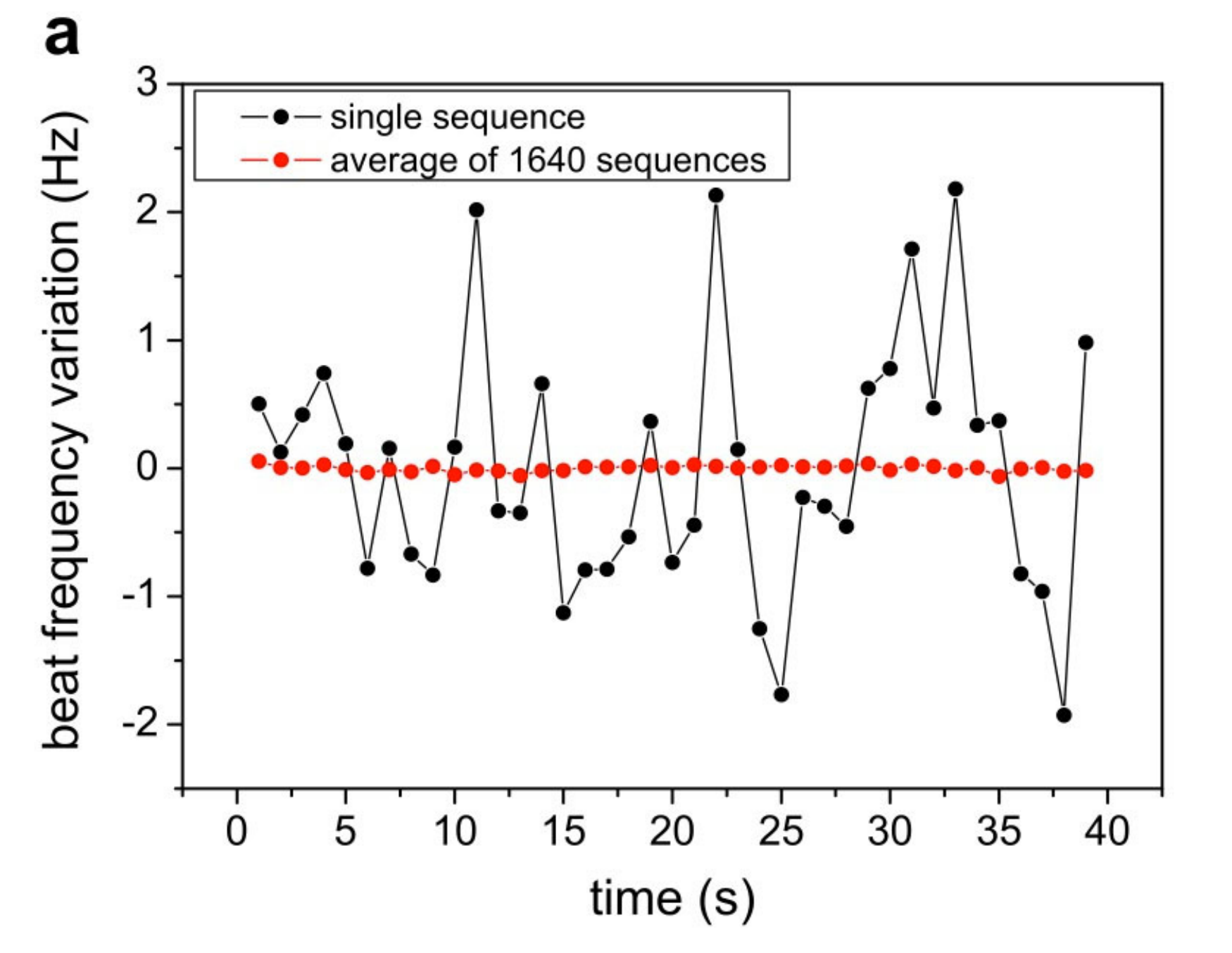}
\end{center}
\caption{\it The experimental frequency shift reported in Fig.9(a)
of ref.\cite{crossed} (courtesy Optics Communications). The black
dots give the instantaneous signal, the red dots give the signal
averaged over 1640 sequences. For a laser frequency $\nu_0=2.8\cdot
10^{14}$ Hz a $\Delta \nu=\pm 1$ Hz corresponds to a fractional
value $\Delta \nu/\nu_0$ of about $\pm 3.5 \cdot 10^{-15}$.}
\label{Figcrossed}
\end{figure}

As anticipated, after averaging many observations, the best limit
which is reported for measurements with vacuum resonators is a
residual $\langle \Delta \nu/\nu_0\rangle \lesssim 10^{-18}$
\cite{schiller2015}. This just reflects the very irregular nature of
the signal because its typical magnitude $|\Delta \nu|/\nu_0\sim
10^{-15}$ is about 1000 times larger, see Fig.\ref{Figcrossed} or
panel b) of Fig.\ref{turbulence_signal}.

The most interesting aspect however is that this $10^{-15}$
instantaneous signal, found in the room-temperature experiments of
refs.\cite{crossed} and \cite{schiller2015}, was also found in
ref.\cite{newberlin} where the solid parts of the vacuum resonators
were made of different material and even in ref.\cite{cpt2013} were
the apparatus was operating in the {\it cryogenic} regime. Since it
is very unlike that spurious effects (e.g. thermal noise
\cite{numata}) remain the same for experiments operating in so
different conditions, one can meaningfully explore the possibility
that such an irregular $10^{-15}$ signal admits a physical
interpretation.

In the same model discussed for the classical experiments, we are
then lead to the concept of a refractive index ${\cal N}_v= 1+
\epsilon_v$ for the vacuum or, more precisely, for the physical
vacuum established in an optical cavity, as in Fig.\ref{CAVITY2},
when this is placed on the Earth surface. The refractivity
$\epsilon_v$ should be at the $10^{-9}$ level, in order to give $
{{|\Delta\bar{c}_\theta|}\over{c}}\sim \epsilon_v~(v^2/c^2)~ \sim
10^{-15}$ and thus would fit with the original idea of \cite{gerg}
where, for an apparatus placed on the Earth's surface, a vacuum
refractivity $\epsilon_v\sim (2G_NM/c^2R) \sim 1.39\cdot 10^{-9}$
was considered, $G_N$ being the Newton constant and $M$ and $R$ the
mass and radius of the Earth. Since this idea will sound
unconventional to many readers, we have first to recall the main
motivations.

An effective refractivity for the physical vacuum becomes a natural
idea when adopting a different view of the curvature effects
observed in a gravitational field. In General Relativity these
curvature effects are viewed as a fundamental modification of
Minkowski space-time. However, it is an experimental fact that many
physical systems for which, at a fundamental level, space-time is
exactly flat are nevertheless described by an effective curved
metric in their hydrodynamic limit, i.e. at length scales much
larger than the size of their elementary constituents. For this
reason, several authors, see e.g. \cite{barcelo1,barcelo2,volo},
have explored the idea that Einstein gravity might represent an
emergent phenomenon and started to considered those gravity-analogs
(moving fluids, condensed matter systems with a refractive index,
Bose-Einstein condensates,...) which are known in flat space.

The main ingredient of this approach consists in the introduction of
some background fields $s_k(x)$ in flat space expressing the
deviations of the effective metric $g_{\mu\nu}(x)$ from the
Minkowski tensor $\eta_{\mu\nu}$, i.e.
\begin{equation} g_{\mu\nu}(x)-\eta_{\mu\nu}=\delta g_{\mu\nu}[s_k(x)]
\end{equation} with $\delta g_{\mu\nu}[s_k=0]=0$. In this type
of approach, to (partially) fill the conceptual gap with classical
General Relativity, as in the original Yilmaz derivation
\cite{yilmaz}, one could impose that Einstein's equations for the
metric become {\it algebraic identities} which follow directly from
the equations of motion for the $s_k$'s in flat space, after
introducing a suitable stress tensor $t^\mu_\nu(s_k)$ \footnote{In
the simplest, original Yilmaz approach \cite{yilmaz} there is only
one inducing-gravity field $s_0(x)$ which plays the role of the
Newtonian potential. Introducing its stress tensor $t^\mu_\nu(s_0)=
-\partial^\mu s_0\partial_\nu s_0 +
1/2\delta^\mu_\nu~\partial^\alpha s_0\partial_\alpha s_0$, to match
the Einstein tensor, produces differences from the Schwarzschild
metric which are beyond the present experimental accuracy. }.

As an immediate consequence, suppose that the $s_k$'s represent {\it
excitations} of the physical vacuum which therefore vanish
identically in the equilibrium state. Then, if curvature effects are
only due to departures from the lowest-energy state, one could
immediately understand \cite{volo} why the huge condensation energy
of the unperturbed vacuum plays no role and thus obtain an intuitive
solution of the cosmological-constant problem found in connection
with the vacuum energy \footnote{This is probably the simplest way
to follow Feynman's indication: ``The first thing we should
understand is how to formulate gravity so that it doesn't interact
with the vacuum energy'' \cite{rule}.}.

Here, in our context of the ether-drift experiments, we will limit
ourselves to explore some phenomenological consequence of this
picture. To this end, let us assume a zeroth-order model of gravity
with a scalar field $s_0(x)$ which behaves as the Newtonian
potential (at least on some coarse-grained scale and consistently
with the experimental verifications of the 1/r law at the
sub-millimeter level \cite{eotwash}). How could the effects of
$s_0(x)$ be effectively re-absorbed into a curved metric structure?
At a pure geometrical level and regardless of the detailed dynamical
mechanisms, the standard basic ingredients would be: 1) space-time
dependent modifications of the physical clocks and rods and 2)
space-time dependent modifications of the velocity of light
\footnote{This point of view can be well represented by some
citations. For instance, ``It is possible, on the one hand, to
postulate that the velocity of light is a universal constant, to
define {\it natural} clocks and measuring rods as the standards by
which space and time are to be judged and then to discover from
measurement that space-time is {\it really} non-Euclidean.
Alternatively, one can {\it define} space as Euclidean and time as
the same everywhere, and discover (from exactly the same
measurements) how the velocity of light and natural clocks, rods and
particle inertias {\it really} behave in the neighborhood of large
masses'' \cite{atkinson}. Or ``Is space-time really curved? Isn't it
conceivable that space-time is actually flat, but clocks and rulers
with which we measure it, and which we regard as perfect, are
actually rubbery? Might not even the most perfect of clocks slow
down or speed up and the most perfect of rulers shrink or expand, as
we move them from point to point and change their orientations?
Would not such distortions of our clocks and rulers make a truly
flat space-time appear to be curved? Yes.''\cite{thorne}.}.

Within this interpretation, one could thus try to check the
fundamental assumption of General Relativity that, even in the
presence of gravity, the velocity of light in vacuum $c_\gamma$ is a
universal constant, namely it remains the same, basic parameter $c$
entering Lorentz transformations. Notice that, here, we are not
considering the so called coordinate-dependent speed of light.
Rather, our interest is focused on the value of the true, physical
$c_\gamma$ as, for instance, obtained from experimental measurements
in vacuum optical cavities placed on the Earth surface.

To spell out the various aspects, a good reference is Cook's article
``Physical time and physical space in general relativity''
\cite{cook}. This article makes extremely clear which definitions of
time and length, respectively $dT$ and $d L$, are needed if all
observers have to measure the same, universal speed of light
(``Einstein postulate''). For a static metric, these definitions are
$dT^2=g_{00} dt^2$ and $dL^2=g_{ij}dx^i dx^j$. Thus, in General
Relativity, the condition $ds^2=0$, which governs the propagation of
light, can be expressed formally as \BE ds^2= c^2dT^2-dL^2=0 \EE
and, by construction, yields the same universal speed $c=dL/dT$.

For the same reason, however, if the physical units of time and
space were instead defined to be $d\hat T$ and $d\hat L$ with, say,
$dT = q~ d\hat T$ and $dL=p~ d\hat L$, the same condition \BE ds^2=
c^2q^2d \hat T^2-p^2 d\hat L^2=0 \EE would now be interpreted in
terms of the different speed \BE c_\gamma={{ d\hat L}\over{ d\hat T
}}=c ~{{q}\over{p}} \equiv {{c}\over { {\cal N}_v }} \EE The
possibility of different standards for space-time measurements is
thus a simple motivation for an effective vacuum refractive index
${\cal N}_v\neq 1$.

With these premises, the unambiguous point of view of Special
Relativity is that the right space-time units are those for which
the speed of light in the vacuum $c_\gamma$, when measured in an
inertial frame, coincides with the basic parameter $c$ entering
Lorentz transformations. However, inertial frames are just an
idealization. Therefore the appropriate realization is to assume
{\it local} standards of distance and time such that the
identification $c_\gamma=c$ holds as an asymptotic relation in the
physical conditions which are as close as possible to an inertial
frame, i.e. {\it in a freely falling frame} (at least by restricting
light propagation to a space-time region small enough that tidal
effects of the external gravitational potential $U_{\rm ext}(x)$ can
be ignored). Note that this is essential to obtain an operational
definition of the otherwise {\it unknown} parameter $c$.

As already discussed in ref.\cite{gerg}, light propagation for an
observer $S$ sitting on the Earth's surface can then be described
with increasing degrees of accuracy starting from step i), through
ii) and finally arriving to iii):

\begin{figure}
\begin{center}
\includegraphics[scale=0.35]{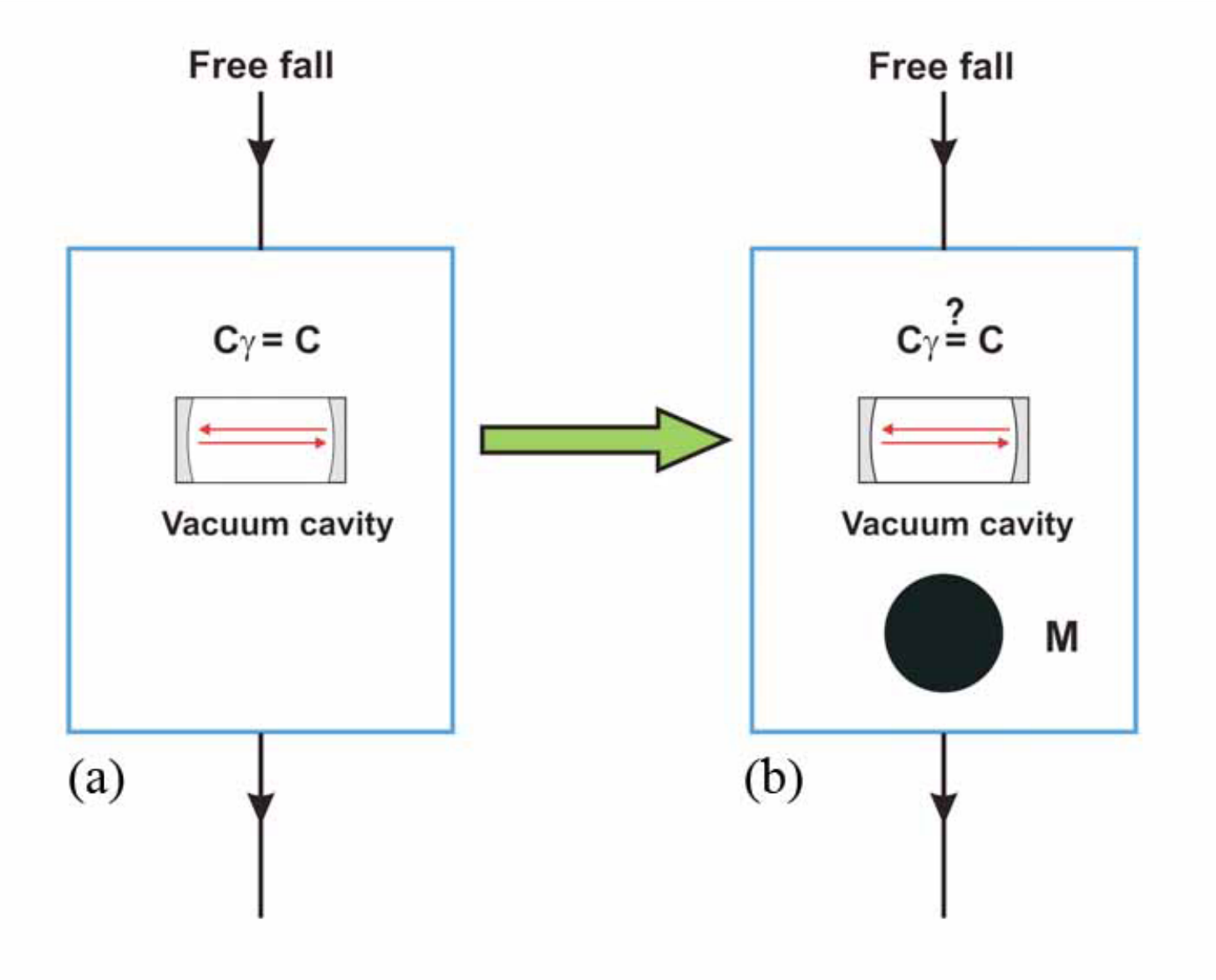}
\end{center}
\caption{ {\it A pictorial representation of the effect of a heavy
mass $M$ carried on board of a freely-falling system, case (b). With
respect to the ideal case (a), the mass $M$ modifies the local
space-time units and could introduce a vacuum refractivity so that
now $c_\gamma \neq c$.}
 } \label{freefall}
\end{figure}

~~~i) $S$ is considered a freely falling frame. This amounts to
assume $c_\gamma=c$ so that, given two events which, in terms of the
local space-time units of $S$, differ by $(dx, dy, dz, dt)$, light
propagation is described by the condition (ff='free-fall')
\begin{equation} \label{zero1} (ds^2)_{\rm ff}=c^2dt^2- (dx^2+dy^2+dz^2)=0~\end{equation}
 \vskip 10 pt ~~~ii) To a
closer look, however, an observer  $S$ placed on the Earth surface
can only be considered a freely-falling frame up to the presence of
the Earth gravitational field. Its inclusion can be estimated by
considering $S$ as a freely-falling frame, in the same external
gravitational field described by $U_{\rm ext}(x)$, that however is
also carrying on board a heavy object of mass $M$ (the Earth mass
itself) which affects the local space-time structure, see
Fig.\ref{freefall}. To derive the required correction, let us denote
by $\delta U$ the extra Newtonian potential produced by the heavy
mass $M$ at the experimental set up where one wants to describe
light propagation. According to General Relativity, and to first
order in $\delta U$, light propagation for the $S$ observer is now
described by
\begin{eqnarray} \label{gr}
ds^2=c^2dt^2 (1-2{{|\delta U|}\over{c^2}})
-(dx^2+dy^2+dz^2)(1+2{{|\delta U|}\over{c^2}}) \equiv c^2 dT^2 -
dL^2=0
\end{eqnarray}
where $dT^2=(1-2{{|\delta U|}\over{c^2}})dt^2 $ and
$dL^2=(1+2{{|\delta U|}\over{c^2}})(dx^2+dy^2+dz^2) $ are the
physical units of General Relativity in terms of which one obtains
the universal value $dL/dT=c_\gamma=c$.

Though, to check {\it experimentally} the assumed identity $c_\gamma
=c$ one should compare with a theoretical prediction for
$(c-c_\gamma)$ and thus {\it necessarily} modify some formal
ingredient of General Relativity. As a definite possibility, let us
maintain the same definition of the unit of length $d\hat L= dL$ but
change the unit of time from $dT$ to $d\hat T$. The reason derives
from the observation that physical units of time scale as inverse
frequencies and that the measured frequencies $\hat \omega$ for $
\delta U \neq 0$, when compared to the corresponding value $\omega$
for $\delta U = 0$, are {\it red-shifted} according to
\begin{equation} \hat \omega= (1-{{|\delta U|}\over{c^2}})~ \omega
\end{equation}
Therefore, rather than the {\it natural} unit of time
$dT=(1-{{|\delta U|}\over{c^2}})dt $ of General Relativity, one
could consider the alternative, but natural (see our footnote $^s$),
unit of time \begin{equation} d\hat T=(1+{{|\delta
U|}\over{c^2}})~dt
\end{equation} Then, to reproduce $ds^2=0$, we can introduce a
vacuum refractive index
\begin{equation} \label{lambda}  {\cal N}_v\sim 1+2{{|\delta U|}\over{c^2}}
\end{equation}
so that the {\it same} Eq.(\ref{gr}) takes now the form
\begin{equation} \label{iso}ds^2 ={{c^2d\hat T^2}\over{{\cal
N}^2_v }}- d \hat L^2=0~\end{equation} This gives
 $d\hat L/d \hat T= c_\gamma= {{c}\over{{\cal N}_v}}$ and, for an observer
placed on the Earth's surface, a refractivity
\begin{equation} \label{refractive0} \epsilon_v= {\cal N}_v - 1 \sim
{{2G_N M}\over{c^2R}} \sim 1.39\cdot 10^{-9}\end{equation} $M$ and
$R$ being respectively the Earth mass and radius.

Notice that, with this natural definition $d\hat T$, the vacuum
refractive index associated with a Newtonian potential is the same
usually reported in the literature, at least since Eddington's 1920
book \cite{eddington}, to explain in flat space the observed
deflection of light in a gravitational field. The same expression is
also suggested by the formal analogy of Maxwell equations in General
Relativity with the electrodynamics of a macroscopic medium with
dielectric function and magnetic permeability \cite{volkov}
$\epsilon_{ik}=\mu_{ik}=\sqrt{-g}~{{(-g^{ik})}\over{g_{00} }}$.
Indeed, in our case, from the relations $g^{il}g_{lk}= \delta ^i_k $
, $(-g^{ik}) \sim \delta ^i_k ~g_{00}$ ,
$\epsilon_{ik}=\mu_{ik}=\delta ^i_k {\cal N}_v$ , we obtain
\begin{equation}   {\cal N}_v \sim \sqrt{-g} \sim \sqrt{
(1-2{{|\delta U|}\over{c^2}})(1+2{{|\delta U|}\over{c^2}})^3} \sim
1+2{{|\delta U|}\over{c^2}} \end{equation} A difference is found
with Landau's and Lifshitz' textbook \cite{landaufield} where the
vacuum refractive index entering the constitutive relations is
instead defined as ${\cal N}_v \sim {{1}\over{\sqrt{g_{00} }}}\sim
1+{{|\delta U|}\over{c^2}}$. Concerning, these two possible
definitions of ${\cal N}_v$, we address the reader to Broekaert's
article \cite{broekaert}, see his footnote 3, where a very complete
set of references for the vacuum refractive index in gravitational
field is reported. However, this difference of a factor of 2 is not
really essential and can be taken into account as a theoretical
uncertainty. The main point is that $c_\gamma$, as measured in a
vacuum cavity on the Earth's surface (panel {\bf (b)} in our
Fig.\ref{freefall}), could differ at a fractional level $10^{-9}$
from the ideal value $c$, as operationally defined with the same
apparatus in a true freely-falling frame (panel {\bf (a)} in our
Fig.\ref{freefall}). In conclusion, this $c_\gamma - c$ difference
can be conveniently expressed through a vacuum refractivity of the
form
\begin{equation} \label{refractive2} \epsilon_v={\cal N}_v - 1 \sim
{{\chi}\over{2}}~ 1.39\cdot 10^{-9} \end{equation} where the factor
$\chi/2$ (with $\chi$= 1 or 2) takes into account the mentioned
theoretical uncertainty. \vskip 10 pt ~~~iii) Could one check
experimentally if ${\cal N}_v \neq$ 1? Today, the speed of light in
vacuum is assumed to be a fixed number with no error, namely 299 792
458 m/s. Thus if, for instance, this estimate were taken to
represent the value measured on the Earth surface, in an ideal
freely-falling frame there could be a slight increase, namely
$+{{\chi}\over{2}}(0.42)$ m/s with $\chi=$ 1 or 2. It seems hopeless
to measure unambiguously such a difference because the uncertainty
of the last precision measurements performed before the `exactness'
assumption had precisely this order of magnitude, namely $\pm 4\cdot
10^{-9}$ at the 3-sigma level or, equivalently, $\pm 1.2$ m/s
\cite{nist}.

Therefore, as pointed out in ref.\cite{gerg}, an experimental test
cannot be obtained from the value of the isotropic speed in vacuum
but, rather, from its possible {\it anisotropy}. In fact, with a
preferred frame and for ${\cal N}_v\neq 1$, an isotropic propagation
as in Eq.(\ref{iso}) would only be valid for a special state of
motion of the Earth laboratory. This provides the definition of
$\Sigma$ while for a non-zero relative velocity there would be off
diagonal elements $g_{0i}\neq 0$ in the effective metric
\cite{volkov}. If $\Sigma$ exists, we would then expect a light
anisotropy ${{|\Delta \bar{c}_\theta| }\over{c}}\sim \epsilon_v
(v/c)^2 \sim 10^{-15}$, consistently with the presently measured
value.

\subsection{Some important technical aspects}

\begin{figure}
\begin{center}
\includegraphics[width=8.5 cm]{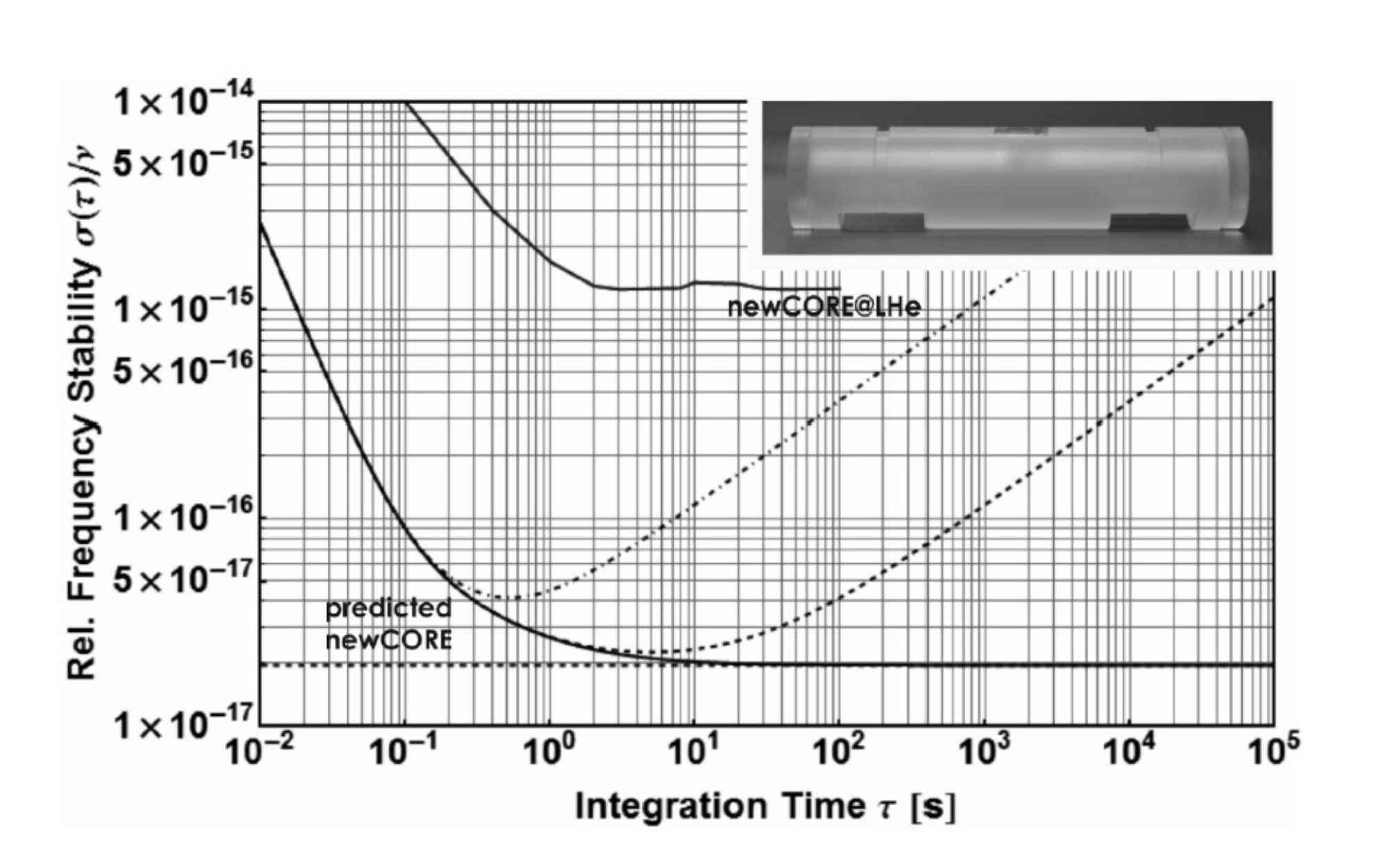}
\end{center}
\caption{\it The stability of the fractional frequency shift of
ref.\cite{cpt2013} at different integration times. The upper solid
curve, denoted as `newCORE', reports the actual measurements with
the cryogenic apparatus in 2013. The lower solid, dashed and
dot-dashed curves, denoted as `predicted newCORE', indicate future
stability limits $(2\div 4) \cdot 10^{-17}$ that could be foreseen
at that time. } \label{singola}
\end{figure}

Before considering the experiments, however, a rather technical
discussion is necessary for an in-depth comparison with the data. In
the mentioned cryogenic experiment of ref.\cite{cpt2013}, the
instantaneous signal is not shown explicitly. However, its magnitude
can be deduced from its typical variation observed over a
characteristic time of 1$\div$2 seconds, see Fig.\ref{singola}. For
a very irregular signal, in fact, this typical variation, of about
$10^{-15}$, gives the magnitude of the instantaneous signal itself
and, indeed, it is in good agreement with the mentioned
room-temperature measurements.

\begin{figure}[h]
\begin{center}
\includegraphics[width=8.0 cm]{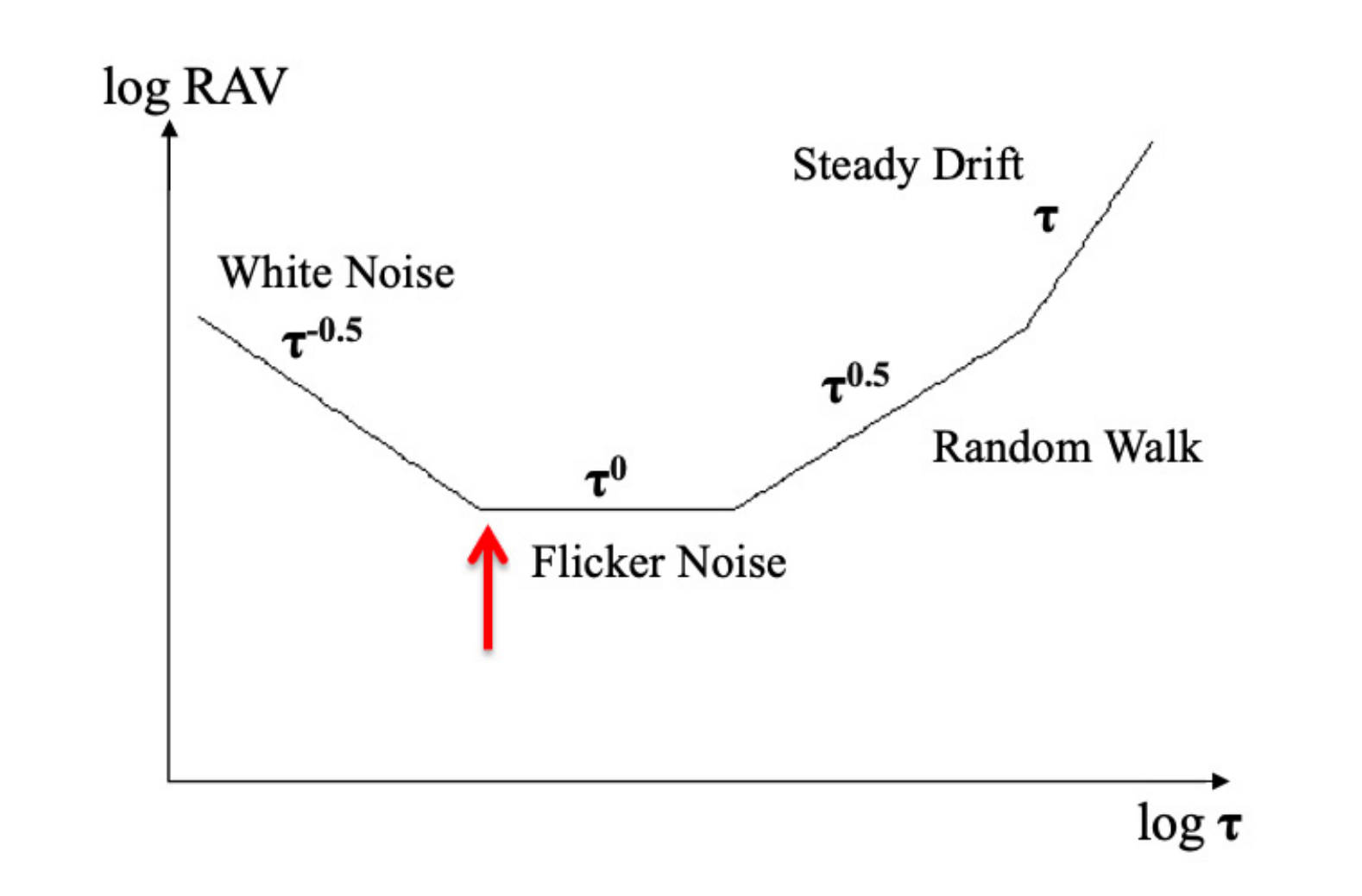}
\end{center}
\caption{\it The typical trend of the RAV for a signal in various
regimes. The minimum of the white-noise trend $\tau^{-0.5}$ defines
the value  $\tau=\bar \tau$ indicated by the arrow. }
\label{RAV-tau}
\end{figure}
The quantity which is reported in Fig.\ref{singola} is the Root
Square of the Allan Variance (RAV) of the fractional frequency
shift. In general, the RAV describes the variation obtained by
sampling a function $f=f(t)$ at steps of time $\tau$. By defining
\BE {\overline f}(t_i;\tau)={{1}\over{\tau }} \int^{t_i+\tau }_{t_i}
dt~f(t)\equiv {\overline f}_i \EE one generates a $\tau-$dependent
distribution of ${\overline f}_i$ values. In a large time interval
$\Lambda= M\tau$, the RAV is then defined as \BE RAV(f,\tau)=
\sqrt{RAV^2(f,\tau)} \EE where \BE RAV^2(f,\tau)= {{1}\over{2(M-1)}}
\sum^{M-1}_{i=1} \left({\overline f}_i-{\overline f}_{i+1} \right)^2
\EE  and the factor of 2 is introduced to obtain the standard
variance $\sigma(f)$ for uncorrelated data with zero mean, as for a pure white-noise
signal.

Note that the actual measurements in Fig.\ref{singola} are indicated
by the upper solid curve denoted as `newCORE'. These were obtained
with the cryogenic apparatus in 2013 (CORE=Cryogenic Optical
REsonators) and were giving a stability at the level of about
$1.2\cdot 10^{-15}$. The lower solid, dashed and dot-dashed curves,
denoted as `predicted newCORE', indicate instead possible improved
limits $(2\div 4) \cdot 10^{-17}$ that could be foreseen at that
time. As a matter of fact, these limits have not yet been achieved
because the highest stability limits are still larger by an order of
magnitude. This persistent signal, which is crucial for our work,
does not depend on the absolute temperature and/or the
characteristics of the optical cavities \cite{nagelthesis}.

After this preliminaries, we then arrive at our main point. As
anticipated, numerical simulations in our stochastic model indicate
that our basic signal has the same characteristics as a universal
white noise. This means that it should be compared with the
frequency shift of two optical resonators at the largest integration
time $\bar \tau$ {\it where the pure white-noise component is as
small as possible} but other disturbances, that can affect the
measurements, are not yet important, see Fig.\ref{RAV-tau}.  In the
experiments we are presently considering this $\bar\tau$ is
typically $1\div 2$ seconds so that one gets the relation with the average magnitude of the
instantaneous signal
\BE 
RAV(\Delta \nu,\bar\tau) \sim  \sigma(\Delta \nu)  \sim \langle|\Delta \nu|\rangle_{\rm stat}
\label{averagedelta}
\EE

\subsection{Comparing our model with experiments in vacuum}

\begin{figure}[h]
\begin{center}
\includegraphics[scale=0.30]{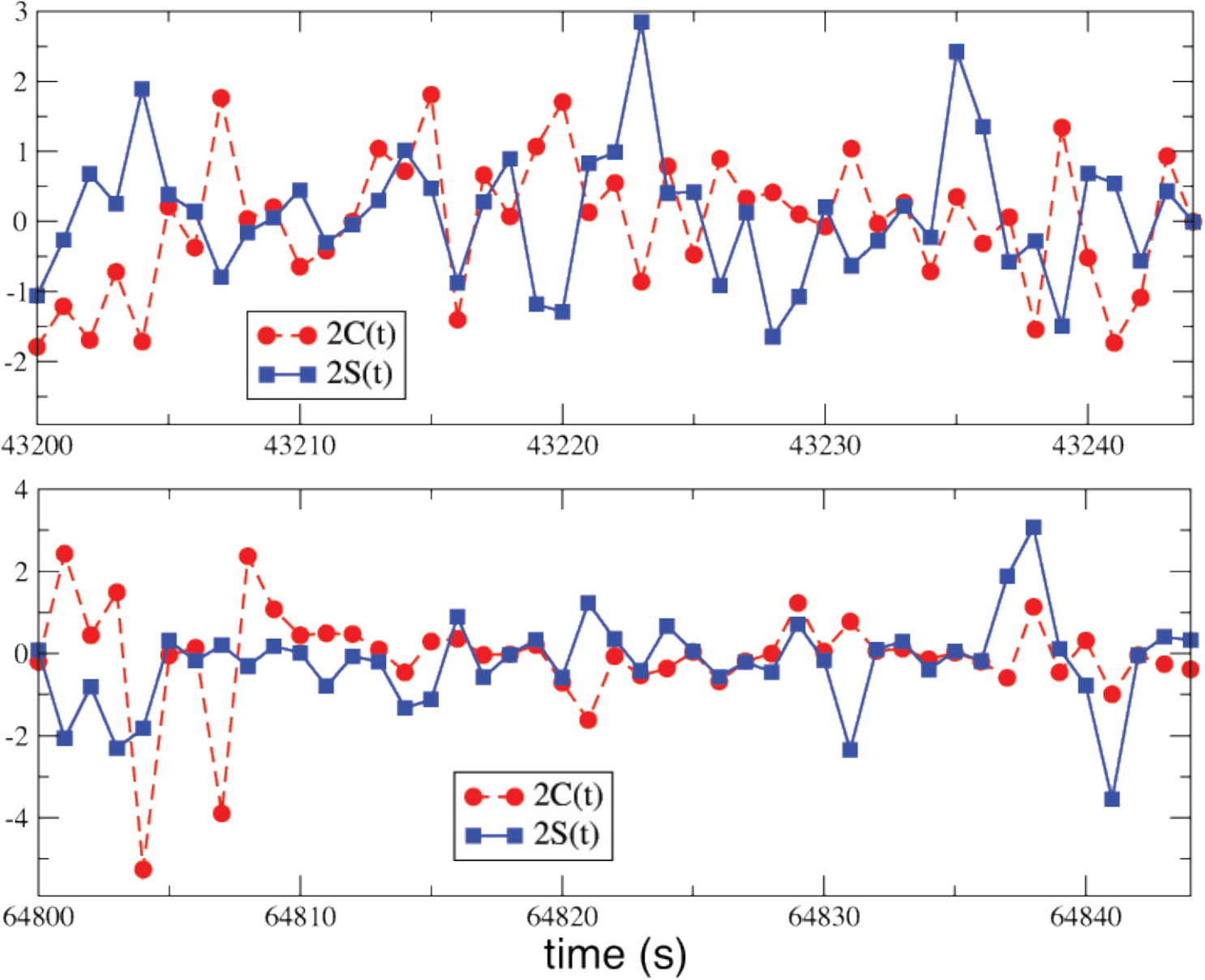}
\end{center}
\caption{\it For $\epsilon_v$ as in Eq.(\ref{refractive2}) and
$\chi=2$, we report a simulation of two sequence of 45 seconds for
the functions $2C(t)$ and $2S(t)$ Eqs.(\ref{amplitude10old}). Units
are $10^{-15}$ and the two sets belong to the same random sequence
for two sidereal times that differ by 6 hours. The boundaries of the
stochastic velocity components, Eqs.(\ref{vx}) and (\ref{vy}) of the
Appendix, are controlled by $(V,\alpha,\gamma)_{\rm CMB}$ through
Eqs.(\ref{projection}) and (\ref{isot}). For a laser frequency of
$2.8\cdot 10^{14}$ Hz, the range $\pm 3.5\cdot 10^{-15}$ corresponds
to a typical frequency shift $\Delta \nu$ in the range $\pm 1$ Hz,
as in our Fig.\ref{Figcrossed}.} \label{rotation}
\end{figure}

We will now compare with the type of signal observed in
\cite{crossed,schiller2015} in vacuum at room temperature. To this
end, we will use the relation which connects the frequency shift
between two orthogonal resonators
${\Delta\nu(\theta;t)=\nu_1(\theta;t) - \nu_2(\theta+\pi/2;t)}$ to
the angular dependence of the velocity of light, namely see
(\ref{basictext}) \BE {{\Delta\nu(\theta;t)}\over{\nu_0}} = {{\Delta
\bar{c}_\theta(t) } \over{c}} = 2{S}(t)\sin 2\theta +
      2{C}(t)\cos 2\theta \end{equation}
where $S(t)$ and  $C(t)$ are given in Eqs.(\ref{amplitude10old}). As
in the case of the classical experiments, the velocity components
$v_x(t)$ and $v_y(t)$ will be expressed as random Fourier series
through the Eqs.(\ref{vx}) and (\ref{vy}) of the Appendix. A
simulation of two short-time sequences of $2C(t)$ and $2S(t)$ is
shown in Fig.\ref{rotation}.

For a quantitative test, we concentrated on the observed value of
the RAV of the frequency shift at the end point $\bar \tau= 1\div 2$
seconds of the white-noise branch of the spectrum, see Fig.3, bottom
part of \cite{schiller2015} . This has a value
\begin{equation} \left[ RAV(\Delta \nu,\bar \tau) \right]_{\rm
exp}= (0.20 \div 0.24)~ {\rm Hz}
\end{equation}
or, in units of the reference frequency $\nu_0=2.8\cdot 10^{14}$ Hz~~  \cite{schiller2015}
\begin{equation} \label{ravschiller2015}\left[RAV(\frac{ \Delta \nu}{\nu_0},\bar\tau) \right]_{\rm exp} =(7.8 \pm 0.7) \cdot 10^{-16}~~~{\rm Vacuum-room~temperature}
\end{equation}

As anticipated, our instantaneous, stochastic signal for $\Delta
\nu(t)$ is, to very good approximation, a pure white noise for which
the RAV coincides with the standard variance. At the same time, for
a very irregular signal with zero mean of the type shown in Fig.
\ref{rotation}, but whose magnitude can have a long-term time dependence, one should replace in Eq. (\ref{averagedelta})  $\langle|\Delta \nu|\rangle_{\rm stat} \to \langle|\Delta \nu(t)|\rangle_{\rm stat}$ and evaluate the RAV in the corresponding temporal range. Therefore,
from ${{\Delta \nu }\over{\nu_0}} =
      {{\Delta \bar{c}_\theta } \over{c}} \sim \epsilon_v \cdot{{v^2
} \over{c^2 }}$, we arrive at our prediction
\begin{equation} \label{deltanuth}
\left[RAV(\frac{ \Delta \nu(t)}{\nu_0},\bar\tau) \right]_{\rm
theor}\sim \epsilon_v~ \frac{\langle v^2_x(t)+ v^2_y(t)\rangle_{\rm
stat}}{c^2} ~\sim {{\pi^2 } \over{18 }} \cdot \epsilon_v \cdot{{V^2
} \over{c^2 }} \sin^2 z(t)
\end{equation}
Then, by using Eq.(\ref{refractive2}), for the projection $\tilde{v}(t) = V|\sin z(t)| = $ $250\div370$ Km/s used for the classical
experiments, our prediction for the RAV can be expressed as
\begin{equation} \label{allanth}\left[RAV(\frac{ \Delta \nu}{\nu_0},\bar\tau) \right]_{\rm
theor}~\sim {{\chi}\over{2}}\cdot(8.5 \pm 3.5) \cdot 10^{-16}
\end{equation}
with $\chi=$ 1 or 2.
By comparing with the experimental Eq.(\ref{ravschiller2015}), the
data favour $\chi=2$, which is the only free parameter of our
scheme. Also, the good agreement with our theoretical value indicates
that, at the end point of the white-noise part of the signal, the
corrections to our simplest model should be small.

Notice, however, that the range in Eq.(\ref{allanth}) is not a
theoretical uncertainty but reflects the daily variations of $
V^2\sin^2 z(t)$ in Eq.(\ref{deltanuth}). This means that, depending
on the sidereal time, the measurements of the RAV  at the
white-noise end point $\tau=\bar\tau$ should exhibit definite daily
variations in the range (for $\chi=2$)\BE\label{dailyrav} 5\cdot
10^{-16} \lesssim~\left[ RAV(\frac{ \Delta
\nu}{\nu_0},\bar\tau)\right]_{\rm theor} \lesssim~ 12\cdot 10^{-16}
\EE Thus it becomes crucial to understand whether these variations
can be observed.

\subsection{Comparing our model with experiments in solids}

To consider modern experiments in solid dielectrics, we will compare
with the very precise work of ref.\cite{nagelnature}. This is a
cryogenic experiment, with microwaves of 12.97 GHz, where almost all
electromagnetic energy propagates in a medium, sapphire, with
refractive index of about 3 (at microwave frequencies). As
anticipated, with a thermal interpretation of the residuals in
gaseous media, we expect that the fundamental $10^{-15}$ vacuum
signal considered above, with very precise measurements, should also
become visible here. In particular, the large refractivity of the
solid ${\cal N}_{\rm solid} -1=$ O(1) should play no role.

Following refs.\cite{plus2,book,universe}, we first observe that for
${\cal N}_v=1 +\epsilon_v$ there is a very tiny difference between
the refractive index defined relatively to the ideal vacuum value
$c$ and the refractive index relatively to the physical isotropic
vacuum value $c/{\cal N}_v$ measured on the Earth surface. The
relative difference between these two definitions is proportional to
$\epsilon_v \lesssim 10^{-9} $ and, for all practical purposes, can
be ignored. All materials would now exhibit, however, the same
background vacuum anisotropy. To this end, let us replace the
average isotropic value \BE {{c}\over { {\cal N}_{\rm solid}}} \to
{{c}\over { {\cal N}_v {\cal N}_{\rm solid} }}\EE and then  use
Eq.(\ref{nbartheta}) to replace ${\cal N}_v$ in the denominator with
its $\theta-$dependent value \BE \bar {\cal N}_v(\theta) \sim  1+
\epsilon_v\beta^2(1+\cos^2\theta)\EE This is equivalent to define a
$\theta-$dependent refractive index for the solid dielectric
\begin{eqnarray}\label{6introsolid}
        {{  \bar{\cal N}_{\rm solid}(\theta)}\over { {\cal N}_{\rm solid}}} \sim  1+\epsilon_v \beta^2(1
        +
       \cos^2\theta)
\end{eqnarray}
so that
\begin{equation}
\label{refractivetheta1} \left[ {\bar c_\gamma (\theta)}
\right]_{\rm solid}={{c}\over{\bar{\cal N}_{\rm
solid}(\theta)}} \sim {{c}\over { {\cal N}_{\rm solid}}} \left[ 1-
\epsilon_v \beta^2 (1 +
       \cos^2\theta)\right]
\end{equation}
with an anisotropy
\begin{equation}
\label{anysolid} {{ \left[\Delta\bar{c}_\theta\right]_{\rm solid}}
\over {\left[ c/ {\cal N}_{\rm solid}\right] }} \sim \epsilon_v
\beta^2 \cos2\theta \sim 10^{-15}
\end{equation}
In this way, a genuine $10^{-15}$ vacuum effect, if there, could
also be detected in a solid dielectric thus implying the same
prediction Eq.(\ref{allanth}).

\begin{figure}[h]
\begin{center}
\includegraphics[scale=0.30]{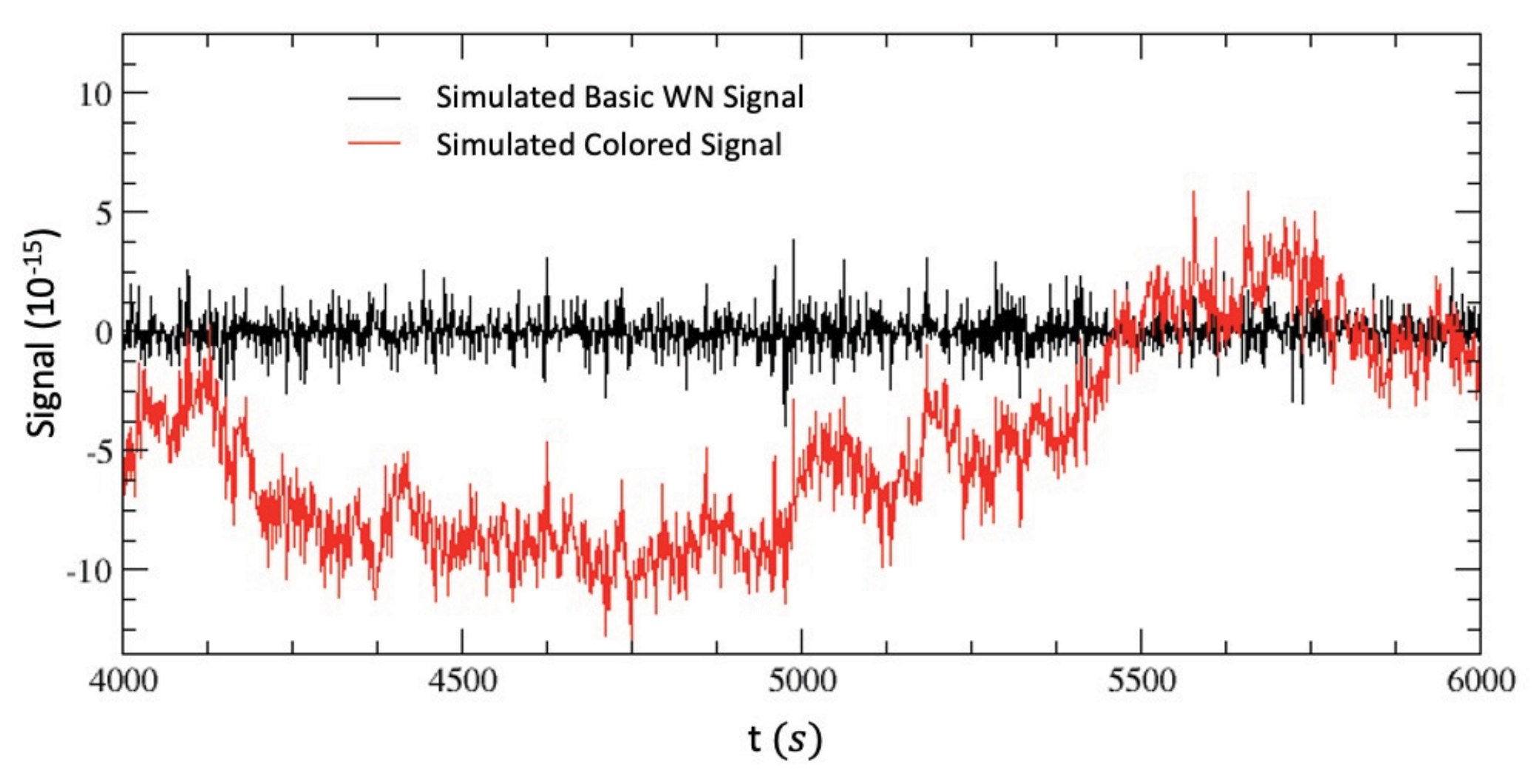}
\end{center}
\caption{\it We report two typical sets of 2000 seconds for our
basic white-noise (WN) signal and its colored version obtained by
Fourier transforming the spectral amplitude of
ref.\cite{nagelnature}. The boundaries of the random velocity
components Eqs.(\ref{vx}) and (\ref{vy}) were defined by
Eq.(\ref{isot}) by plugging in Eq.(\ref{projection}) the CMB
kinematical parameters, for a sidereal time $t=4000-6000$
seconds and for the latitude of Berlin-Duesseldorf, see the
Appendix. The figure is taken from ref.\cite{universe}.}
\label{colored}
\end{figure}

In ref.\cite{universe}, a detailed comparison with
\cite{nagelnature} was performed. First, from Figure 3(c) of
\cite{nagelnature}, see also panel b) of our
Fig.\ref{turbulence_spectrum}, it is seen that the spectral
amplitude of this particular apparatus becomes flat at frequencies
$\omega \ge 0.5$ Hz indicating that the end-point of the white-noise
branch of the signal is at an integration time $\bar\tau \sim 1\div 2$
seconds. The data for the spectral amplitude were then fitted to an
analytic, power-law form to describe the lower-frequency part 0.001
Hz $\leq \omega \leq 0.5$ Hz which reflects apparatus-dependent
disturbances. This fitted spectrum was then used to generate a
signal by Fourier transform. Finally, very long sequences of this
signal were stored to produce ``colored'' version of our basic
white-noise signal.

\begin{figure}[h]
\begin{center}
\includegraphics[scale=0.30]{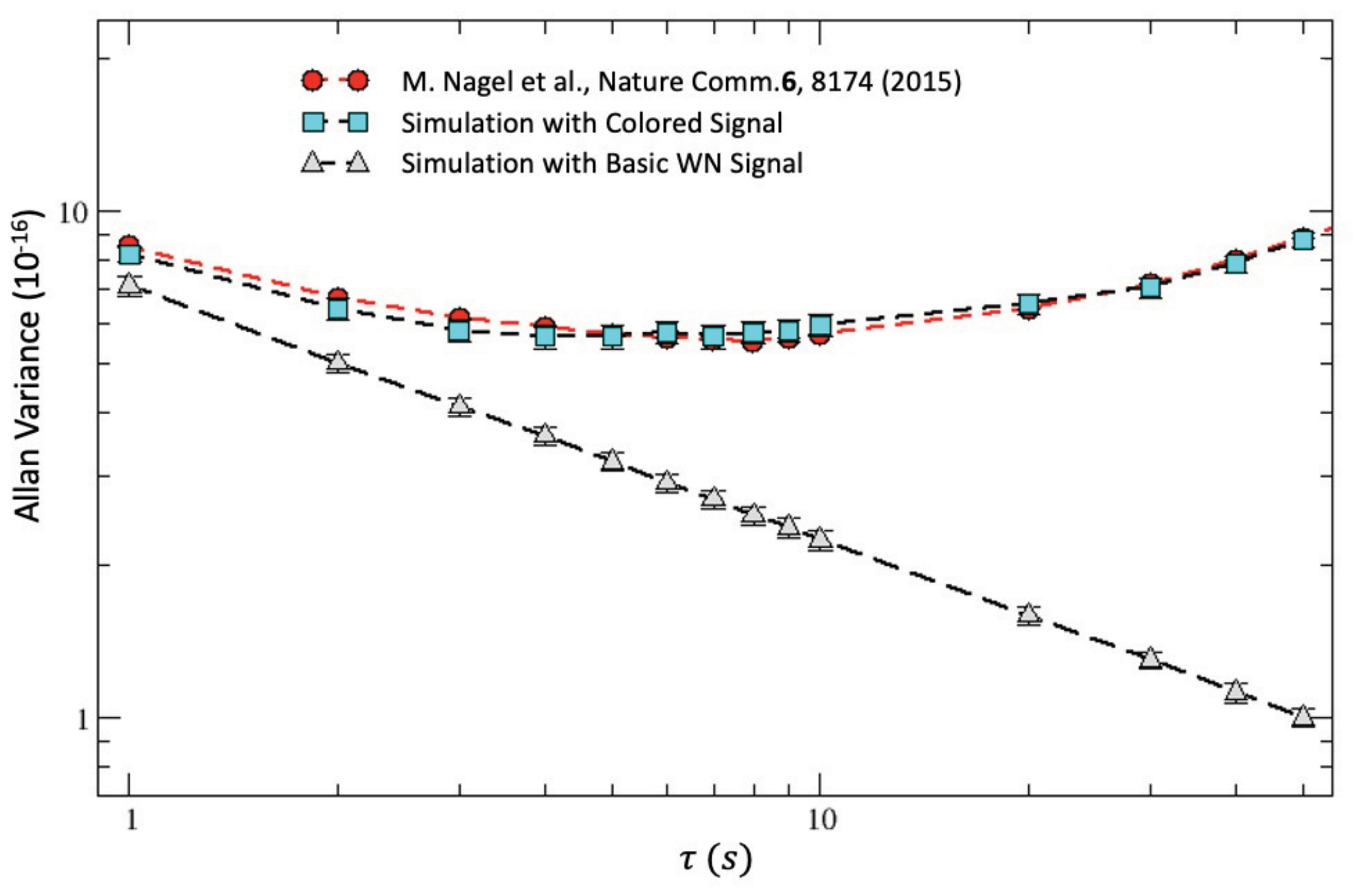}
\end{center}
\caption{\it We report the RAV for the fractional frequency shift
obtained from many simulations of sequences of 2000 seconds for our
basic white-noise (WN) signal (decreasing as $\tau^{-0.5}$) and for
its colored version, see Fig.\ref{colored}. The direct experimental
results of ref.\cite{nagelnature}, for the non-rotating setup, are
also shown as red dots. The figure is taken from
ref.\cite{universe}.} \label{allancomparison}
\end{figure}

To get a qualitative impression of the effect, we report in
Fig.\ref{colored} a sequence of our basic simulated white-noise signal and a
sequence of its colored version. By averaging over many 2000-second
sequences of this type, the corresponding RAV's for the two simulated signals
are then reported in Fig.\ref{allancomparison}. The experimental RAV
extracted from Figure 3(b) of ref.\cite{nagelnature} is also
reported (for the non-rotating setup). At this stage, the agreement
of our simulated, colored signal with the experimental data remains
satisfactory only up $\tau=$ 50 seconds. Reproducing the signal at
larger $\tau$'s would have required further efforts but this is not
relevant, our scope being just to understand the modifications of
our stochastic signal near the endpoint of the white-noise spectrum.

As one can check from Fig.3(b) of ref.\cite{nagelnature}, see also
the red dots in our Fig.\ref{allancomparison}, the experimental RAV
for the fractional frequency shift, at the white-noise end point
$\bar\tau \sim 1\div 2$ second, is in the range $(6.8\div 8.6)\cdot
10^{-16}$, say~~ \cite{nagelnature}
\BE \label{ravnagelnature} \left[RAV(\frac{ \Delta
\nu}{\nu_0},\bar\tau) \right]_{\rm exp} =(7.7 \pm 0.9) \cdot
10^{-16} ~~~~~{\rm Solid-cryogenic}
\EE

As such, it coincides with
Eq.(\ref{ravschiller2015}) that we extracted from
ref.\cite{schiller2015} after normalizing their experimental result
$RAV(\Delta \nu,\bar\tau)_{\rm exp}=0.20\div 0.24 $ Hz to their
laser frequency $\nu_0=2.8 \cdot 10^{14}$ Hz. At the same time, it
is well consistent with our theoretical prediction
Eq.(\ref{allanth}) for $\chi=2$. Therefore this beautiful agreement,
between ref.\cite{schiller2015} (a vacuum experiment at room
temperature) and ref.\cite{nagelnature} (a cryogenic experiment in a
solid dielectric), on the one hand, and with our theoretical
prediction Eq.(\ref{allanth}), on the other hand, confirms our
interpretation of the data in terms of a stochastic signal
associated with the Earth cosmic motion within the CMB and
determined by the vacuum refractivity $\epsilon_v$
Eq.(\ref{refractive2}), for $\chi=2$.

Two ultimate experimental checks still remain. First, as
anticipated, one should try to detect our predicted, daily
variations Eq.(\ref{dailyrav}). Due to the excellent systematics,
these variations should remain visible with both experimental
setups. Second, one more complementary test should be performed by
placing the vacuum (or solid dielectric) optical cavities on board
of a satellite, as in the OPTIS proposal \cite{optis}. In this ideal
free-fall environment, as in panel (a) of our Fig.\ref{freefall},
the typical instantaneous frequency shift should be much smaller (by
orders of magnitude) than the corresponding $10^{-15}$ value
measured with the same interferometers on the Earth surface.

\section{Summary and outlook}

In this paper, we started from the present, basic ambiguity concerning the version of relativity
which is {\it physically realized} in nature, namely Einstein Special Relativity vs. 
a Lorentzian formulation with a preferred reference frame $\Sigma$. 
This ambiguity is usually presented by
a two-step argument. First, the basic quantitative 
ingredients, namely Lorentz transformations, are the same in both formulations. 
Second, even in a Lorentzian formulation, Michelson-Morley experiments can only
produce null results. Therefore, rather than introducing an
experimentally unobservable and logically superfluous entity, it
seemed more satisfactory to adopt the point of view of Special
Relativity where those effects (length contraction and time dilation), that
were at the base of the original Lorentzian formulation, so to speak, 
become part of the kinematics. In this way, relativity
becomes axiomatic and extendable beyond the original domain 
of the electromagnetic phenomena. This wider perspective has been the main reason for the 
traditional supremacy given to Einstein's view. 

However, discarding all historical aspects, it was emphasized by Bell that a change of perspective, from
Special Relativity to a Lorentzian formulation, could be crucial to
reconcile hypothetical faster-than-light signals with causality, as
with the apparent non-local aspects of the Quantum Theory. In
addition, the present view of the lowest-energy state as a Bose
condensate of elementary quanta (Higgs particles, quark-antiquark
pairs, gluons...), indicates a vacuum structure with some degree of
substantiality which could characterize non trivially the form of
relativity which is physically realized in nature. So, there may be
good reasons for a preferred reference frame but, without the possibility of detecting 
experimentally an `ether wind' in laboratory, the difference between the two formulations remains
a philosophical problem. 

This impossibility-in-principle, however, is somewhat mysterious.
While it is certainly true that evidence for both the
undulatory and corpuscular aspects of radiation has substantially
modified the consideration of an underlying ethereal medium, yet, if
$\Sigma$ exists, only a `conspiracy' of relativistic effects would
make undetectable our motion with respect to it. But this conspiracy
works exactly if the velocity of light $c_\gamma$ propagating
in the various interferometers, or more precisely its two-way
combination  $\bar c_\gamma$, coincides with the basic parameter $c$
entering Lorentz transformations. Therefore if $\bar c_\gamma\neq
c$, as for instance in the presence of matter, where light gets
absorbed and then re-emitted, nothing would really prevent an
angular dependence $\Delta \bar c_\theta=\bar c_\gamma(\pi/2
+\theta) - \bar c_\gamma(\theta)\neq 0$. If an angular dependence can be
detected, and correlated with the cosmic motion of the Earth, the
long sought $\Sigma$ tight to the CMB could finally emerge.

We have thus recalled the two key
points of our extensive work. First, one should impose that all
measurable effects vanish exactly in the $\bar c_\gamma \to c$
limit, i.e. in the {\it ideal} vacuum limit of a refractive index
${\cal N}=1$.  Instead, in
the infinitesimal region ${\cal N}=1 + \epsilon$ simple symmetry
arguments  lead to the relation ${{|\Delta \bar
c_\theta|}\over{c}}\sim \epsilon(v^2/c^2)$. For a typical cosmic
$v\sim$ 300 km/s and $\epsilon=2.8\cdot10^{-4}$ , for air, or
$\epsilon=3.3\cdot10^{-5}$,  for gaseous helium, this reproduces the
order of magnitude of the effects observed in the classical
experiments.

The other peculiar aspect of our analysis concerns the observed,
irregular character of the data that, giving often substantially different
directions of the drift at the same hour on consecutive days, were
contradicting the traditional expectation of a regular phenomenon
completely determined by the cosmic motion of the Earth. As we have
emphasized, here again, there may be a logical gap. The relation between
the macroscopic motion of the Earth and the microscopic propagation
of light in a laboratory depends on a complicated chain of effects
and, ultimately, on the physical nature of the vacuum. By comparing
with the motion of a body in a fluid, the standard view corresponds
to a form of regular, laminar flow where the projection $\tilde v_\mu(t)$ of the global, cosmic velocity, at the site 
of the experiment, coincides with the local $ v_\mu(t)$ that
determines the signal in the plane of the interferometer. Instead,
some general arguments and some experimental analogies suggest that
the {\it physical vacuum} might rather resemble a turbulent fluid
where large-scale and small-scale flows are only related {\it
indirectly}. In this different perspective, with forms of
turbulence that, as in most models, become statistically isotropic
at small scales, the local $v_\mu(t)$ would fluctuate randomly
within boundaries fixed by
the global $\tilde v_\mu(t)$ (see the Appendix). Therefore, one
should analyze the data in phase and amplitude (giving respectively
the instantaneous direction and magnitude of the drift) and
concentrate on the latter which is positive definite and remains
non-zero under any averaging procedure. In this way, by restricting
to the amplitudes, experiments always believed in contradiction with
each other, as Miller vs. Piccard-Stahel, become consistent, see
Fig.\ref{overlap}. Most notably, by adopting the parameters
$(V,\alpha,\gamma)_{\rm CMB}$ to fix the boundaries of the local
random $v_\mu(t)$ in our stochastic model, one finds a good description of the 
irregular behaviour of the amplitudes extracted from Joos' very precise observations
 (see Figs.\ref{joos-comparison} and
\ref{joos-comparison-errors}). Viceversa, by fitting Joos' amplitudes with
Eqs.(\ref{nassau1}) and (\ref{projection}), one
finds a right ascension $\alpha({\rm fit-Joos})= (168 \pm 30)$
degrees and an angular declination $\gamma({\rm fit-Joos})= (-13 \pm
14)$ degrees which are well consistent with the present values
$\alpha({\rm CMB}) \sim$ 168 degrees and $\gamma({\rm CMB}) \sim -$7
degrees. The summary of all classical experiments 
given in Table \ref{summary} shows the complete consistency
with our theoretical predictions. 

To conclude our analysis of the classical experiments in gaseous
systems, we have emphasized that our basic relation
${{|\Delta\bar{c}_\theta|}\over{c}}\sim \epsilon_{\rm gas}
(v^2/c^2)$ derives from general, symmetry arguments and does {\it
not} explain the ultimate origin of the tiny observed residuals. Due
to the consistency with the velocity of 370 km/s, a plausible
explanation consists in a collective interaction of gaseous matter
with the CMB radiation. This could bring the gas out of equilibrium
as if there were an effective temperature difference, $|\Delta T^{\rm
gas}(\theta)|= 0.2\div 0.3$ mK, in the gas along the two optical
paths. This magnitude is slightly smaller than the value of
about 1 mK considered by Joos and Shankland and, being just a small fraction  of the whole $\Delta T^{\rm
CMB}(\theta) = \pm 3.3$ mK in Eq.(3), indicates the weakness of the collective gas-CMB interactions. 
Most notably, the thermal interpretation leads to an important prediction. In fact, it implies
that if a physical signal could definitely be detected in vacuum
then, with very precise measurements, the same signal should also
show up in a {\it solid dielectric} where disturbing temperature
differences of a fraction of millikelvin become irrelevant.
Detecting such `non-thermal' light anisotropy, through the combined
analysis of the modern experiments in vacuum and in solid
dielectrics, for the same cosmic motion indicated by the classical
experiments, is thus necessary to confirm the idea of a fundamental
preferred frame.

Despite the much higher precision of modern experiments, the
assumptions behind the analysis of the data are basically the same
as in the classical experiments. A genuine signal is assumed to
be a regular phenomenon, depending deterministically on the Earth
cosmic motion, so that averaging more and more observations is
considered a way of improving the accuracy. But the classical experiments
indicate genuine physical fluctuations which are not spurious noise and, instead, 
express how the cosmic motion of the Earth is actually seen in a detector. 
Therefore, the present quoted
average, namely ${{\langle \Delta \bar c_\theta \rangle}\over{c}}
\lesssim 10^{-18}$, could just reflect the very irregular nature of
the signal. Indeed, its typical instantaneous magnitude in vacuum
${{|\Delta \bar c_\theta|}\over{c}}
\sim 10^{-15}$ is about 1000 times larger, see
Fig.\ref{Figcrossed} or panel b) of Fig.\ref{turbulence_signal}.

To understand if this vacuum signal can
admit a physical interpretation, a crucial
observation is that the same $10^{-15}$ magnitude is found in
measurements where the resonators are made of different materials,
in measurements at room-temperature and also in the {\it cryogenic}
regime. Since it is very unlike that spurious effects remain the same in so different
conditions, in the same model used for the classical experiments we are driven to the idea of a refractive index ${\cal N}_v= 1+
\epsilon_v$ for the vacuum or, more precisely, for the physical
vacuum established in an optical cavity placed on the Earth
surface. The refractivity $\epsilon_v$ should be at the $10^{-9}$
level, in order to give $ {{|\Delta\bar{c}_\theta|}\over{c}}\sim
\epsilon_v~(v^2/c^2)~ \sim 10^{-15}$ and thus would fit with the
original idea of ref.\cite{gerg}. The motivation was that, if
Einstein's gravity is a phenomenon which emerges, at some small
length scale, from a fundamentally flat space, for an apparatus
placed on the Earth surface (which is in free fall with respect to
all masses in the Universe but not with respect to the Earth, see
Fig.\ref{freefall}) there should be a tiny vacuum refractivity
$\epsilon_v\sim (2G_NM/c^2R) \sim 1.39\cdot 10^{-9}$, where $G_N$ is
the Newton constant and $M$ and $R$ are the mass and radius of the
Earth. This is the same type of refractivity considered by
Eddington, or much more recently by Broekaert, to explain in flat
space the deflection of light in a gravitational field. Therefore
Michelson-Morley experiments, by detecting a light anisotropy $
{{|\Delta\bar{c}_\theta|}\over{c}}\sim \epsilon_v~(v^2/c^2)~ \sim
10^{-15}$, can also resolve this other ambiguity.

With this identification of $\epsilon_v$, we first compared qualitatively the
observed signal, in Fig.\ref{Figcrossed} or in panel b) of
Fig.\ref{turbulence_signal}, with simulations in our stochastic
model, see Figs.\ref{rotation} and \ref{colored}. For a more
quantitative analysis, we then considered the value of a particular
statistical indicator which is used nowadays, namely the Allan
Variance of the fractional frequency shift $RAV(\frac{ \Delta
\nu}{\nu_0},\tau)$ as function of the integration time $\tau$. Since
the irregular signal of our stochastic model has the characteristics of a universal white noise
and should represent an irreducible component, we have thus
compared with the $RAV$ measured at the {\it end point} of the
white-noise branch of the spectrum. This is defined as the largest
integration time $\bar \tau$ where the white-noise component is
as small as possible but other spurious disturbances, that can
affect the measurements, are not yet important, see
Fig.\ref{RAV-tau}. In this way, for the same velocity range $\tilde
v = 250\div 370$ km/s used for the classical
experiments, our theoretical prediction Eq.(\ref{allanth}) (for
$\chi=2$) is in very good agreement with the results of the most
precise experiment in vacuum Eq.(\ref{ravschiller2015}).

But, then, the second crucial test. As anticipated, if this $10^{-15}$ signal
observed in vacuum has a real physical meaning, the same effect should also be detected with a
very precise experiment in
a solid dielectric, see Eq.(\ref{anysolid}). This expectation is
confirmed by the extraordinary agreement between
Eq.(\ref{ravnagelnature}) and  Eq.(\ref{ravschiller2015}).
Note that the two experiments are completely different because in
ref.\cite{nagelnature} light propagates in a solid in the cryogenic
regime and in ref.\cite{schiller2015} light propagates in vacuum at
room temperature. As such, there is a plenty of systematic
differences. Yet, the two experiments give exactly the same signal
{\it at the white-noise end point}. Therefore, there must be an
ubiquitous form of white noise that admits a definite physical
interpretation. Our theoretical prediction Eq.(\ref{allanth}) is, at
present, the only existing explanation. Together with the classical experiments, we thus conclude
that there is now an alternative scheme challenging the traditional `null
interpretation' of Michelson-Morley experiments, always presented as a  self-evident scientific
truth.

We have also discussed two further experimental tests. First, one
should try to detect our predicted, daily variations
Eq.(\ref{dailyrav}). Second, one should also try to place the optical cavities on a satellite, as
in the OPTIS proposal \cite{optis}. In this ideal free-fall
environment, as in panel (a) of our Fig.\ref{freefall}, the typical
instantaneous frequency shift should be much smaller (by orders of
magnitude) than the corresponding $10^{-15}$ value measured with the
same interferometers on the Earth's surface.

\vskip 25 pt

\centerline{\Large \bf Appendix} \vskip 10 pt In this appendix, we
will summarize the stochastic model used in
refs.\cite{plus,plus2,book,universe} to compare with experiments. To
make explicit the time dependence of the signal let us first
re-write Eq.(\ref{bbasic2new}) as \begin{equation} \label{basic2}
     {{\Delta \bar{c}_\theta(t) } \over{c}}
    \sim
 \epsilon {{v^2(t) }\over{c^2}}\cos 2(\theta
-\theta_2(t)) \end{equation}  where $v(t)$ and $\theta_2(t)$
indicate respectively the instantaneous magnitude and direction of
the drift in the $(x,y)$ plane of the interferometer. This can also
be re-written as
\begin{equation} \label{basic3} {{\Delta \bar{c}_\theta(t) } \over{c}}\sim
2{S}(t)\sin 2\theta +
      2{C}(t)\cos 2\theta \end{equation} with \begin{equation} \label{amplitude10}
       2C(t)= \epsilon~ {{v^2_x(t)- v^2_y(t)  }
       \over{c^2}}~~~~~~~2S(t)=\epsilon ~{{2v_x(t)v_y(t)  }\over{c^2}}
\end{equation}  and $v_x(t)=v(t)\cos\theta_2(t)$, $v_y(t)=v(t)\sin\theta_2(t)$

As anticipated in Sect.3, the standard assumption to analyze the
data has always been based on the idea of regular modulations of the
signal associated with a cosmic Earth velocity. In general, this is
characterized by a magnitude $V$, a right ascension $\alpha$ and an
angular declination $\gamma$. These parameters can be considered
constant for short-time observations of a few days where there are
no appreciable changes due to the Earth orbital velocity around the
sun. In this framework, where the only time dependence is due to the
Earth rotation, the traditional identifications are $v(t)\equiv
\tilde v(t)$ and $\theta_2(t)\equiv\tilde\theta_2(t)$ where $\tilde
v(t)$ and $\tilde\theta_2(t)$ derive from the simple application of
spherical trigonometry \cite{nassau}
\begin{equation} \label{nassau1}
       \cos z(t)= \sin\gamma\sin \phi + \cos\gamma
       \cos\phi \cos(\tau-\alpha)
\end{equation} \begin{equation} \label{projection}
       \tilde {v}(t) =V \sin z(t)
\end{equation} \begin{equation} \label{nassau2}
    \tilde{v}_x(t) = \tilde{v}(t)\cos\tilde\theta_2(t)= V\left[ \sin\gamma\cos \phi -\cos\gamma
       \sin\phi \cos(\tau-\alpha)\right]
\end{equation} \begin{equation} \label{nassau3}
      \tilde{v}_y(t)= \tilde{v}(t)\sin\tilde\theta_2(t)= V\cos\gamma\sin(\tau-\alpha) \end{equation}
 Here $z=z(t)$ is the zenithal distance of
${\bf{V}}$, $\phi$ is the latitude of the laboratory,
$\tau=\omega_{\rm sid}t$ is the sidereal time of the observation in
degrees ($\omega_{\rm sid}\sim {{2\pi}\over{23^{h}56'}}$) and the
angle $\theta_2$ is counted conventionally from North through East
so that North is $\theta_2=0$ and East is $\theta_2=90^o$. With the
identifications $v(t)\equiv \tilde v(t)$ and
$\theta_2(t)\equiv\tilde\theta_2(t)$, one thus arrives to the simple
Fourier decomposition \begin{equation} \label{amorse1}
      S(t)\equiv {\tilde S}(t) =S_0+
      {S}_{s1}\sin \tau +{S}_{c1} \cos \tau
       + {S}_{s2}\sin(2\tau) +{S}_{c2} \cos(2\tau)
\end{equation}
\begin{eqnarray}
 \label{amorse2}
       C(t)\equiv {\tilde C}(t)=
       {C}_0 + {C}_{s1}\sin \tau +{C}_{c1} \cos \tau
       + {C}_{s2}\sin(2 \tau) +{C}_{c2} \cos(2 \tau)
\end{eqnarray}
where the  $C_k$ and $S_k$ Fourier coefficients depend on the three
parameters $(V,\alpha,\gamma)$ and are given explicitly in
refs.\cite{plus,book}.

Though, the identification of the instantaneous quantities $v_x(t)$
and $v_y(t)$ with their counterparts $\tilde{v}_x(t)$ and
$\tilde{v}_y(t)$ is not necessarily true. As anticipated in Sect.3,
one could consider the alternative situation where the velocity
field is a non-differentiable function and adopt some other
description, for instance a formulation in terms of random Fourier
series \cite{onsager,landau,fung}. In this other approach, the
parameters of the macroscopic motion are used to fix the typical
boundaries for a microscopic velocity field which has an intrinsic
non-deterministic nature.

The model adopted in refs.\cite{plus,plus2,book,universe}
corresponds to the simplest case of a turbulence which, at small
scales, appears homogeneous and isotropic. The analysis can then be
embodied in an effective space-time metric for light propagation
\begin{equation} \label{random} g^{\mu\nu}(t) \sim \eta^{\mu\nu} + 2
\epsilon v^\mu(t) v^\nu(t) \end{equation} where $v^\mu(t)$ is a
random 4-velocity field which describes the drift and whose
boundaries depend on a smooth field $\tilde{v}^\mu(t)$ determined by
the average Earth motion.

For homogeneous turbulence a series representation, suitable for
numerical simulations of a discrete signal, can be expressed in the
form
\begin{equation} \label{vx} v_x(t_k)= \sum^{\infty}_{n=1}\left[
       x_n(1)\cos \omega_n t_k + x_n(2)\sin \omega_n t_k \right] \end{equation}
\begin{equation} \label{vy} v_y(t_k)= \sum^{\infty}_{n=1}\left[
       y_n(1)\cos \omega_n t_k + y_n(2)\sin \omega_n t_k \right] \end{equation}
Here $\omega_n=2n\pi/T$ and T is the common period of all Fourier
components. Furthermore, $t_k= (k-1) \Delta t$, with $k=1, 2...$,
and $\Delta t$ is the sampling time. Finally, $x_n(i=1,2)$ and
$y_n(i=1,2)$ are random variables with the dimension of a velocity
and vanishing mean. In our simulations, the value $T=T_{\rm day}$=
24 hours and a sampling step $\Delta t=$ 1 second were adopted.
However, the results would remain unchanged by any rescaling $T\to s
T$ and $\Delta t\to s \Delta t$.

In general, we can denote by $[-d_x(t),d_x(t)]$ the range for
$x_n(i=1,2)$ and by $[-d_y(t),d_y(t)]$ the corresponding range for
$y_n(i=1,2)$. Statistical isotropy would require to impose $d_x(t)=
d_y(t)$. However, to illustrate the more general case, we will first
consider $d_x(t) \neq d_y(t)$.

If we assume that the random values of $x_n(i=1,2)$ and $y_n(i=1,2)$
are chosen with uniform probability, the only non-vanishing
(quadratic) statistical averages are
\begin{equation} \label{quadratic} \langle x^2_n(i=1,2)\rangle_{\rm
stat}={{d^2_x(t) }\over{3 ~n^{2\eta}}}~~~~~~~~~~~\langle
y^2_n(i=1,2)\rangle_{\rm stat}={{d^2_y(t) }\over{3 ~n^{2\eta}}}
\end{equation} Here, the exponent $\eta$ ensures
finite statistical averages  $\langle v^2_x(t)\rangle_{\rm stat}$
and $\langle v^2_y(t)\rangle_{\rm stat}$ for an arbitrarily large
number of Fourier components. In our simulations, between the two
possible alternatives $\eta=5/6$ and $\eta=1$ of ref.\cite{fung}, we
have chosen $\eta=1$ that corresponds to the Lagrangian picture in
which the point where the fluid velocity is measured is a wandering
material point in the fluid.

Finally, the connection with the Earth cosmic motion is obtained by
identifying $d_x(t)=\tilde v_x(t)$ and $d_y(t)=\tilde v_y(t)$ as
given in Eqs. (\ref{nassau1})$-$(\ref{nassau3}). If, however, we
require statistical isotropy, the relation
\begin{equation} \label{correct} \tilde{v}^2_x(t) +
\tilde{v}^2_y(t)=\tilde{v}^2(t)\end{equation} requires the
identification
\begin{equation} \label{isot} d_x(t)=d_y(t)={{ \tilde{v}(t)}\over{\sqrt{2} }} \end{equation}
 For such isotropic model, by combining
Eqs.(\ref{vx})$-$(\ref{isot}) and in the limit of an infinite
statistics, one gets
\begin{eqnarray}
\label{vanishing} \langle v^2_x(t)\rangle_{\rm stat}=\langle
v^2_y(t)\rangle_{\rm stat}={{\tilde{v}^2(t)}\over{2}}~{{1}\over{3}}
\sum^{\infty}_{n=1} {{1}\over{n^2}}= {{\tilde{v}^2(t)}\over{2}}~
{{\pi^2}\over{18}}\nonumber \\ \langle v_x(t)v_y(t)\rangle_{\rm
stat}=0
\end{eqnarray}
and  vanishing statistical averages
\begin{equation} \label{vanishing2}\langle C(t)\rangle_ {\rm
stat}=0~~~~~~~~~~~~~~~~~~\langle S(t)\rangle_ {\rm stat}=0
\end{equation} at {\it any} time $t$, see Eqs.(\ref{amplitude10}).
Therefore, by construction, this model gives a definite non-zero
signal but, if the same signal were fitted with Eqs.(\ref{amorse1})
and (\ref{amorse2}), it would also give average values $(C_k)^{\rm
avg}=0$, $(S_k)^{\rm avg}=0$ for the Fourier coefficients.

To understand how radical is the modification produced by
Eqs.(\ref{vanishing2}), we recall the traditional procedure adopted
in the classical experiments. One was measuring the fringe shifts at
some given sidereal time on consecutive days so that changes of the
orbital velocity were negligible. Then, see Eqs.(\ref{newintro1})
and (\ref{basic3}), the measured shifts at the various angle
$\theta$ were averaged
\begin{equation} \label{averagefringe}\langle{{\Delta
\lambda(\theta;t)}\over{\lambda}}\rangle_ {\rm stat}=
{{2D}\over{\lambda}} \left[2\sin 2\theta~\langle S(t)\rangle_ {\rm
stat} + 2\cos 2\theta~\langle C(t)\rangle_ {\rm stat} \right]
\end{equation} and finally these average values were compared with
models for the Earth cosmic motion.

However if the signal is so irregular that, by increasing the number
of measurements, $\langle C(t)\rangle_ {\rm stat} \to 0$ and
$\langle S(t)\rangle_ {\rm stat} \to 0$ the averages
Eq.(\ref{averagefringe}) would have no meaning. In fact, these
averages would be non vanishing just because the statistics is
finite. In particular, the direction $\theta_2(t)$ of the drift
(defined by the relation $\tan2\theta_2(t)= S(t)/C(t)$) would vary
randomly with no definite limit.

This is why one should concentrate the analysis on the 2nd-harmonic
amplitudes
\begin{equation} \label{AA} A_2(t)={{2D}\over{\lambda}}~
2\sqrt{S^2(t) + C^2(t)} \sim {{2 D }\over{\lambda}}~ \epsilon {{
v^2_x(t)+v^2_y(t)} \over{c^2 }}
\end{equation}
which are positive-definite and remain non-zero under the averaging
procedure. Moreover, these are rotational-invariant quantities and
their statistical average \BE \label{amplitude10001finalapp1}
\langle A_2(t)\rangle_{\rm stat} \sim  {{2 D }\over{\lambda}} \cdot
{{\pi^2 } \over{18 }}\cdot \epsilon\cdot \frac{V^2 \sin^2 z(t)}{c^2}
\EE would remain unchanged in the isotropic model Eq.(\ref{isot}) or
with the alternative choice $d_x(t)\equiv \tilde v_x(t)$ and
$d_y(t)\equiv \tilde v_y(t)$. Analogous considerations hold for the
modern experiments where $\frac{\Delta \bar{c}_\theta(t)}{c}$ is
extracted from the frequency shift of two optical resonators. Again,
the $C(t)$ and $S(t)$ obtained, through Eq.(\ref{basic3}), from the
very irregular, measured signal (see e.g. Fig.\ref{Figcrossed}), are
compared with the slowly varying parameterizations
Eqs.(\ref{amorse1}) and (\ref{amorse2}) to extract the $C_k$ and
$S_k$ Fourier coefficients. Then, by comparing with our simulation
of the $C(t)$ and $S(t)$ in Fig.\ref{rotation}, it is no surprise if
the average values $(C_k)^{\rm avg} \to 0$, $(S_k)^{\rm avg} \to 0$
by simply increasing the number of observations.


\end{document}